\documentclass[review, sort&compress, 1p]{elsarticle}

\usepackage{amsmath}
\usepackage{amsfonts,color}
\usepackage{amssymb,float}
\usepackage{mathtools}
\usepackage[utf8]{inputenc}
\usepackage{graphicx}
\usepackage{wasysym}
\usepackage{tabularx}
\usepackage{accents}
\usepackage{graphicx}
\usepackage{color}
\usepackage[dvipsnames]{xcolor}
\usepackage{hyperref}
\usepackage{ulem}

\hypersetup{
    colorlinks=true,
    linkcolor=blue,
    filecolor=magenta,      
    citecolor=red
}

\usepackage{subfigure} 
\usepackage{hyperref} 
\usepackage{multirow}

\journal{}

\begin{document}

\begin{frontmatter}

\title{Parametric and nonparametric methods hint dark energy evolution}


\author[a]{Reginald Christian Bernardo\corref{mycorrespondingauthor}}
\address[a]{Institute of Physics, Academia Sinica, Taipei 11529, Taiwan}
\cortext[mycorrespondingauthor]{Corresponding author}
\ead{reginaldchristianbernardo@gmail.com}

\author[b]{Daniela Grand\'on}
\address[b]{Grupo de Cosmolog\'ia y Astrof\'isica Te\'orica, Departamento de F\'isica, FCFM, Universidad de Chile, Blanco Encalada 2008, Santiago, Chile}
\ead{daniela.grandon@ug.uchile.cl}

\author[c,d]{Jackson Levi Said}
\ead{jackson.said@um.edu.mt}
\address[c]{Institute of Space Sciences and Astronomy, University of Malta, Malta, MSD 2080}
\address[d]{Department of Physics, University of Malta, Malta, MSD 2080}

\author[e]{V\'ictor H. C\'ardenas}
\ead{victor.cardenas@uv.cl}
\address[e]{Instituto de F\'isica y Astronom\'ia, Universidad de Valpara\'iso, Av. Gran Bretaña 1111, Valpara\'iso, Chile}

\begin{abstract}
We study dark energy through the viewpoints of parametric and nonparametric analyses of late-time cosmological data. We consider four Hubble parameter priors reflecting the Hubble tension and make use of two phenomenological functions, namely, a normalized dark energy density and a compactified dark energy equation of state. We predict the shape of both functions and present new constraints on the dark energy equation of state. The results hint at dark energy evolution regardless of the choice of the method and of the priors. The fact that similar evolutions for the dark energy densities are found through drastically different approaches suggests that the features found in this paper are driven by the data, and are not artifact of the reconstruction methods applied.
\end{abstract}


\end{frontmatter}



\section{Introduction}
\label{sec:intro}

The $\Lambda$CDM model is the parametrically simplest, and arguably the most successful, cosmological model to date \cite{Smith:2007rg,2011PhRvL.107b1301D,2012PhRvL.109d1101H,Martin:2012bt}. However, on the fundamental side, there are concerns, for one, relying on a cosmological constant $\Lambda$ to support the current accelerating phase of cosmic expansion whereas its theoretical value differs from the observed one by several order of magnitude \cite{RevModPhys.61.1}. It also leaves unanswered: ``What is the origin of the cosmological constant $\Lambda$?'' \cite{Dymnikova:2000gnk, Mukhopadhyay:2007ed} and ``Why is this constant coincidentally of the same order of magnitude as the matter contribution today?'' \cite{Zlatev:1998tr, Malquarti:2003hn}. In addition to these longstanding theoretical problems, the tension between the early \cite{Aghanim:2018eyx, DES:2021wwk} and late \cite{Scolnic:2017caz, Riess:2019, Riess:2019cxk, Riess:2020fzl} Universe measurements of the cosmological parameters now has become an open problem prompting to revisit the cornerstones of modern cosmology such as the cosmological principle \cite{Krishnan:2021dyb} and the general theory of relativity \cite{DiValentino:2021izs, Schoneberg:2021qvd}. With these considerations, it becomes more important to openly explore the alternatives to the standard model of cosmology such as when $\Lambda$ is replaced by an evolving dark energy (DE) component, one whose nature remains yet to be determined.

There are various frameworks in which DE can be inherently dynamical. Most of these fall under the wing of scalar-tensor theories wherein a scalar field acts as both DE in the cosmological arena and a measure of the deviation of the underlying physical model from general relativity \cite{Clifton:2011jh, DeFelice:2011bh, Kase:2018aps, Kobayashi:2019hrl,CANTATA:2021ktz,Bahamonde:2021gfp}. In some models, a vector may also take the place of the scalar \cite{DeFelice:2016yws,Nicosia:2020egv}, or sometimes, scalar and vector fields may coexist, their interplay determining the overall cosmological dynamics \cite{Skordis:2020eui,Skordis:2021vuk}.

Given these models, a reasonable way to proceed is therefore to test each one with observational data. And, indeed, a lot of progress have been made in this direction leading to a selection of observationally competitive theories of DE \cite{Renk:2017rzu, Peirone:2017vcq, Peirone:2019aua, Frusciante:2019puu, Aoki:2020oqc, Anagnostopoulos:2021ydo, Atayde:2021pgb,Schoneberg:2021qvd}. However, the number of viable theoretical models may, in principle, also be quite large and testing new models is often taxing and computationally expensive. Theory-agnostic frameworks, on the other hand, offer a refreshing take to studying DE phenomenology \cite{2010PhRvD..81h3537S, Haridasu:2018gqm, Liao:2020zko, LHuillier:2018rsv, Teng:2021cvy}, and is the direction to be considered in this work. Adding new substance to this approach, it was recently shown how modified gravity can lead to preferred directions in the parameter space when viewed as dynamical DE \cite{Wen:2021bsc}.

Parametric approaches in cosmology are well known to be sourced by proposals of new physics such as modified gravity and dark matter or DE models. Instead, nonparametric approaches offer a new way by which a physics independent setup can be established where elements in a data set are related together statistically. Thus, instead of using data sets to fit an a priori model, we now use observational data to train a statistical model to eventually offer reconstructions of cosmological functions from which cosmological parameters can be inferred. In this work, we consider a joint parametric and nonparametric statistical analysis of late-time cosmological observations \cite{Rani:2015lia, Jesus:2017cyo,Li:2014yza,2014RAA....14.1221Z}. This naturally ties in with the theme of theory-agnostic frameworks as both approaches can be implemented without specifying a cosmological model. We look at DE in light of recent data using both implementations, and analyze whether there are quantifiable departures from the $\Lambda$CDM model that emerge regardless of the stark differences of the methods applied. Our goal is to therefore establish a baseline where a sensible assessment of both can be made, and, most importantly, to draw conclusions about the nature of DE from both methods.

For our parametric analysis (Section \ref{subsec:parametric}), we consider a form of the DE function $[ X(z)=\rho_{\text{DE}}(z)/\rho_{\text{DE}}(0) ]$ in terms of a set of free parameters to be constrained using observational data \cite{Wang:2001ht, Wang_2004, Wang:2004ru, Cardenas:2014jya, Wang:2018fng}. This is an alternative to the traditional parametric methods beginning with the DE equation of state (DE EoS) $w(z)$ \cite{Chevallier:2000qy, Linder:2002et} but with the advantage of better preserving the dynamical information in DE specifically since the luminosity distance is two integrals away from $w(z)$. Once the best fit is obtained, we reconstruct the DE phenomenological functions such as $X(z)$ and the DE equation of state (DE EoS) which quantify the deviation of DE from otherwise being a stale, constant $\Lambda$. We invoke both quadratic and cubic parametrizations of the DE function that have been studied recently in Ref. \cite{Grandon_2021}.

On the other hand, we tackle the problem of elucidating the nature of DE by using Gaussian processes (GP) \cite{rasmussen:2003} (Section \ref{subsec:nonparametric}), our representative nonparametric approach that utilizes a covariance function (or the kernel) that relates observable points in the data to make predictions on an entire range of points. This method has been widely used in cosmology \cite{Seikel2012,Seikel:2013fda, Cai:2015zoa, Cai:2015pia, Gomez-Valent:2018hwc,Yennapureddy:2017vvb,Li:2019nux, Belgacem:2019zzu,Moore:2015sza,Canas-Herrera:2021qxs, Briffa:2020qli,Cai:2019bdh, LeviSaid:2021yat, Reyes:2021owe, Bernardo:2021qhu, Bernardo:2021mfs, Bengaly:2021wgc,Benisty:2020kdt, Keeley:2020aym} principally because it is data-driven and makes no assumption about the underlying cosmological model.

The overall motivation of this work is then this. Given the results obtained with both parametric and nonparametric approaches, we compare them and analyze the trends obtained for each method. This part is interesting because it can shed some light on DE dynamics if the features found when using GP match the ones obtained with the parametric forms. In addition, we employ four different $H_0$ priors reported in the literature \cite{Aghanim:2018eyx, Freedman:2019jwv, Anand:2021sum, Riess:2021jrx} that represent the current Hubble tension. In this way, our results may also add further insight on the current tension by analyzing its impact on DE evolution.

The rest of this paper proceeds as follows. We introduce the observational data under study and briefly review both parametric and nonparametric methods to be considered (Section \ref{sec:statistical_analysis}). Then, we discuss our main results jointly coming from parametric and nonparametric analyses. First, using the base Hubble data, we discuss two phenomenological functions capable of describing dynamical DE, a normalized DE density (Section \ref{subsec:de_from_Hz}) and a compactified DE equation of state (Section \ref{subsec:compactified_de_eos}). We then extend the analysis by including supernovae observations (Section \ref{sec:extended_analysis}) and put together our constraints on the DE equation of state in Table \ref{tab:w_de_constraints} (Section \ref{sec:constraints_de_eos}). We summarize our work in Section \ref{sec:conclusions} and pave the road for further studies on the subject. Our computations are transparently presented as jupyter notebooks and can be downloaded from our GitHub repository\footnote{Link to GitHub here: \href{https://github.com/reggiebernardo/notebooks}{https://github.com/reggiebernardo/notebooks}.}.

\section{Statistical analysis of cosmological data}
\label{sec:statistical_analysis}

We introduce the data sets that form the core of our study. Then, we review model-independent, parametric and nonparametric approaches to studying DE evolution.

\subsection{Late-time cosmic data}
\label{subsec:late_time_cosmic_data}

We consider measurements of the Hubble function $H(Z)$ at a number of redshifts $Z$ from cosmic chronometers (CC) and baryon acoustic oscillations (BAO). The CC data set consists of 31 data points \cite{Moresco:2016mzx, Moresco:2015cya, 2014RAA....14.1221Z, 2010JCAP...02..008S, 2012JCAP...08..006M,Ratsimbazafy:2017vga} obtained through a differential aging method involving adjacent and passively evolving galaxies. This relies on measurements of the \textit{temporally} adjacent galaxies' ages and redshifts through which the Hubble function at a redshift $z = Z$ can be approximately obtained as $H(z) = \dot{a}/a \sim \left( \Delta z/\Delta t \right)/ \left(1 + z\right)$. The second part of this Hubble data, namely BAO, comes from fluctuations of the baryon density in the early Universe which leave observational imprints on the sky often referred to as standard rulers. The BAO directly measures the combination $H(z) r_d$ where $r_d$ is the radius of the sound horizon during baryon drag. This additionally relies on the $\Lambda$CDM model which sets the scale of the sound horizon ($r_d = 147.74$ Mpc) during baryon drag and supplements 26 more points to the Hubble data \cite{BOSS:2014hwf, 2012MNRAS.425..405B, Chuang:2013hya, BOSS:2013igd, Bautista:2017zgn, Gaz34, Oka37, Wang33, Chuang28, Alam38, Anderson32, Busca36}. The compiled Hubble data from CC and BAO form the base of our study and is also important for establishing a sensible assessment that transcends the technical differences between parametric and nonparametric approaches.

For this paper, we do not take into account the recently proposed covariance matrix of the CC \cite{Moresco:2020fbm}, but rather we consider only uncorrelated points at various redshifts often considered for parametric and nonparametric analyses of the expansion data. Obviously, however, this full covariance matrix should be considered in a future work building on this preliminary assessment. The expansion data used in this work is presented on Table \ref{tab:expansion_data}.

We additionally take into account the 1048 supernovae (SNe) type Ia observations using the Pantheon data set \cite{Scolnic:2017caz}. This provides a measurement of the supernovae apparent magnitudes at their brightest states and comes with a full covariance matrix relating the measurements at various redshifts ranging from $0.01 < z < 2.3$. We also make use of the SNe measurements of $E(z)$ of the CANDELS and CLASH Multi-Cycle Treasury data (MCT) \cite{Riess:2017lxs}. This compression of the Pantheon data in terms of the normalized expansion function is utilized for the GP which together with an $H_0$ prior can be used to reconstruct $H(z) = H_0 E(z)$. We do so by using the covariance matrix of the samples and considering only five out of six points due to the non-Gaussian nature of the last point (which makes it incompatible with the GP) \cite{Gomez-Valent:2018hwc, Briffa:2020qli}. When considering the SNe observations, we marginalize analytically over the nuisance parameter, which in this case is the SNe absolute magnitude, as detailed in Ref. \cite{SNLS:2011lii}. The $H_0$ priors are then treated together with the SNe observations in statistically the same manner as with only the base Hubble data.

\begin{figure}[h!]
\center
	\subfigure[]{
		\includegraphics[width = 0.5 \textwidth]{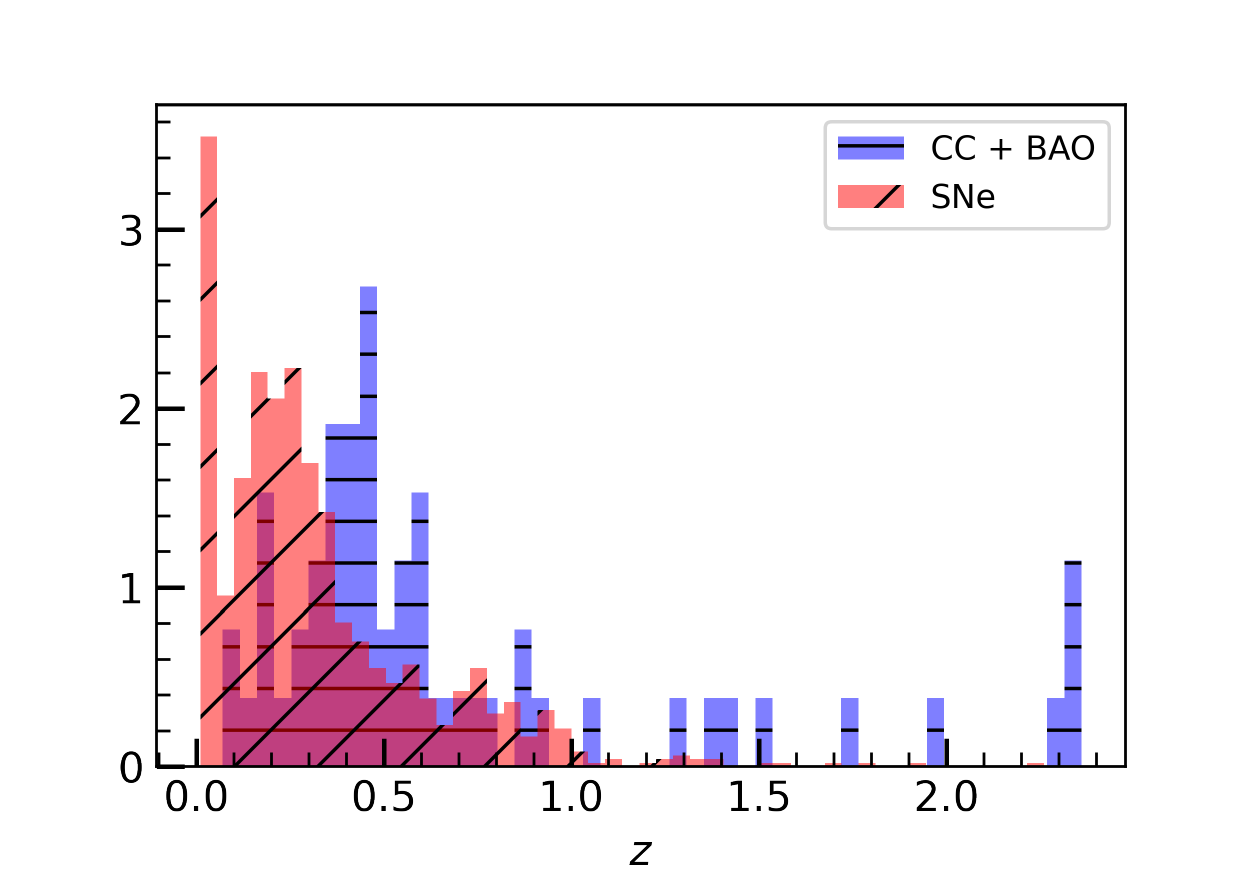}
		}
	\subfigure[]{
		\includegraphics[width = 0.42 \textwidth]{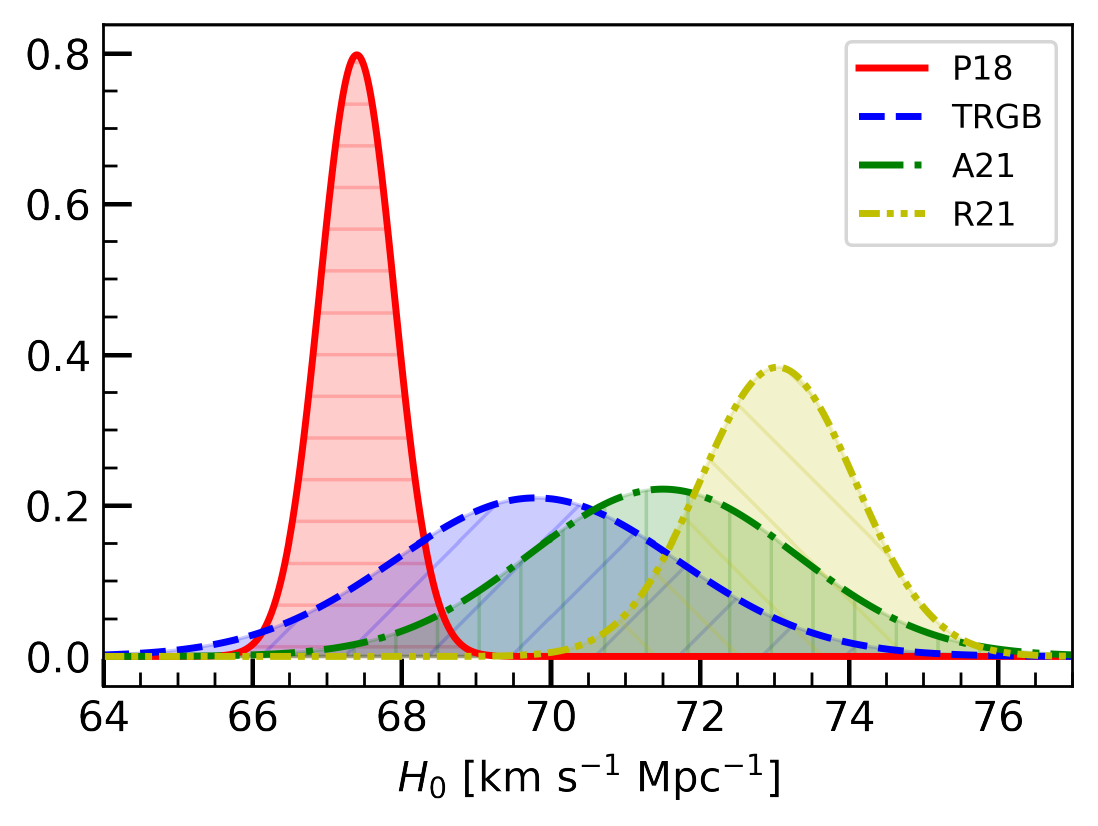}
		}
\caption{(a) Redshift distributions of the base Hubble data coming from CC and BAO, and SNe observations. (b) $H_0$ priors considered in this paper: $H_0^{\text{P18}} = 67.4 \pm 0.5$ km s$^{-1}$Mpc$^{-1}$, $H_0^{\text{TRGB}} = 69.8 \pm 1.9$ km s$^{-1}$Mpc$^{-1}$, $H_0^{\text{A21}} = 71.5 \pm 1.8$ km s$^{-1}$Mpc$^{-1}$, and $H_0^{\text{R21}} = 73.04 \pm 1.04$ km s$^{-1}$Mpc$^{-1}$.}
\label{fig:histogram_data}
\end{figure}

The redshift distribution of the base Hubble data (Table \ref{tab:expansion_data}) and SNe observations considered in this work is shown in Figure \ref{fig:histogram_data}(a). These data sets both take on the form of a Gamma distribution \cite{Dialektopoulos:2021wde}.

The use of priors on the Hubble constant $H_0$ helps to reduce the uncertainties in the reconstruction. Also, performing the statistical analysis with different $H_0$ priors makes the results reflective of any possible influence from the Hubble tension. Keeping this in mind, we consider three priors on $H_0$ that have been reported in the literature, namely the Riess (R21) prior which is $H_0^{\text{R21}} = 73.04 \pm 1.04$ km s$^{-1}$ Mpc$^{-1}$ \cite{Riess:2021jrx}, the Anand (A21) prior $H_0^{\text{R21}} = 71.5 \pm 1.8$ km s$^{-1}$ Mpc$^{-1}$ \cite{Anand:2021sum}, the Carnegie-Chicago (TRGB) Hubble prior $H_0^{\text{TRGB}} = 69.8 \pm 1.9$ km s$^{-1}$ Mpc$^{-1}$ \cite{Freedman:2019jwv}, and the latest value from the Planck collaboration (P18) $H_0^{\text{P18}} = 67.4 \pm 0.5$ km s$^{-1}$ Mpc$^{-1}$ \cite{Aghanim:2018eyx}. These $H_0$ values illustrated in Figure \ref{fig:histogram_data}(b) represent the current Hubble tension and are considered in this analysis to shed more light on this intriguing puzzle \cite{DiValentino:2020zio}.

We emphasize that the usage of parameter priors such as those on $H_0$ compromise the notion of ``model-independence'' depending on which assumptions were considered to obtain the priors in the first place. The local measurements of the expansion rate (R21, A21, and TRGB) for example are arguably cosmology-independent unlike the Planck prior (P18) which necessarily assumes the $\Lambda$CDM model. At the same time, these local distance-ladder values differ in terms of how the supernovae were calibrated, with $H_0^{\text{R21}}$ being calibrated using cepheids while $H_0^{\text{TRGB}}$ using the tip of the red giant branch. $H_0^{\text{A21}}$ is a recent reanalysis of the TRGB prior which lead to a slight positive shift in the estimate of $H_0$. On the other hand, considering these different measurement of $H_0$ in compromised model-independent analyses adds insight as to how the Hubble tension may influence the estimates of cosmological parameters. We hope that such interplay between different $H_0$ values and other cosmological parameters can eventually trace a resolution to the Hubble tension. Obviously, given the significance of $H_0$ (in setting cosmic distance scales) and the Hubble tension, no method of cosmological analysis can be ignored.

On a different note, to keep the parametric and nonparametric treatments identical, we consider the matter fraction prior ($\Omega_{m0} h^2 = 0.1430 \pm 0.0011$) \cite{Aghanim:2018eyx} which is directly measured from the cosmic microwave background using the peak structure in the damping tail. Therefore, even though the methods are intrinsically different, they stand on at least a common ground in this analysis, which is the Planck prior on combination of the matter density and the present expansion rate. This is particularly needed for the GP since unlike parametric methods, the GP does not estimate parameters outside of the information it is provided, but rather it reconstructs a particular data set which it is given. We discuss the methods in detail in the following sections to make this clearer.

\subsection{Parametric methods}
\label{subsec:parametric}

A direct probe to test if a cosmological constant $\Lambda$ drives the evolution, consists in considering an arbitrary function $X(z)$, in the range of the data. As far as we know, this was first proposed in Ref. \cite{Wang:2001ht} assuming a linear interpolation between redshifts, and also a quadratic one in Refs. \cite{Wang_2004, Wang:2004ru}. The results of all these first explorations were that the DE density showed a slight increase with redshift, being consistent with $\Lambda$ at 2$\sigma$. With more and higher quality data, the problem was revisited in Ref. \cite{Cardenas:2014jya} where a quadratic interpolation was used, with data from supernovae, gas mass fraction in galaxy clusters, BAO, and the cosmic microwave background. Surprisingly, the trend obtained was opposite to the previous one, indicating a DE density that decreases with redshift, even giving negative values for $z>1.5$ at 1$\sigma$ (which is consistent with Refs.~\cite{Akarsu:2021fol,Akarsu:2019hmw}). The extension of this work in light of more recent data is presented in Ref. \cite{Grandon_2021}, where evidence for DE evolution using a quadratic and also a cubic interpolation was studied.

The evolution of the dark energy sector enters into the evolution of the cosmology through the normalized Hubble parameter $E(z)=H(z)/H_0$ through the Friedmann equation
\begin{equation}
\label{eq:frd_eq_X}
    E(z)^2 = \Omega_{m0} (1+z)^3 + (1-\Omega_{m0})X(z)\,,
\end{equation}
where a flat background is assumed. The $X(z)$ parametrization then enters all facets of the cosmological evolution. In particular, the luminosity distance is modified through the changes in the evolution of the reduced Hubble parameter in
\begin{equation}
\label{eq:luminosity_distance}
    d_L(z) = \frac{c(1+z)}{H_0}\int_0^z \frac{dz'}{E(z')}\,.
\end{equation}
Thus, we can contain any deviations in the data from a cosmological constant in the evolution of the $X(z)$ parameter. In the work that follows, we aim to explore the space of realisations of $X(z)$ through parametric and nonparametric techniques.

\begin{figure}[h!]
\center
\includegraphics[width = 0.7 \textwidth]{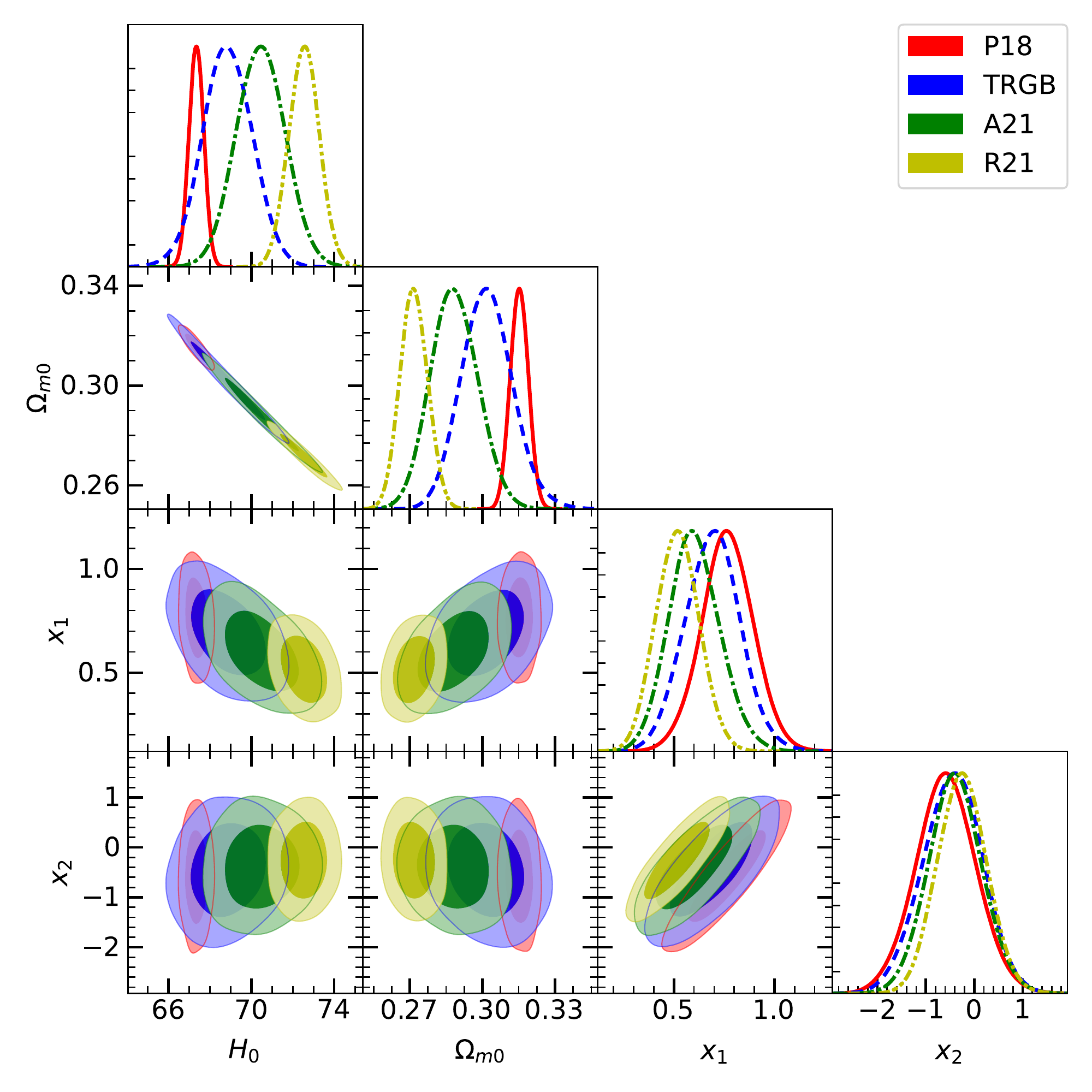}
\caption{The sampled posteriors of the parameters $\left( H_0, \Omega_{m0}, x_1, x_2 \right)$ in quadratic parametrized DE for each $H_0$ prior: $H_0^{\text{P18}} = 67.4 \pm 0.5$ km s$^{-1}$Mpc$^{-1}$, $H_0^{\text{TRGB}} = 69.8 \pm 1.9$ km s$^{-1}$Mpc$^{-1}$, and $H_0^{\text{A21}} = 71.5 \pm 1.8$ km s$^{-1}$Mpc$^{-1}$, and $H_0^{\text{R21}} = 73.04 \pm 1.04$ km s$^{-1}$Mpc$^{-1}$. These were obtained with the base Hubble data (CC + BAO).}
\label{fig:quad_bestfit_zm2}
\end{figure}

For the quadratic parametrization, we use
\begin{equation}
\label{xdz1}
\begin{split}
X(z) = x_0 \dfrac{(z-z_1)(z-z_2)}{(z_0-z_1)(z_0-z_2)} & + x_1
\dfrac{(z-z_0)(z-z_2)}{(z_1-z_0)(z_1-z_2)} + x_2 \dfrac{(z-z_0)(z-z_1)}{(z_2-z_0)(z_2-z_1)}\,,
\end{split}
\end{equation}
where $x_0$, $x_1$ and $x_2$ are constant values of $X(z)$ evaluated at $z_0$, $z_1$ and $z_2$ where we assume that $z_2>z_1>z_0$. Setting $X(z=z_0=0)=1$ by definition, and using $x_1 = X(z_m/2)$ and $x_2=X(z_m)$ with $z_m$ being the maximum redshift in the data set, Eq.~(\ref{xdz1}) reduces to
\begin{equation}\label{inter1}
    X(z)= 1+ \left(4x_1 -x_2-3\right) \left( \dfrac{z}{z_m} \right) - 2 \left(2 x_1-x_2-1 \right) \left( \frac{z}{z_m} \right)^2\,.
\end{equation}
It is worth noting that when all $x_i = 1$, then $X(z) = 1$, or rather, that the model reduces to $\Lambda$CDM in this limit. Substituting Eq. (\ref{inter1}) into the Friedmann equation and then sampling over the parameter space with the Hubble data, we obtain the posteriors shown in Figure \ref{fig:quad_bestfit_zm2}.

It can be seen from this that the parameters are influenced by the choice of prior on $H_0$. This motivates us to further consider them in the analysis to make conclusions that would be impervious to the Hubble tension. The most notable feature of Figure \ref{fig:quad_bestfit_zm2} is that the posteriors of $x_1$ and $x_2$ continue to deviate further from $\Lambda$CDM ($x_1 = x_2 = 1$) for increasing values of $H_0$. Granted, the reason for this hierarchy maybe that because the Planck $\Omega_{m0} h^2$ prior is used. However, even for the Planck prior, the measured $(x_1, x_2)$ turns out to be $x_1 = 0.8 \pm 0.1$ and $x_2 = -0.6 \pm 0.6$, which disfavours $\Lambda$CDM ($x_1 = x_2 = 1$) at 95\% confidence. Nonetheless, these measured values together with their covariances can be used to reconstruct the shape of $X(z)$ and the DE equation of state. Also, Figure \ref{fig:quad_bestfit_zm2} shows that the value of $\Omega_{m0} h^2$ is fairly constant irrespective of the value of priors on $H_0$, while $x_1$ and $x_2$ seem to be correlated giving lower and higher values with respectively higher and lower $H_0$ priors. As one would expect the value of $H_0$ varies with the value of priors on this parameter. Furthermore, having fixed $\Omega_{m0} h^2$ using the cosmic microwave background data, it can be seen that the tension in $H_0$ also reflects as a tension in the matter fraction $\Omega_{m0}$ in the opposite direction in parameter space. This will be discussed in Section \ref{sec:evidence_dynamical_de}. But, for the meantime, we move on a step further to generalize this parametric approach.

\begin{figure}[h!]
\center
\includegraphics[width = 0.85 \textwidth]{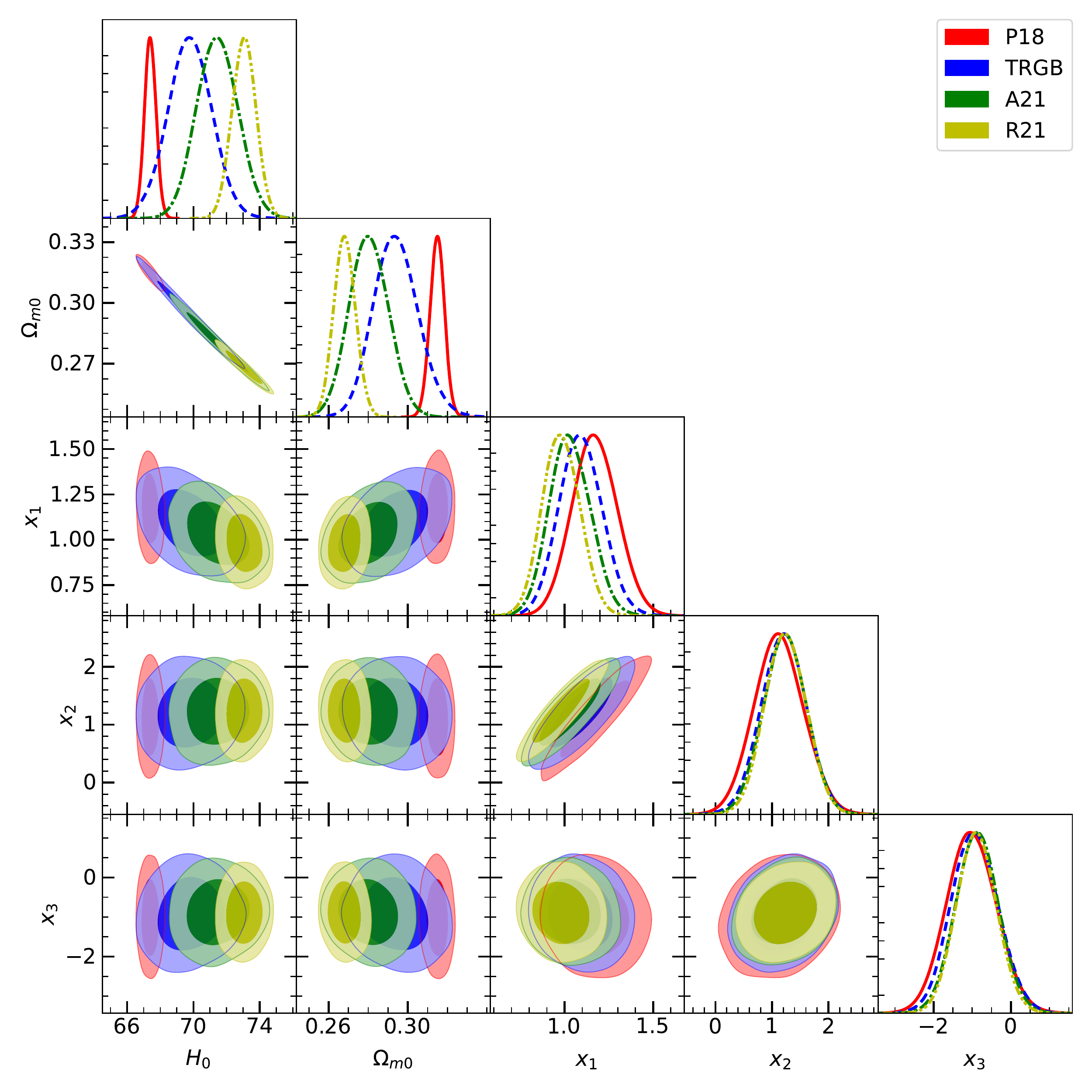}
\caption{The sampled posteriors of the parameters $\left( H_0, \Omega_{m0}, x_1, x_2, x_3 \right)$ in cubic parametrized DE for each $H_0$ prior: $H_0^{\text{P18}} = 67.4 \pm 0.5$ km s$^{-1}$Mpc$^{-1}$, $H_0^{\text{TRGB}} = 69.8 \pm 1.9$ km s$^{-1}$Mpc$^{-1}$, and $H_0^{\text{A21}} = 71.5 \pm 1.8$ km s$^{-1}$Mpc$^{-1}$, and $H_0^{\text{R21}} = 73.04 \pm 1.04$ km s$^{-1}$Mpc$^{-1}$. These were obtained with the base Hubble data (CC + BAO).}
\label{fig:cubic_bestfit_zm2}
\end{figure}

Following the same idea, for a cubic parametrization we get 
\begin{equation}
\label{eq:cubic}
\begin{split}
X(z) = 1
& + \dfrac{1}{2} \left(-11 + 18 x_1 - 9 x_2 + 2 x_3\right) \left( \frac{z}{z_m} \right) \\
& - \dfrac{9}{2} \left(-2 + 5 x_1 - 4 x_2 + x_3 \right) \left( \frac{z}{z_m} \right)^2 + \dfrac{9}{2} \left(-1 + 3 x_1 - 3 x_2 + x_3 \right) \left( \frac{z}{z_m} \right)^3\,.
\end{split}
\end{equation}
As we did before, we set $z_0=0$ in such a way that $x_0=X(z=z_0=0)=1$, and we set $z_3 = z_m$ as the maximum redshift in the data. The other points are $z_2=2z_m/3$ and $z_1=z_m/3$. In summary, the free parameters are $x_1=X(z_1)$, $x_2=X(z_2)$, and $x_3=X(z_3)$. As before, the $\Lambda$CDM limit can be seen to be $x_1=x_2=x_3=1$. Here, the results of the sampling are shown in Figure \ref{fig:cubic_bestfit_zm2}.

Clearly, again, we see the influence of the choice of $H_0$ on the parameters of the model. The hierarchy of deviation from the $\Lambda$CDM model can also be seen with increasing values of $H_0$, i.e., the deviation of the posteriors of $x_1$, $x_2$, and $x_3$ away from $x_1 = x_2 = x_3 = 1$ increases in the order P18, TRGB, A21, and R21, which is most poignant in the $x_3$ parameter. As in the quadratic case, this can be traced to the use of the Planck prior on the combination $\Omega_{m0} h^2$ of the matter fraction and the Hubble constant. \ref{sec:Om0h2} shows the distribution of $\Omega_{m0} h^2$ confirming that the Planck prior on this combination is respected in both the quadratic and cubic methods during the Bayesian analysis. But then again, the inevitable is that even for $H_0^\text{P18}$, a deviation from $\Lambda$CDM cannot be turned away, particularly with the marginalized posterior of $x_3 = -0.9 \pm 0.5$ excluding $x_3 = 1$ at more than $2\sigma$. Also, we again see the correlation feature for the $x_i$ parameters, while the value of $\Omega_{m0}$ reflects the tension on $H_0$ having fixed $\Omega_{m0}$ by the cosmic microwave background damping tail. We shall see this deviation again later by reconstructing the $X(z)$ function itself and the DE equation of state.

A noteworthy observation also emerges. In both Figures \ref{fig:quad_bestfit_zm2} and \ref{fig:cubic_bestfit_zm2}, the Hubble tension is practically only influencing the matter density and \textit{not} the dark energy parameters, i.e., there are no tensions in $x_i$. This means that our conclusions on dark energy evolution would be transparent to the Hubble tension, motivating our use of the various $H_0$ priors.

To conclude this section, we note that polynomials of degree higher than three are not considered in this work since they do not perform significantly different using these data sets. This was discussed in Ref. \cite{Grandon_2021} where higher order polynomial based parametrizations of DE were shown to be disfavored in terms of statistical performance due to additional free parameters. This provides a natural cutoff for which only certain models are favoured. We also emphasize that our conclusions hold regardless of the choice of $z_m$ \cite{Grandon_2021}.

\subsection{Nonparametric reconstruction methods}
\label{subsec:nonparametric}

We provide a brief introduction to nonparametric reconstruction methods and the GP approach in particular \cite{10.5555/971143, 10.5555/1162254}.

Parametric descriptions of cosmological expansion require a fundamental understanding of the gravitational and matter content of the Universe, such as in $\Lambda$CDM. On the other hand, nonparametric techniques provide a physics-independent avenue by which cosmological parameters can be inferred from the reconstructions with a particular confidence in a certain range.

Parametric descriptions of cosmological expansion require a fundamental understanding of the gravitational and matter content of the Universe, such as in $\Lambda$CDM. On the other hand, nonparametric techniques provide a physics-independent avenue by which cosmological quantities can be inferred from the reconstructions with a particular confidence in a certain range. This is important so that we can infer the evolution of cosmological parameters without the need of a prescribed physical description, which is very important in assessing the performance of cosmological models \cite{Busti:2014aoa,Escamilla-Rivera:2021rbe,Gomez-Valent:2018hwc}. Nonparametric reconstruction methods are based purely on learning how elements in a data set are connected together in a statistical way. This requires some approach in which a statistical setup is established. These statistical models are constructed to mimic the behaviour of the natural process from which the data sets are being taken. In this case, we are probing expansion data and considering the GP reconstruction method. GP relies on using a covariance function, or kernel, which represents the way in which the data set elements are related together. The training process is an iterative one in which the kernel hyperparameters (non-physical parameters) are progressively approximated by maximizing the reconstruction likelihood. The kernel can then be used to reconstruct the entire parameter evolution for some limited range (which is normally limited to the range of the data set under consideration).

The GP is an emerging scientific tool in cosmology for the reconstruction of a data set primarily due to its objectivity in making predictions even without a cosmological model \cite{Seikel2012, Seikel:2013fda, Shafieloo:2012ht}. This is a particularly refreshing change of view in analyzing data in light of the cosmological tensions where the very foundations of the field such as the cosmological principle and general relativity are being closely reexamined. Moreover, the ease with which the GP algorithm can be implemented makes it an even more attractive approach. This is summarized in three equations (Eqs. (\ref{eq:gp_ave}), (\ref{eq:gp_cov}), and (\ref{eq:logmlike})) which we turn to next.

Consider an observation of a function $H(z)$ with a covariance matrix $C$ of size $N \times N$ where $N$ is the number of points in the data. In terms of a covariance function $K\left( z^* , \tilde{z}^* \right)$, also often referred to as the kernel, relating the function values at coordinates $z^*$ and $\tilde{z}^* \neq z^*$, the mean and covariance of the GP reconstruction of the $n$th derivative of $H(z)$ are given by
\begin{equation}
\label{eq:gp_ave}
    \langle H^{* (n)} \rangle = K^{(n, 0)} \left( z^*, Z \right) \left[ K\left( Z, Z \right) + C \right]^{-1} H \left( Z \right)\,,
\end{equation}
and
\begin{equation}
\label{eq:gp_cov}
    \text{cov} \left( H^{* (n)} \right) = K^{(n, n)} \left( z^*, z^* \right) - K^{(n, 0)} \left( z^*, Z \right) \left[ K\left(Z, Z\right) + C \right]^{-1} K^{(0, n)} \left(Z, z^*\right)\,,
\end{equation}
respectively, where $Z$ stands the redshifts in the observation and $f^{(n, m)}(x, y)$ refers to the $n$th partial derivative of a function $f$ with respect to its first argument $x$ and the $m$th partial derivative with respect to the second argument $y$. The kernel is then optimized for the input data (or the observation) $H(Z)$ by letting its internal hyperparameters $\theta$ be determined through the marginalization of the likelihood function $\mathcal{L} = p \left( H | Z, \theta \right)$ where
\begin{equation}
\label{eq:logmlike}
\ln \mathcal{L} = -\dfrac{1}{2} H\left(Z\right)^{T} \left[ K\left(Z, Z\right) + C \right]^{-1} H\left(Z\right) - \dfrac{1}{2} \ln | K\left(Z, Z\right) + C | - \dfrac{N}{2} \ln \left(2\pi\right)\,.
\end{equation}
In practice, optimization is usually taken as an efficient substitute to marginalization. Eqs.~(\ref{eq:gp_ave}--\ref{eq:logmlike}) therefore reflect the simplicity of the GP algorithm and why it is often pursued for nonparametric reconstruction not only in cosmology but in others fields as well.

\begin{figure}[h!]
\center
\includegraphics[width = 0.5 \textwidth]{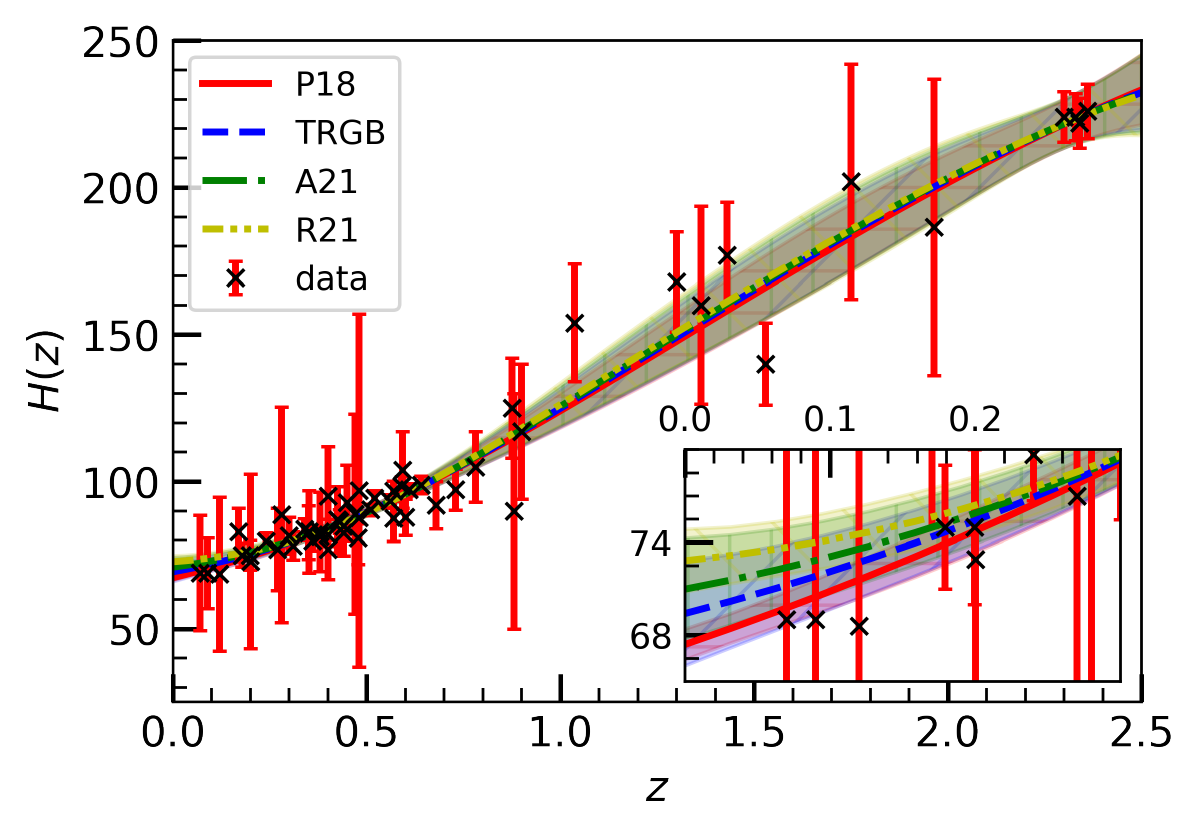}
\caption{The GP reconstructed Hubble function per $H_0$ prior: $H_0^{\text{P18}} = 67.4 \pm 0.5$ km s$^{-1}$Mpc$^{-1}$, $H_0^{\text{TRGB}} = 69.8 \pm 1.9$ km s$^{-1}$Mpc$^{-1}$, $H_0^{\text{A21}} = 71.5 \pm 1.8$ km s$^{-1}$Mpc$^{-1}$, and $H_0^{\text{R21}} = 73.04 \pm 1.04$ km s$^{-1}$Mpc$^{-1}$. The colored and hatched parts show the region within $2\sigma$ of the GP. Hatches: (P18: ``$-$''), (TRGB: ``$/$''), (A21: ``$|$''), (R21: ``\textbackslash''). The inset shows the low redshift region $z \in (0, 0.3)$ of the GP reconstructed Hubble function.}
\label{fig:Hz_rec_per_method}
\end{figure}

Figure \ref{fig:Hz_rec_per_method} shows the GP reconstructed Hubble function provided the compiled Hubble data from the CC and BAO. The GP reconstructed evolution of $H(z)$ provides reasonable confidence levels for the whole range of redshifts of interest. Moreover, in the inset, we show how the different priors on $H_0$ affect the reconstructed $H_0$ values as well as its neighboring low redshift vicinity.

This illustrates the GP algorithm. It predicts the intermediate points within a data set unlike the parametric approaches which make best estimates of the parameters. The GPs shown in Figure \ref{fig:Hz_rec_per_method} also clearly reflect the Hubble tension coming from the choice of an $H_0$ prior. The inset exemplifies this by showing the low redshift region of the reconstruction.

Understandably, the GP also comes with quirks, the most notable of these are overfitting \cite{Escamilla-Rivera:2021rbe}, underestimating uncertainties \cite{Colgain:2021ngq}, and kernel selection \cite{10.3389/fbuil.2017.00052, Bernardo:2021mfs}. In addition, it has been shown that features of the reconstruction may depend on the hyperparameter priors for certain data sets \cite{Perenon:2021uom}. For the data set at hand, we optimized the GP starting with a common hyperparameter length of 2 units and an amplitude of 130 units. Later, we shall witness this overfitting in our assessment of the GP and the parametric methods in the next section. On the other hand, the underestimated uncertainties can be seen in Figure \ref{fig:Hz_rec_per_method} where clearly the reconstructed function is generally narrower than the error bars of the data points. The kernel selection problem was also tackled in Refs. \cite{Bernardo:2021mfs} by employing evolutionary algorithms to reduce prejudice in choosing a kernel for a specific problem. In fact, in Figure \ref{fig:Hz_rec_per_method}, we are using the Matern($\nu = 5/2$) kernel which was singled out as preferable by the evolutionary algorithms when using Hubble data from cosmic chronometers and supernovae. We shall continue to rely on this kernel throughout this paper but also note that different kernel choices only lead to statistically consistent results \cite{Bernardo:2021mfs}.

\section{Statistical reconstruction and evidence of dynamical dark energy}
\label{sec:evidence_dynamical_de}

We present our main results, hinting at a preference for a dynamical DE scenario, derived from both parametric and nonparametric analyses of observations in the late Universe.

\subsection{Dark energy from Hubble data}
\label{subsec:de_from_Hz}

We present the reconstructed DE density and assess the performance of each approach implemented in the reconstructions using various statistical metrics.

\begin{figure}[h!]
\center
	\subfigure[ $H_0^{\text{P18}} = 67.4 \pm 0.5$ km s$^{-1}$Mpc$^{-1}$ ]{
		\includegraphics[width = 0.475 \textwidth]{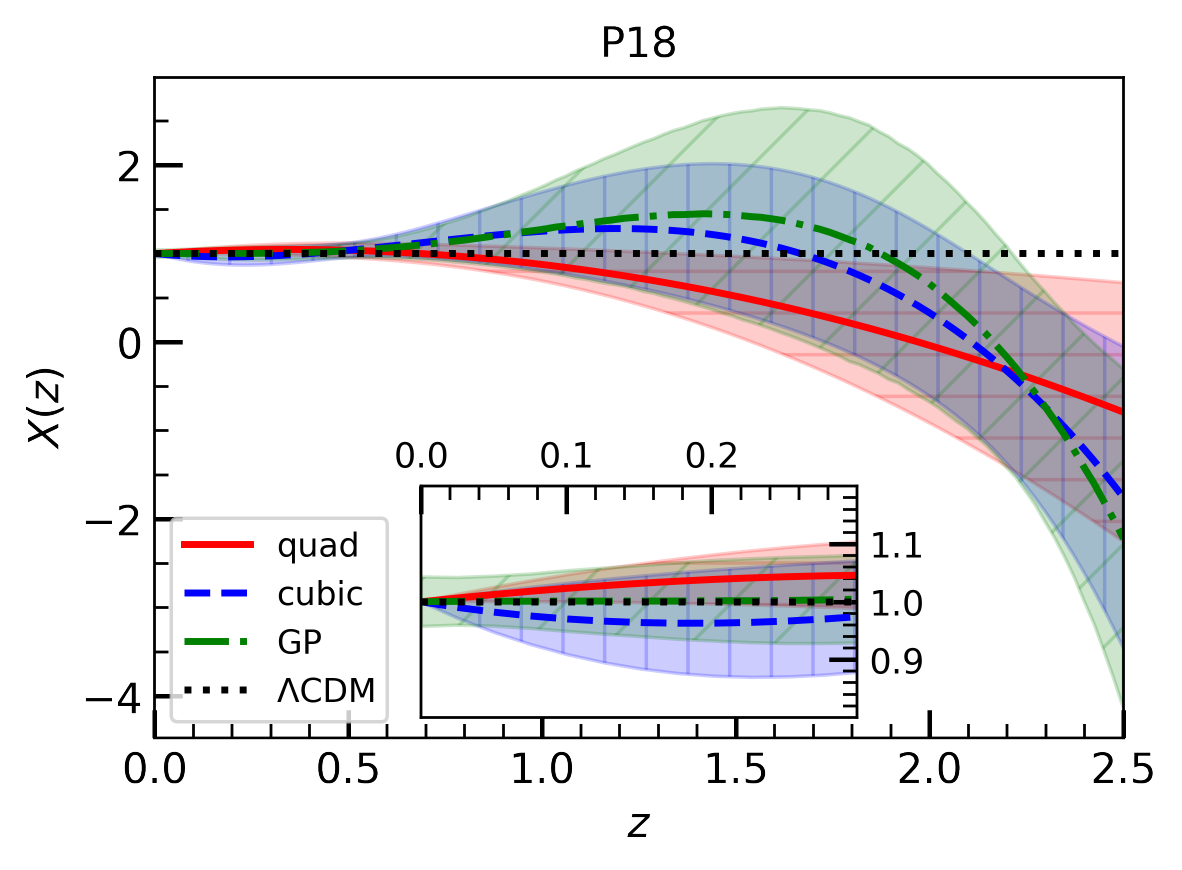}
		}
	\subfigure[ $H_0^{\text{TRGB}} = 69.8 \pm 1.9$ km s$^{-1}$Mpc$^{-1}$ ]{
		\includegraphics[width = 0.475 \textwidth]{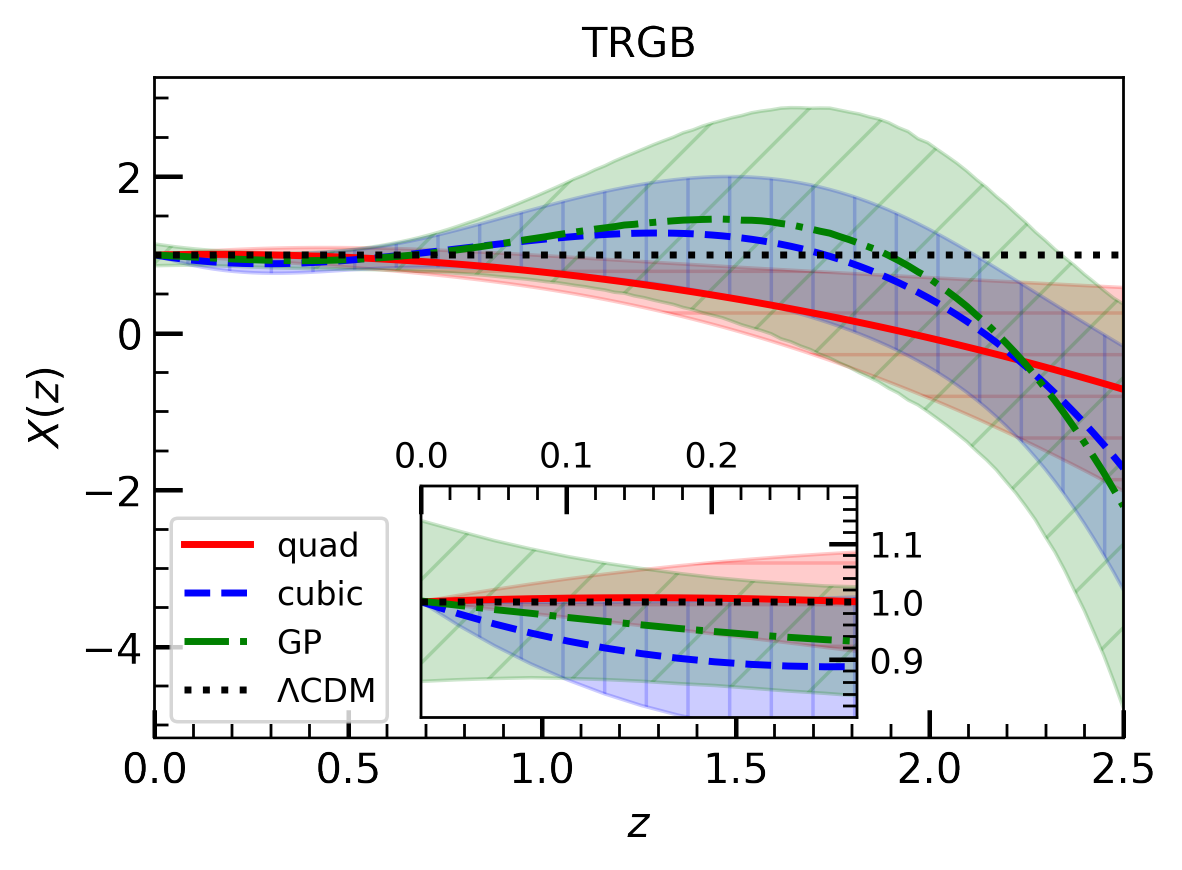}
		}
	\subfigure[ $H_0^{\text{A21}} = 71.5 \pm 1.8$ km s$^{-1}$Mpc$^{-1}$ ]{
		\includegraphics[width = 0.475 \textwidth]{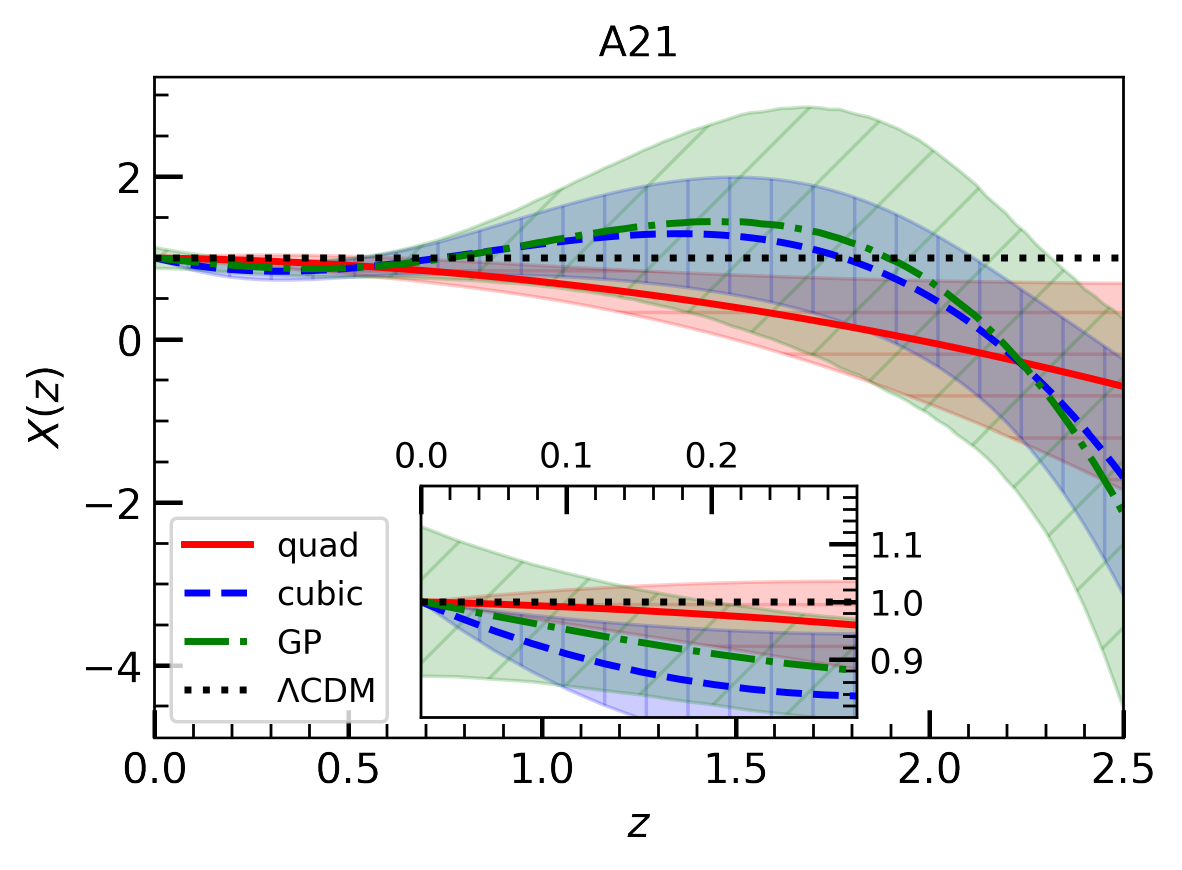}
		}
	\subfigure[ $H_0^{\text{R21}} = 73.04 \pm 1.04$ km s$^{-1}$Mpc$^{-1}$ ]{
		\includegraphics[width = 0.475 \textwidth]{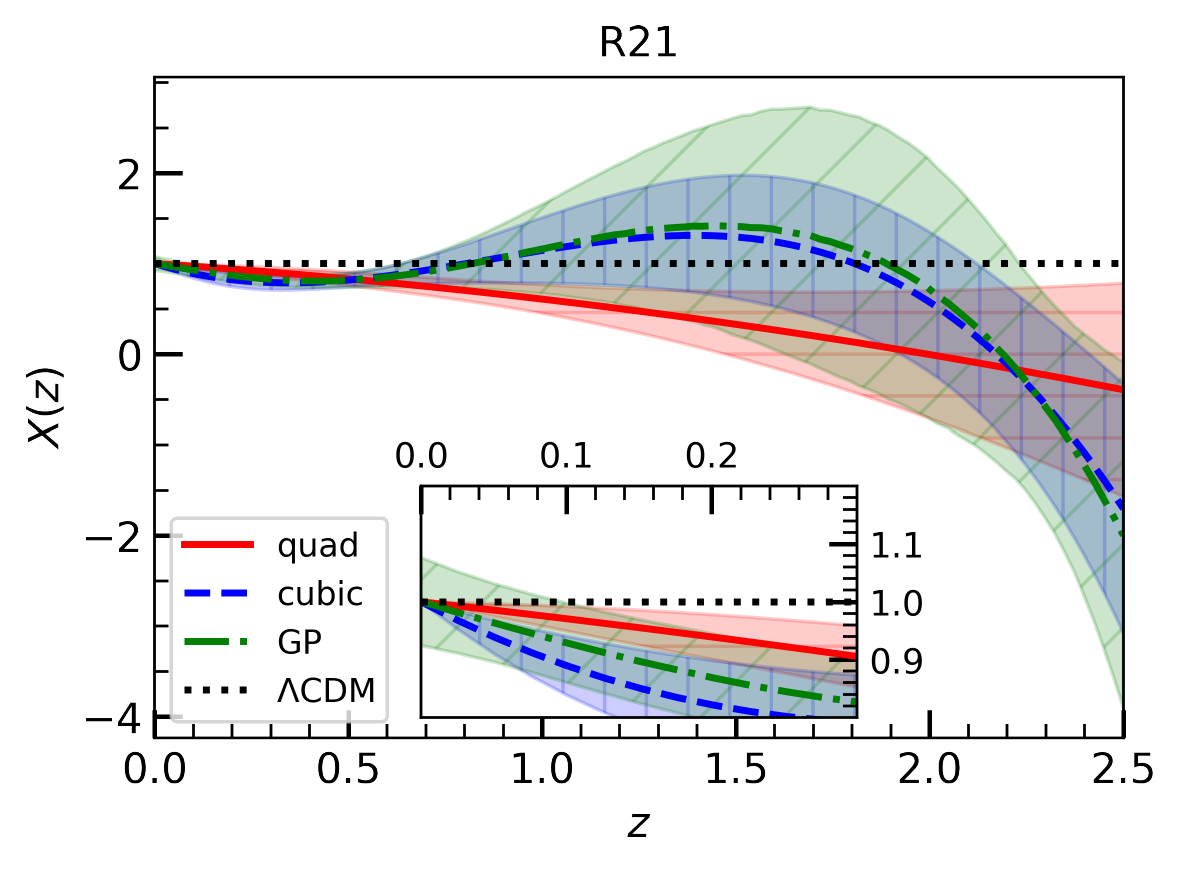}
		}
\caption{The reconstructed normalized DE per method derived from the base Hubble data (CC + BAO) for each $H_0$ prior: (a) P18, (b) TRGB, (c) A21, and (d) R21. Legends: ``quad'' and ``cubic'' stands for the quadratic and cubic parametrized DE, respectively; ``GP'' for the GPs. The colored and hatched regions show the $2\sigma$ confidence interval of the reconstructions. Hatches: (quad: ``$-$''), (cubic: ``$|$''), (GP: ``$/$''). The inset zooms in on the low redshift region $z \in (0, 0.3)$.}
\label{fig:Xz_rec_per_method}
\end{figure}

Figure \ref{fig:Xz_rec_per_method} shows the normalized DE density $X(z)$ obtained using the quadratic and cubic parametric methods as well as the GP for each $H_0$ prior. The $\Lambda$CDM curves appear as horizontal dotted line at $X = 1$.

This reveals that regardless of the $H_0$ prior, the concordance $\Lambda$CDM model is generally supported at low redshifts. However, the situation becomes more nuanced at higher redshifts where the posteriors in all of the methods begin to slip past $X(z) \sim 1$. Generally, it can be seen that the deviation from $\Lambda$CDM happens to be more observable at higher redshifts for the larger $H_0$ priors. For $z \gtrsim 1.5$, the quadratic method disfavours the $\Lambda$CDM line from within its $2\sigma$ region. The cubic method and the GP also supports this beyond $2\sigma$-exclusion of $\Lambda$CDM, albeit starting at a higher redshift $z \sim 2.3$ where the earliest observational data can be found. The influence of the $H_0$ priors also come into play at low redshifts ($z \lesssim 0.3$). The insets of Figure \ref{fig:Xz_rec_per_method} show that the cubic method and the GP even excludes $\Lambda$CDM at more than $2\sigma$, with the exception of the $H_0^{\text{P18}}$ prior. The exception may be due to the use of a matter fraction prior coming from Planck; however, regardless of this, even the reconstruction of $X(z)$ coming from the $H_0^{\text{P18}}$ prior suggests a deviation from the concordance model at the higher redshifts. This becomes even more notable considering that the Planck constraints assume a constant $\Lambda$ dark energy to support the cosmic acceleration, yet the late Universe spells an inconsistency with this assumption. Overall, all three methods, despite their intrinsic differences, seemingly hint at dynamical DE, or rather an evolving $X(z)$ as shown in Figure \ref{fig:Xz_rec_per_method}. It should be noted that this conclusion holds regardless of the choice of $H_0$ prior, and despite the fact that a minority of the points in the data set are anchored on the $\Lambda$CDM model. 

A notable characteristic of DE which appears in Figure \ref{fig:Xz_rec_per_method} is that $X(z) < 0$ is teased by all reconstructions. This is supported by other recent reports using different approaches \cite{Wang:2018fng, Akarsu:2019hmw, Escamilla:2021uoj}. Imposing a hard prior $X(z) > 0$ would instead put a limit to the constraining capability of data-driven approaches as the theory space which they cover becomes narrower. We draw the reader to the last paragraph of the introduction as well as Appendix A of Ref. \cite{Wang:2018fng} for an elaboration of this point. The possibility of non-positive energy densities allows data-driven approaches to flourish by covering a wider range of phenomenology permissible in modified gravity.

We also highlight the interesting difference between the low redshift behavior of the GP and the parametrized approaches as can be seen in the insets of Figure \ref{fig:Xz_rec_per_method}. In both parametrized methods, it can be seen that as one goes closer to $z = 0$, the posterior shrinks to an infinitely narrower size. This can be traced from the fact that these parametrized methods are an expansion about the redshift, i.e., $X(z) \sim 1 + a z + b z^2 + O(z^3)$ for constants $a, b, \cdots$. The GP, on the other hand, does not share this feature, and continues to be able to make a reasonable prediction for low redshifts even down to $z = 0$ since $X(z=0)=1$ by construction.

The robustness of this result stands on the observation that the intrinsically contrasting parametric and nonparametric approaches somehow agree in their macrophysical implications. Echoing our sentiments in the introduction, each method is unequivocally challenged in its own way, likely leaving traces of nonphysical artefacts in their reconstruction, but when all agree on a conclusion despite this difference, there could at least be a physical picture emerging that transcends such details. This hints to an evolving dark energy picture in this work.

To improve our confidence in their performance, we move on to assess each method. We consider the $\chi^2$ measure in order to assess the quality of the reconstruction together with other measures, where
\begin{equation}
\label{eq:chi2}
    \chi^2 = \sum_z \left( \dfrac{H_{\text{rec}}(z) - H_{\text{obs}}(z)}{ \sigma_{\text{obs}}(z) } \right)^2\,,
\end{equation}
where $H_{\text{rec}}(z)$ is the reconstructed Hubble function while $H_{\text{obs}}(z)$ and $\sigma_{\text{obs}}(z)$ are the mean and uncertainty of the data. This statistic ($\chi^2$) measures how far away a reconstruction is in units of the uncertainty of the data and has the particular advantage for this work that it can be defined for both parametric and nonparametric approaches. This would not be true for the information criterion and the Bayes factor which are associated with parametric methods but do not make sense for a nonparametric analysis. A respectable $\chi^2$ would be close to the size of the data $N$ while overfitting in a parametric sense corresponds to $\chi^2 < N$.

We further consider two statistical measures which have been used previously to compare nonparametric reconstruction methods \cite{Escamilla-Rivera:2021rbe}. These are given by
\begin{equation}
\label{eq:D1}
    \mathcal{D} = \sum_z \left( \dfrac{H_{\text{rec}}(z) - H_{\text{obs}}(z)}{ \sqrt{ \sigma_{\text{rec}}(z)^2 + \sigma_{\text{obs}}(z)^2 } } \right)\,,
\end{equation}
and
\begin{equation}
\label{eq:gamma2}
    \gamma^2 = \sum_z \left( \dfrac{H_{\text{rec}}(z) - H_{\text{obs}}(z)}{ \sqrt{ \sigma_{\text{rec}}(z)^2 + \sigma_{\text{obs}}(z)^2 } } \right)^2\,,
\end{equation}
where $\sigma_{\text{rec}}(z)$ is the uncertainty in the reconstruction. A crucial difference between the familiar $\chi^2$ and the statistics $\mathcal{D}$ and $\gamma^2$ is how they treat the uncertainties in the reconstruction. Notably, $\mathcal{D}$ and $\gamma^2$ consider the uncertainty in a reconstruction on an equal footing with the uncertainty in the data, but $\mathcal{D}$ can be positive or negative (depending on whether the data points lie mostly above/below the best fit) while $\gamma^2$ is strictly positive. Most importantly, all three statistics $\chi^2$, $\mathcal{D}$, and $\gamma^2$ are capable of being defined for parametric and nonparametric methods, which make them suitable for this study. Generally speaking, the smaller $\chi^2$, $\mathcal{D}$, and $\gamma^2$ are, the better a reconstruction is.

The statistics $\chi^2$, $\mathcal{D}$, and $\gamma^2$ measuring the deviation for each method from the Hubble data are presented in Table \ref{tab:performance}. A clear, unequivocal result is that each of the theory-agnostic implementations outperforms the $\Lambda$CDM model. This holds independent of the choice of an $H_0$ prior and can be observed for each of the metrics where the $\Lambda$CDM values always bring the largest deviation throughout. Understandably, this may also be viewed as unsurprising, considering the fact that the parametric methods enjoy more parameters than $\Lambda$CDM while machine learning algorithms such as the GP are prone to overfitting.

\begin{table}[h!]
\center
\caption{An assessment of the performance of each method using the statistics given by Eqs. (\ref{eq:chi2}), (\ref{eq:D1}), and (\ref{eq:gamma2}) with the Hubble data from cosmic chronometers and baryon acoustic oscillations. The Planck prior for the matter fraction $\Omega_{m0} h^2 = 0.1430 \pm 0.0011$ was considered throughout \cite{Aghanim:2018eyx}.}
\begin{tabular}{| c | c | c | c | c |}
\hline

$H_0$ prior & \phantom{ $\dfrac{1}{1}$ } \textit{Method/Model} \phantom{ $\dfrac{1}{1}$ } & \phantom{ $\dfrac{1}{1}$ } $\chi^2$ \phantom{ $\dfrac{1}{1}$ } & \phantom{ $\dfrac{1}{1}$ } $\mathcal{D}$ \phantom{ $\dfrac{1}{1}$ } & \phantom{ $\dfrac{1}{1}$ } $\gamma^2$ \phantom{ $\dfrac{1}{1}$ } \\ \hline \hline

\multirow{4}{*}{P18} 
& \phantom{ $\dfrac{1}{1}$ } $\Lambda$CDM \phantom{ $\dfrac{1}{1}$ } & $37.6$ & $-3.11$ & $30.7$ \\

& \phantom{ $\dfrac{1}{1}$ } \textit{Parametric} (quadratic) \phantom{ $\dfrac{1}{1}$ } & $30.6$ & $-3.06$ & $7.01$ \\ 

& \phantom{ $\dfrac{1}{1}$ } \textit{Parametric} (cubic) \phantom{ $\dfrac{1}{1}$ } & $27.0$ & $-3.08$ & $14.3$ \\

& \phantom{ $\dfrac{1}{1}$ } \textit{Nonparametric} (GP) \phantom{ $\dfrac{1}{1}$ } & $26.6$ & $-2.19$ & $25.2$ \\ \hline \hline

\multirow{4}{*}{TRGB} 
& \phantom{ $\dfrac{1}{1}$ } $\Lambda$CDM \phantom{ $\dfrac{1}{1}$ } & $38.6$ & $1.27$ & $24.6$ \\

& \phantom{ $\dfrac{1}{1}$ } \textit{Parametric} (quadratic) \phantom{ $\dfrac{1}{1}$ } & $32.1$ & $-2.28$ & $6.15$ \\ 

& \phantom{ $\dfrac{1}{1}$ } \textit{Parametric} (cubic) \phantom{ $\dfrac{1}{1}$ } & $26.1$ & $-1.69$ & $10.8$ \\

& \phantom{ $\dfrac{1}{1}$ } \textit{Nonparametric} (GP) \phantom{ $\dfrac{1}{1}$ } & $26.1$ & $-1.03$ & $24.4$ \\ \hline \hline

\multirow{4}{*}{A21} 
& \phantom{ $\dfrac{1}{1}$ } $\Lambda$CDM \phantom{ $\dfrac{1}{1}$ } & $42.8$ & $4.29$ & $25.2$ \\

& \phantom{ $\dfrac{1}{1}$ } \textit{Parametric} (quadratic) \phantom{ $\dfrac{1}{1}$ } & $33.8$ & $-1.84$ & $6.99$ \\ 

& \phantom{ $\dfrac{1}{1}$ } \textit{Parametric} (cubic) \phantom{ $\dfrac{1}{1}$ } & $25.7$ & $-1.13$ & $11.3$ \\

& \phantom{ $\dfrac{1}{1}$ } \textit{Nonparametric} (GP) \phantom{ $\dfrac{1}{1}$ } & $25.8$ & $-0.25$ & $24.0$ \\ \hline \hline

\multirow{4}{*}{R21} 
& \phantom{ $\dfrac{1}{1}$ } $\Lambda$CDM \phantom{ $\dfrac{1}{1}$ } & $60.0$ & $15.5$ & $34.8$ \\

& \phantom{ $\dfrac{1}{1}$ } \textit{Parametric} (quadratic) \phantom{ $\dfrac{1}{1}$ } & $36.4$ & $-0.97$ & $8.16$ \\ 

& \phantom{ $\dfrac{1}{1}$ } \textit{Parametric} (cubic) \phantom{ $\dfrac{1}{1}$ } & $25.3$ & $-0.72$ & $13.3$ \\

& \phantom{ $\dfrac{1}{1}$ } \textit{Nonparametric} (GP) \phantom{ $\dfrac{1}{1}$ } & $25.6$ & $0.53$ & $23.6$ \\ \hline \hline
\end{tabular}
\label{tab:performance}
\end{table}

Recalling that our Hubble data consists of $N = 57$ points coming from CC (31 points) and BAO (26 points), a good $\chi^2$ can be recognized as $\chi^2 \sim 57$. Table \ref{tab:performance} therefore shows which of the methods overfit the data. However, it must be pointed out that the best fit $\Lambda$CDM model also tends toward this direction for any of the $H_0$ priors. On the other hand, all of the model-independent approaches predict a $\chi^2 < \chi^2_{\Lambda\text{CDM}} < N$ where $\chi^2_{\Lambda\text{CDM}}$ is the corresponding value from the best fit $\Lambda$CDM model. In terms of relative sizes of $\chi^2$, we find that for the quadratic method, $\chi^2_{\text{P18}} < \chi^2_{\text{TRGB}} < \chi^2_{\text{R21}} < N$, while for both the cubic method and the GP, $\chi^2_{\text{R21}} < \chi^2_{\text{TRGB}} < \chi^2_{\text{P18}} < N$. It is also worth noting that the $\chi^2$ for the quadratic method is generally larger than those of the cubic method and the GP which are coincidentally of comparable sizes to within a few percent. It should be further noted that as far as the $\chi^2$ values can be trusted, the results with the cubic parametrized DE and the GP are less sensitive to the choice of the $H_0$ prior. Extremely exemplifying this, the $| \Delta \chi^2 | = 22.4$ between the P18 and R21 $H_0$ priors within $\Lambda$CDM while the corresponding values are $\Delta \chi^2 = 1.70$ and $1.00$ for the cubic method and the GP, respectively.

Now, for the statistic $\mathcal{D}$, the general result is that $\mathcal{D} < 0$, implying that the mean of the data points is larger than the mean of the reconstructions for most of the redshifts in the data set. Another way of saying this is that most of the data points can be found above the best fit line. Furthermore, it can be seen that $| \mathcal{D} |$ is the smallest for the GP while the cubic method closely trails behind. The results for $\gamma^2$ also turn out to be interesting, suggesting that the quadratic method, the visually less flexible of all three, performs better than the cubic method and the GP. A more consistent trend for each method can be observed for $\gamma^2$, that $\gamma^2_{\text{quad}} < \gamma^2_{\text{cubic}} < \gamma^2_{\text{GP}} < \gamma^2_{\Lambda\text{CDM}}$. We remind that $\gamma^2$ considers the uncertainty of the reconstruction and the data on an equal footing. This explains the general edge of the parametrized implementations compared to the GP since most of the data points can be found at the low redshifts where the posteriors of both quadratic and cubic methods shrink inevitably by design (recall the insets of Figure \ref{fig:Xz_rec_per_method} embodying this feature). Comparing the parametrized methods, the slight edge turned out to favor the quadratic method which had smaller uncertainties at low redshifts than the cubic method.

In all this, the highlight is that all of the methods perform better than the $\Lambda$CDM model (Table \ref{tab:performance}), and that all hint at DE evolution (Figure \ref{fig:Xz_rec_per_method}). This is further supported by reconstructions of a diagnostic function $X'(z)$ in \ref{sec:diagnostic}.

\subsection{The compactified dark energy equation of state}
\label{subsec:compactified_de_eos}

Another commonly used measure of DE is its equation of state $w(z) = P(z)/\rho(z)$ where $\rho$ and $P$ are the density and pressure of DE, respectively. However, one of the significant challenges pertaining to its reconstruction at higher redshifts is due to the fact that the DE density $\rho(z)$ changes sign \cite{Dutta:2018vmq}, or rather crosses the zero mark, at a certain redshift $z \sim 1$. Consequently, $w \rightarrow \infty$ at some point. This physically corresponds to a loss of predictability, blocking our knowledge of DE a few redshifts away. Computationally, it means that the error bars diverge in the vicinity of a critical redshift regardless of available computational resources. The \textit{compactified} DE equation of state $\arctan \left( 1 + w(z) \right)$ introduced in Ref. \cite{Bernardo:2021qhu} was considered to overcome this challenge for studying DE.

\begin{figure}[h!]
\center
	\subfigure[ ]{
		\includegraphics[width = 0.475 \textwidth]{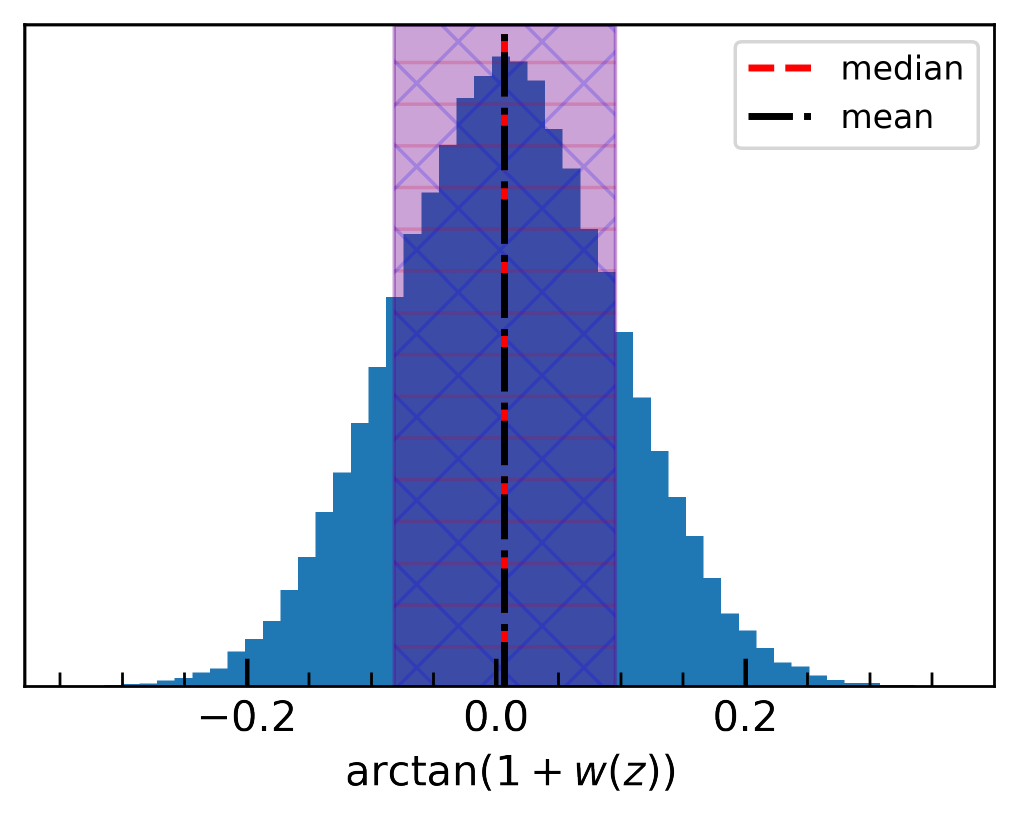}
		}
	\subfigure[ ]{
		\includegraphics[width = 0.475 \textwidth]{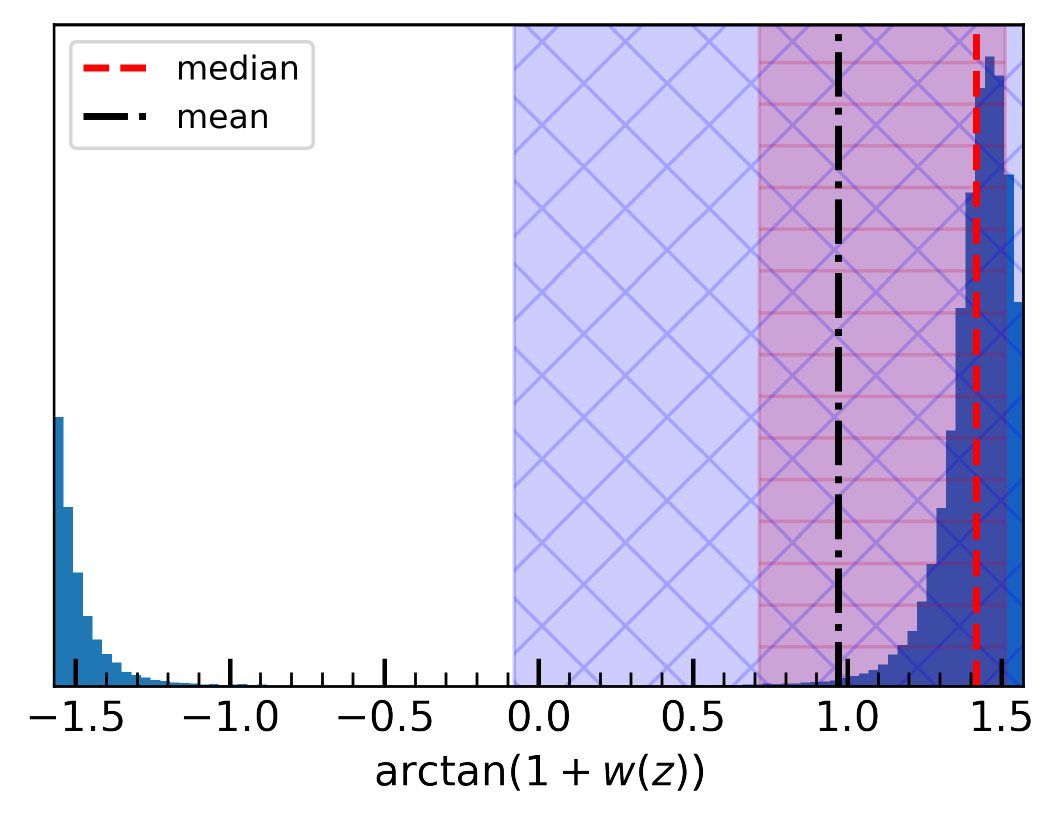}
		}
\caption{Histograms of the compactified DE equation of state $\arctan \left( 1 + w(z) \right)$ (a) at $z = 0$ when it is nearly Gaussian-distributed and (b) at $z \sim 2.3$ when it is bimodal. The $H_0^{\text{P18}}$ prior is used for this reconstruction. The red dashed and black dash-dotted vertical lines show the median and the mean, respectively. The red $'-'$-hatched region shows the 34.1\% probability mass surrounding the median above and below while the blue $'\times'$-hatched region shows the $1\sigma$ confidence interval from the mean where $\sigma$ is the standard deviation, or the second moment, of the distribution.}
\label{fig:dist_examples}
\end{figure}

In contrast with the bare $w(z)$, its compactified version $\arctan \left( 1 + w(z) \right)$ easily converges for any redshift and the posteriors can be defined even beyond the region earlier than when the DE density vanishes. Needless to say, the convergence of a distribution is very important in order for a computed quantity to have physical meaning. Its usage in the previous work highlighted the important result that the DE equation of state tilts further away from a Gaussian posterior for higher redshifts; but most importantly, it points to the fact that the distribution can even be bimodal particularly within the temporal vicinity of the singularity of $w(z)$. Understandably, this comes with some statistical quirks, a crucial one being that since the distributions evolve from being Gaussian at lower redshifts to generally bimodal at higher ones. This is illustrated in Figure \ref{fig:dist_examples} where the distribution themselves are taken from the GP reconstructed compactified DE equation of state at $z = 0$ and at $z \sim 2.3$ with the $H_0^{\text{P18}}$ prior. Taking in this insight, it was argued in Ref. \cite{Bernardo:2021qhu} that the median surrounded by $34.1\%$ of its probability mass from above and below can be considered as a reasonable generalization to the Gaussian-anchored mean and sigma statistics. 

\begin{figure}[h!]
\center
	\subfigure[ $H_0^{\text{P18}} = 67.4 \pm 0.5$ km s$^{-1}$Mpc$^{-1}$ ]{
		\includegraphics[width = 0.475 \textwidth]{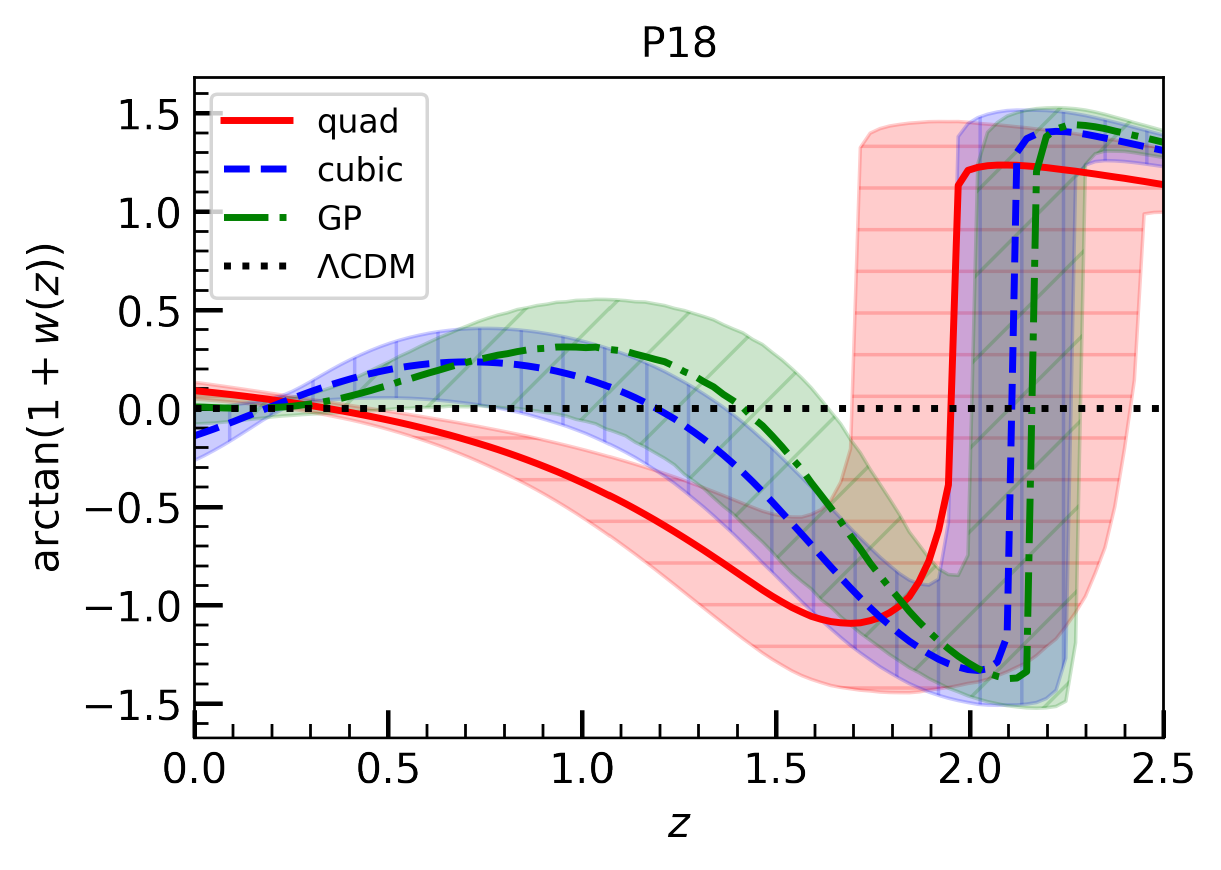}
		}
	\subfigure[ $H_0^{\text{TRGB}} = 69.8 \pm 1.9$ km s$^{-1}$Mpc$^{-1}$ ]{
		\includegraphics[width = 0.475 \textwidth]{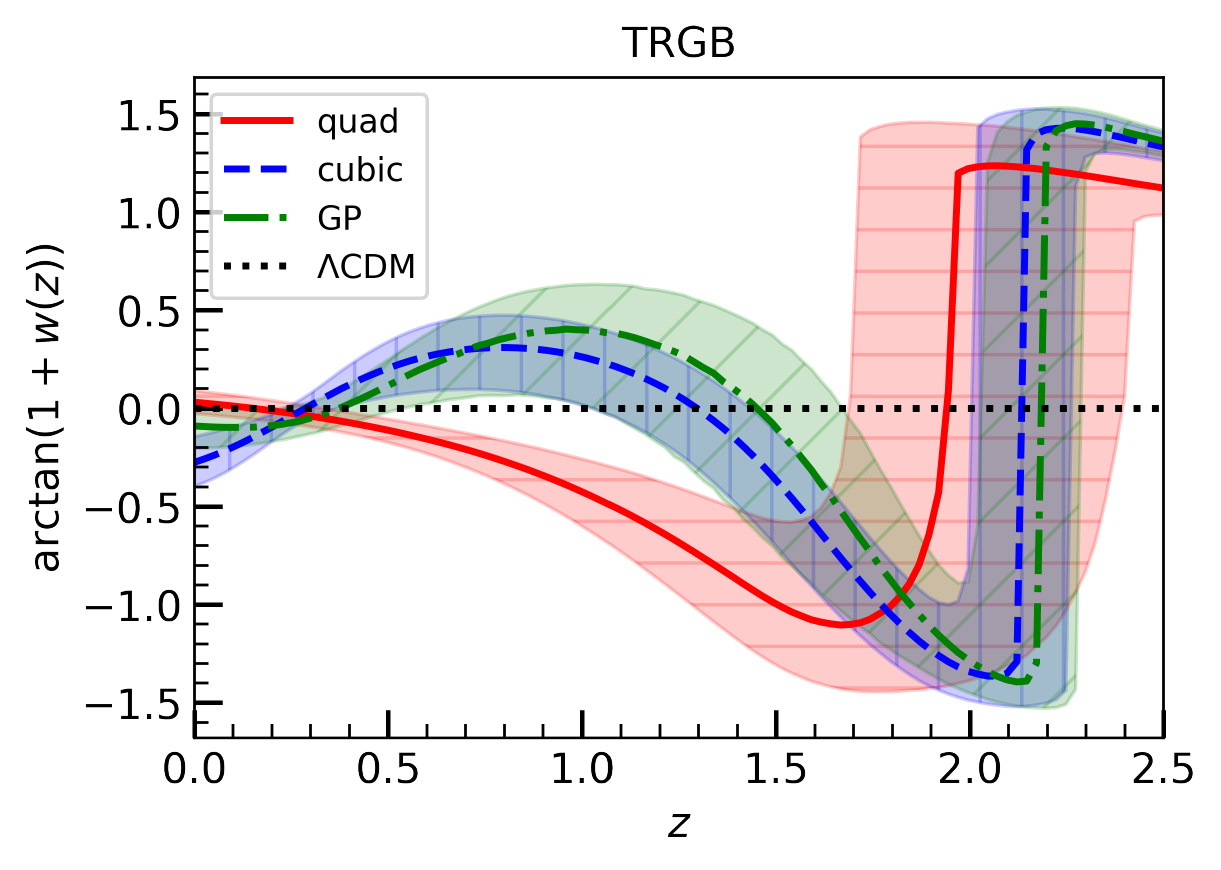}
		}
	\subfigure[ $H_0^{\text{A21}} = 71.5 \pm 1.8$ km s$^{-1}$Mpc$^{-1}$ ]{
		\includegraphics[width = 0.475 \textwidth]{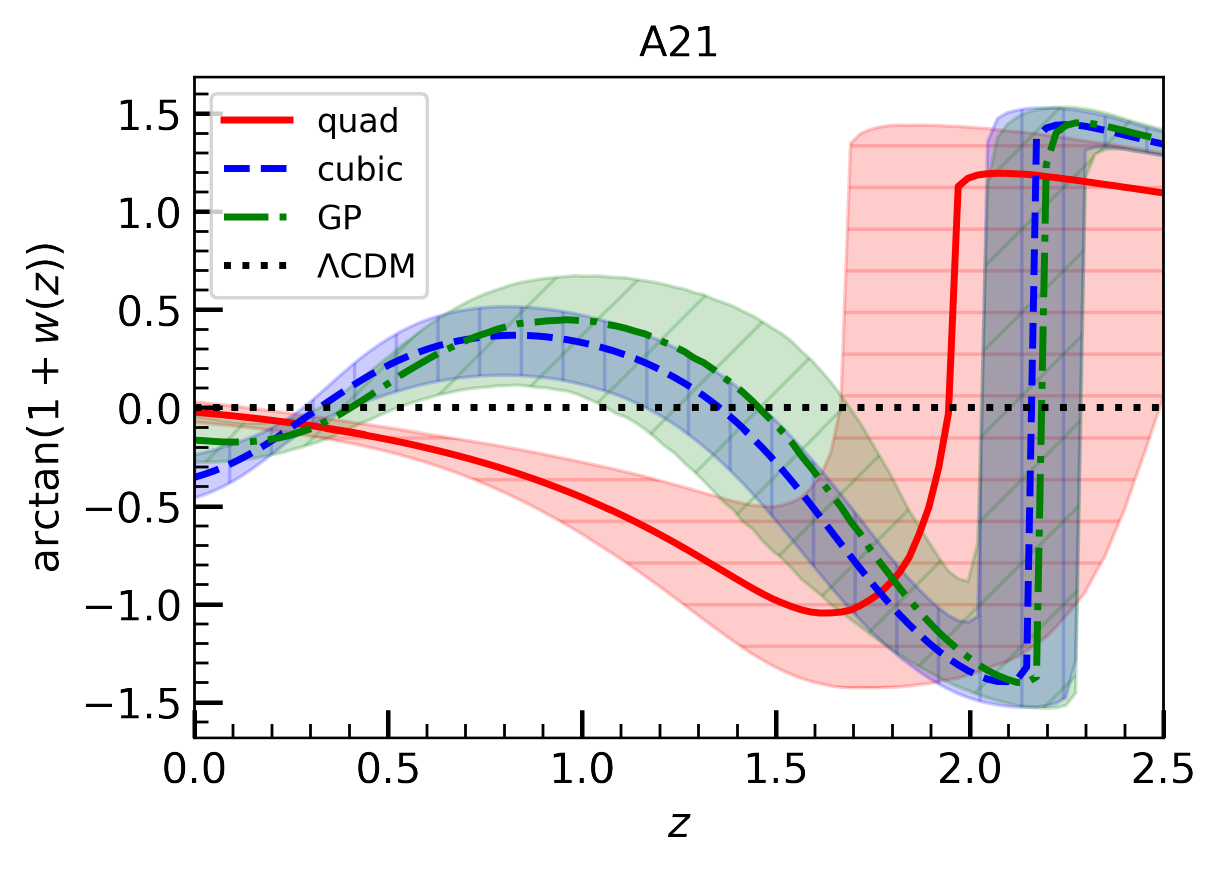}
		}
	\subfigure[ $H_0^{\text{R21}} = 73.04 \pm 1.04$ km s$^{-1}$Mpc$^{-1}$ ]{
		\includegraphics[width = 0.475 \textwidth]{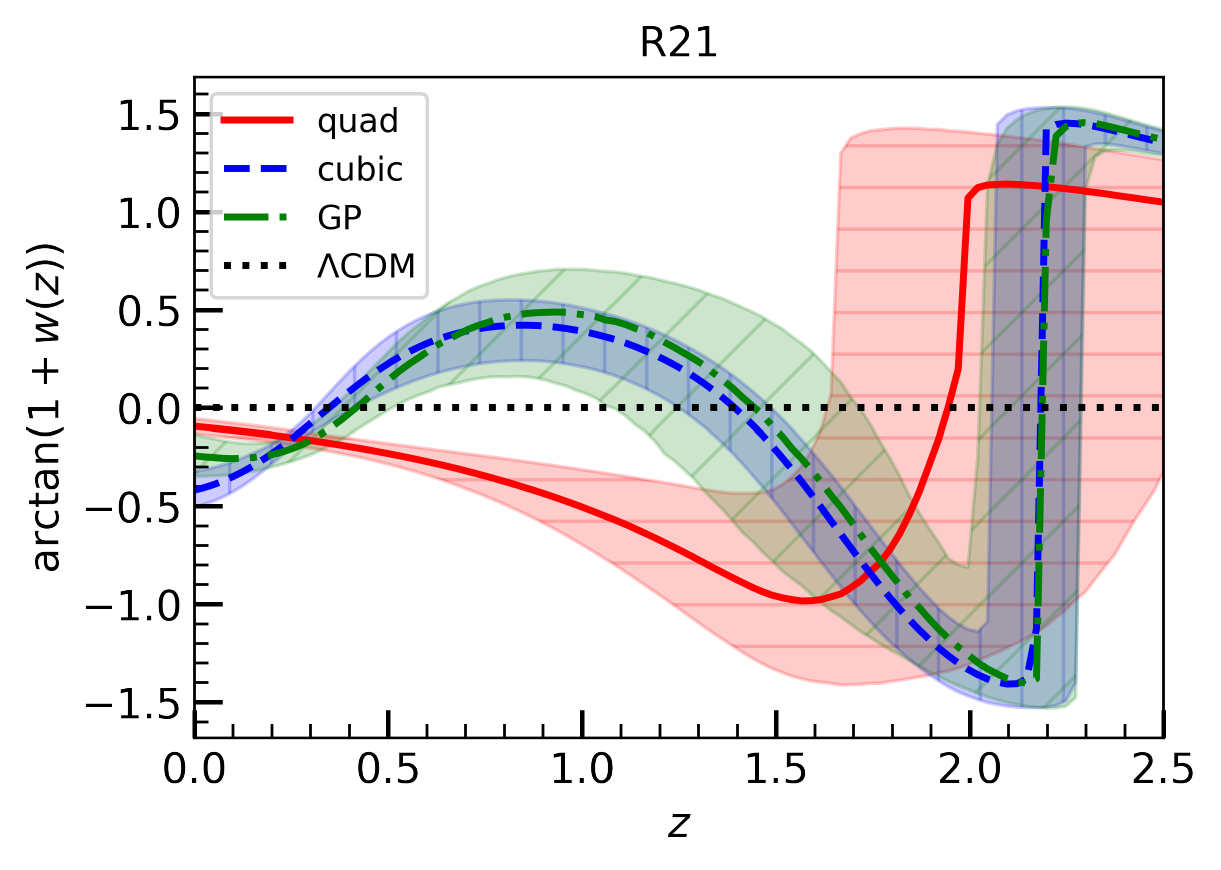}
		}
\caption{The reconstructed compactified DE equation of state per method derived from the base Hubble data (CC + BAO) for each $H_0$ prior: (a) P18, (b) TRGB, (c) A21, and (d) R21. Legends: ``quad'' and ``cubic'' stands for the quadratic and cubic parametrized DE, respectively; ``GP'' for the Gaussian processes. The colored-hatched regions show the median and the surrounding $34.1\%$ probability mass above and below. Hatches: (quad: ``$-$''), (cubic: ``$|$''), (GP: ``$/$'').}
\label{fig:awz_rec_per_method}
\end{figure}

Figure \ref{fig:dist_examples} illustrates this. When the posteriors are reliably Gaussian (e.g., Figure \ref{fig:dist_examples}(a)), the median together with $34.1\%$ of the probability mass surrounding it effectively reduces to the one-sigma probability density. However, when the distribution is bimodal (e.g., Figure \ref{fig:dist_examples}(b)), as is generally the case for the DE equation of state, it can be seen that the one-sigma region no longer captures the essence of the actual posterior, and even misleads to values outside of the range of the random variable. On the other hand, the median and its surrounding $34.1\%$ of the probability mass always capture the place in probability space where the density is localized regardless of the true shape of the distribution and the domain of the random variable. The standard Gaussian distribution is undoubtably an excellent approximation to true posteriors in light of the central limit theorem. However, it may also happen to be an oversimplification in special cases, including the case of the DE equation of state, where the distribution cannot be accurately described any longer by only the first two moments. Thus, we rely instead on the generalized statistic of the median and its surrounding mass in presenting our reconstructed DE equation of state for each of the methods considered in this work. The results are shown in Figure \ref{fig:awz_rec_per_method} for each of the methods and prior $H_0$ values. The $\Lambda$CDM curves arise as horizontal dotted lines.

We find again that the cubic method and the GP more or less share the same shape in terms of $\arctan \left( 1 + w(z) \right)$ while the quadratic method follows a stiffer trend. This is likely due to the quadratic method having one less parameter than the cubic case, and so is naturally the more rigid parametric method. On the other hand, the GP is inherently flexible owing to its roots in machine learning. Most importantly, all of the method agree about deviating from the $\Lambda$CDM model, suggesting a dynamical DE. When using the $H_0^{\text{P18}}$ prior (Figure \ref{fig:awz_rec_per_method}(a)), this deviation can be seen at low, intermediate, and high redshifts. However, it becomes particularly noticeable at the higher redshifts close to the singularity of $w(z)$ when the median of $\arctan(1 + w(z))$ is near zero. Calling this redshift $z_\text{tr} > 0$ corresponding to the transition of the DE density from a positive to a negative value, i.e., $w(z_\text{tr}) = -1$ or $\arctan \left( 1 + w(z_\text{tr}) \right) = 0$, the following can be observed consistently: $z_\text{tr}^\text{quad} < z_\text{tr}^\text{cubic} < z_\text{tr}^\text{GP}$ where $z_\text{tr}^{i}$ is the transition redshift for method $i$. It should be noted that $z_\text{tr}$ is additionally marked by the largest uncertainty. The trend extends to the other $H_0$ priors, suggesting that the deviation from $\Lambda$CDM goes beyond the methodology.

The deviation from the $\Lambda$CDM model becomes even stronger with an $H_0$ prior further away from the Planck $H_0$ prior. Figures \ref{fig:awz_rec_per_method}(b-d) show this at the higher redshifts ($z \gtrsim 2$) when the $\Lambda$CDM model line is just far away from the mass of the distribution. Interestingly, a deviation is also reflected at redshifts close to $z = 0$, suggesting that the $\Lambda$CDM model is disfavored by the analyses for the largest $H_0^{\text{R21}}$ prior (Figure \ref{fig:awz_rec_per_method}(d)) for generally any method. The reason for witnessing this low redshift deviation from $\Lambda$CDM for the larger $H_0$ priors could be due to the use of a common matter fraction prior $\Omega_{m0}^{\text{P18}}$ for all of the methods. We remind that $\Omega_{m0}^{\text{P18}} h^2$ was considered in order to make a sensible assessment that works for both parametric and nonparametric approaches. Without such a prior, it is not possible to obtain $X(z)$ from the GP which only directly reconstructs the data set it is given, in this case $H(z)$ data. However, while this may be true, there is a clear deviation from even within the Planck $H_0$ prior that transcends this reasoning, e.g., Figure \ref{fig:dist_examples}(b), the actual distribution of the samples at $z \sim 2.3$ for the GP with $H_0^{\text{P18}}$, supports this deviation from the concordance model.

In this analysis, all three approaches agree that there is some deviation from the standard model which is an intriguing result. This appears for all priors on $H_0$ despite the reliance on the Planck value of $\Omega_{m0} h^2$ which is obtained from the cosmic microwave background $\Lambda$CDM constraint. As already discussed, these priors have a compromising impact on the analysis since they may have been obtained in conjunction with some reliance on $\Lambda$CDM cosmology. Nonetheless, the analysis in each case points to a possible deviation from $\Lambda$CDM. More data may reveal further deviations as what may happen as more prior values are reported in the literature.

\section{Extended analysis with supernovae and Horndeski priors}
\label{sec:extended_analysis}

In Sections \ref{subsec:de_from_Hz} and \ref{subsec:compactified_de_eos}, we considered the base Hubble data alone in order to make an assessment of parametric and nonparametric methods. We now examine the robustness of the previous results by including supernovae observations in the analysis. This is done using the full Pantheon sample \cite{Scolnic:2017caz} for the parametric methods and through the CANDELS and CLASH Multi-Cycle Treasury (MCT) data \cite{Riess:2017lxs} for the GP.

In practice, as mentioned in Section \ref{subsec:late_time_cosmic_data}, for the parametric methods, we consider the 1048 SNe observations from Pantheon \cite{Scolnic:2017caz} and sum up the log-likelihoods for the Hubble data and SNe in the Bayesian analysis. On the other hand, for the GP, we take the compressed $E(z)$ measurements from the CANDELS and CLASH MCT \cite{Riess:2017lxs} and use the $H_0$ priors to convert this into $H(z) = H_0 E(z)$ measurements, which are then subsequently used together with the base Hubble data \cite{Gomez-Valent:2018hwc, Briffa:2020qli}.

\begin{figure}[h!]
\center
	\subfigure[ $H_0^{\text{P18}} = 67.4 \pm 0.5$ km s$^{-1}$Mpc$^{-1}$ ]{
		\includegraphics[width = 0.475 \textwidth]{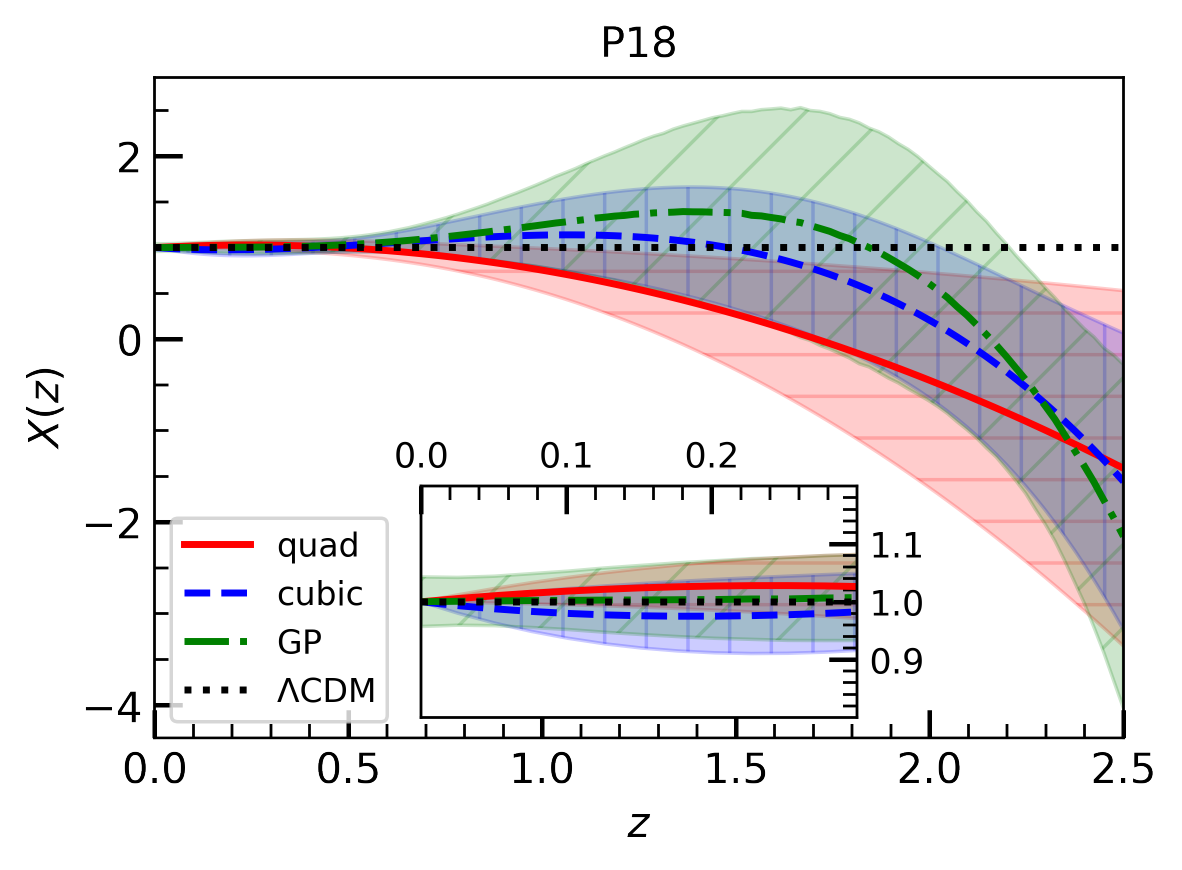}
		}
	\subfigure[ $H_0^{\text{TRGB}} = 69.8 \pm 1.9$ km s$^{-1}$Mpc$^{-1}$ ]{
		\includegraphics[width = 0.475 \textwidth]{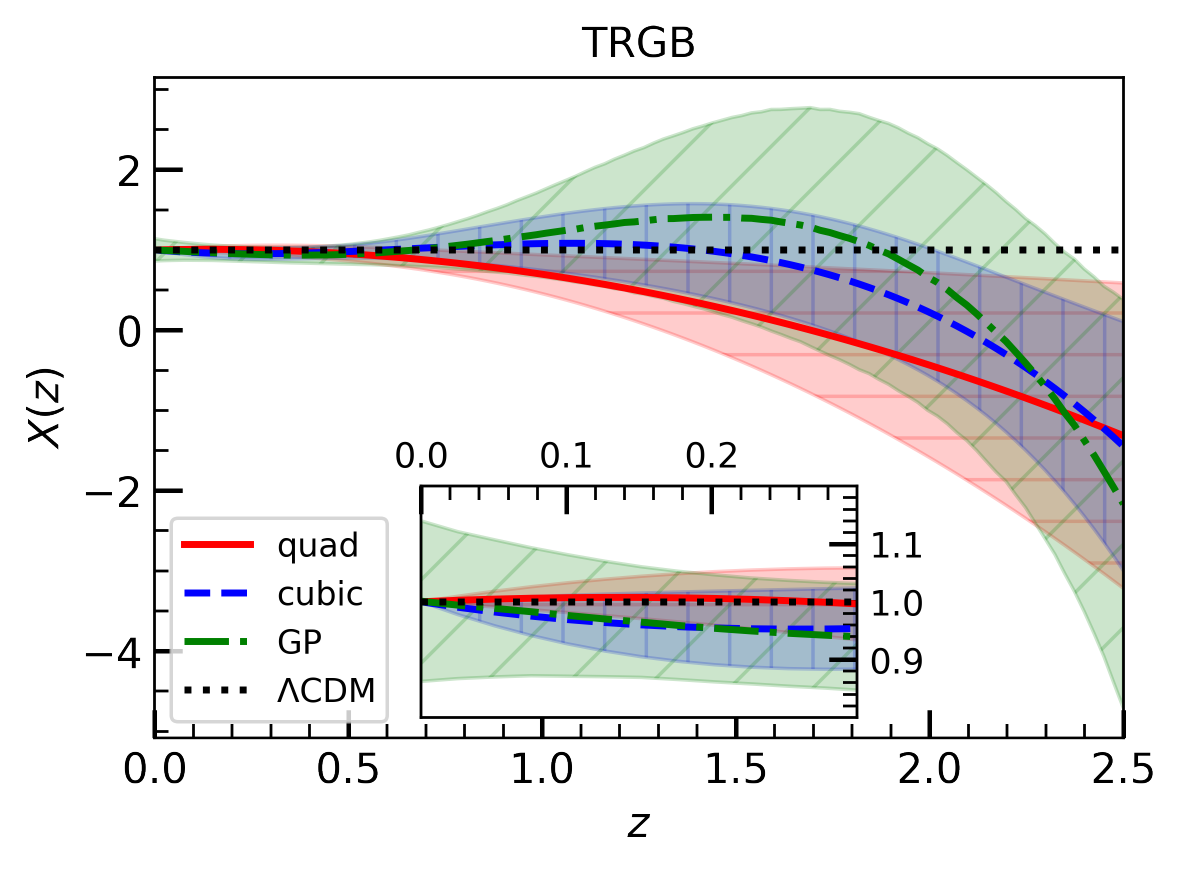}
		}
	\subfigure[ $H_0^{\text{A21}} = 71.5 \pm 1.8$ km s$^{-1}$Mpc$^{-1}$ ]{
		\includegraphics[width = 0.475 \textwidth]{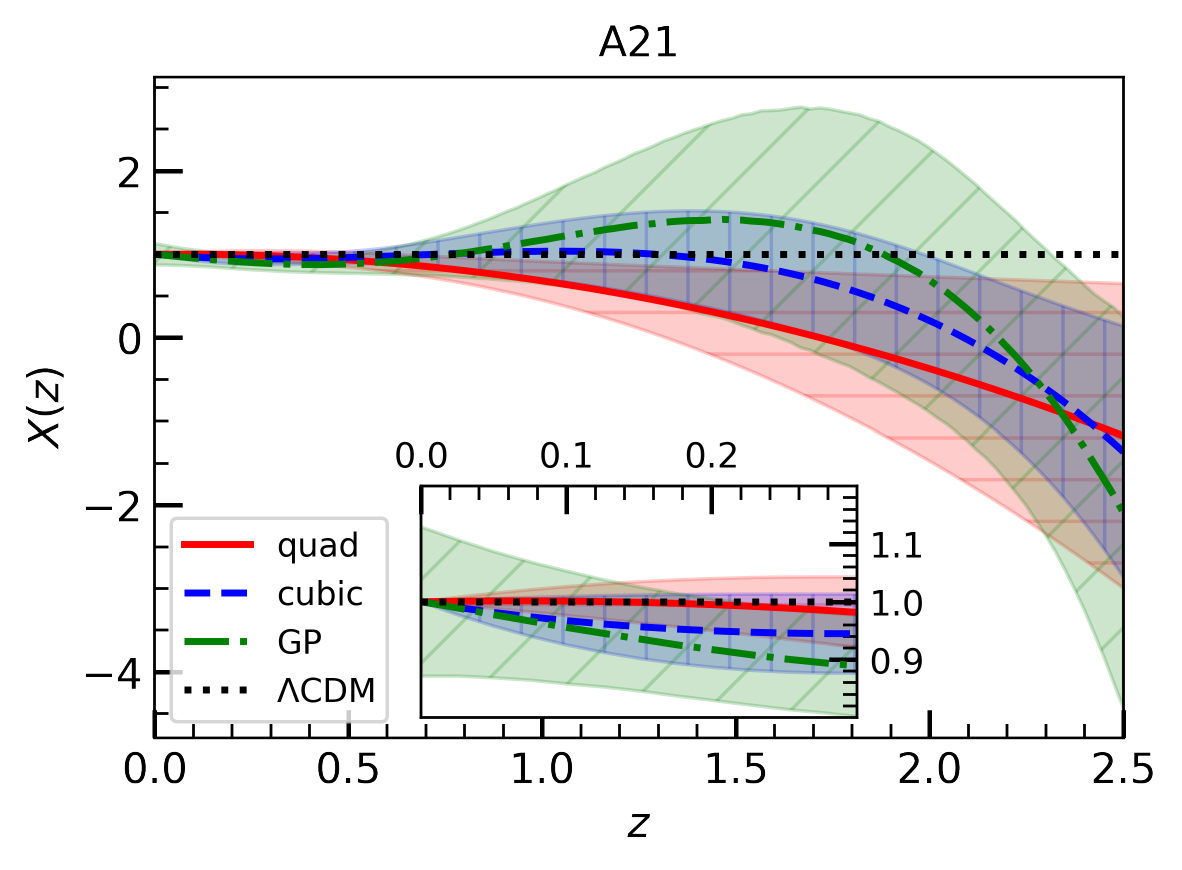}
		}
	\subfigure[ $H_0^{\text{R21}} = 73.04 \pm 1.04$ km s$^{-1}$Mpc$^{-1}$ ]{
		\includegraphics[width = 0.475 \textwidth]{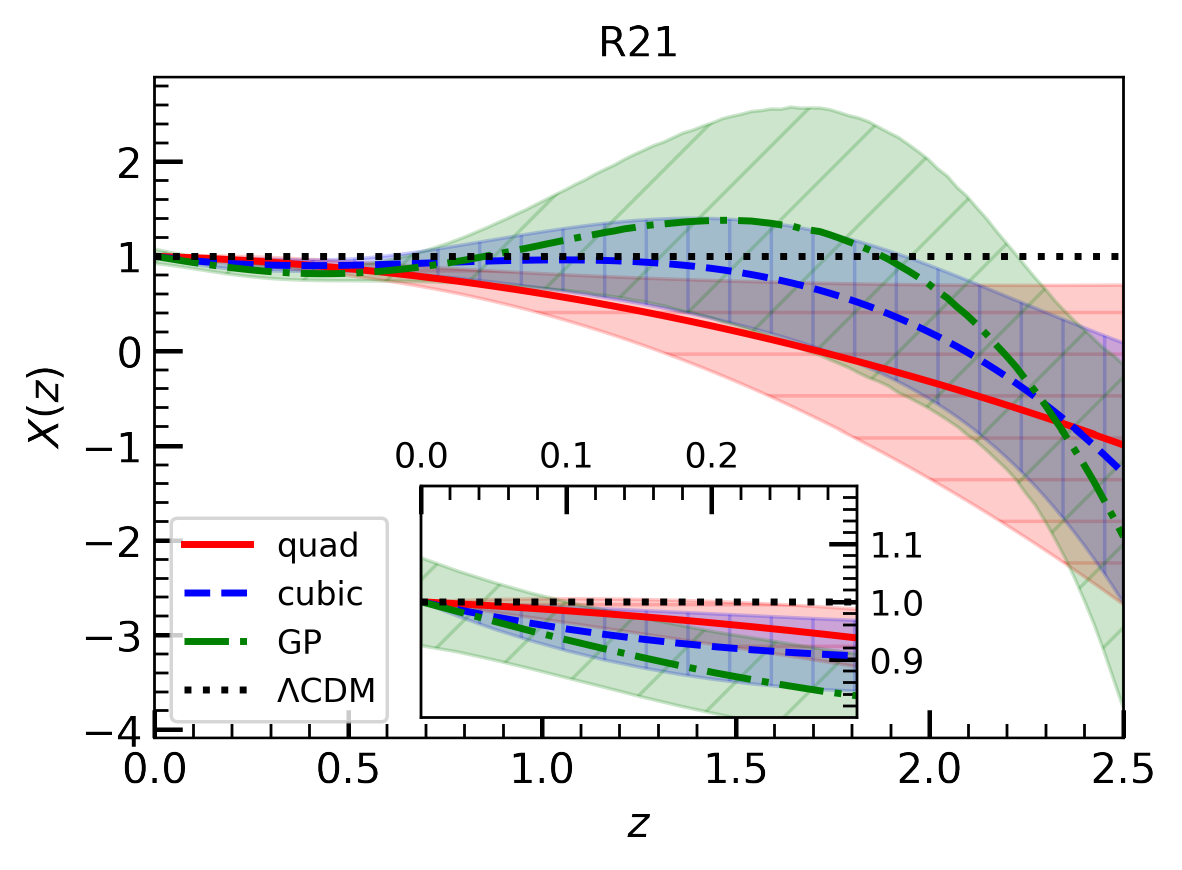}
		}
\caption{The reconstructed normalized DE density per method derived from the base Hubble data (CC + BAO) and supernovae observations (Pantheon/MCT) for each $H_0$ prior: (a) P18, (b) TRGB, (c) A21, and (d) R21. Legends: ``quad'' and ``cubic'' stands for the quadratic and cubic parametrized DE, respectively; ``GP'' for the Gaussian processes. The colored-hatched regions show the $2\sigma$-region of the reconstructions while the insets reveal a magnified view of the low redshift region $z \in (0, 0.3)$. Hatches: (quad: ``$-$''), (cubic: ``$|$''), (GP: ``$/$'').}
\label{fig:Xz_rec_wSNe}
\end{figure}

\begin{figure}[h!]
\center
	\subfigure[ $H_0^{\text{P18}} = 67.4 \pm 0.5$ km s$^{-1}$Mpc$^{-1}$ ]{
		\includegraphics[width = 0.475 \textwidth]{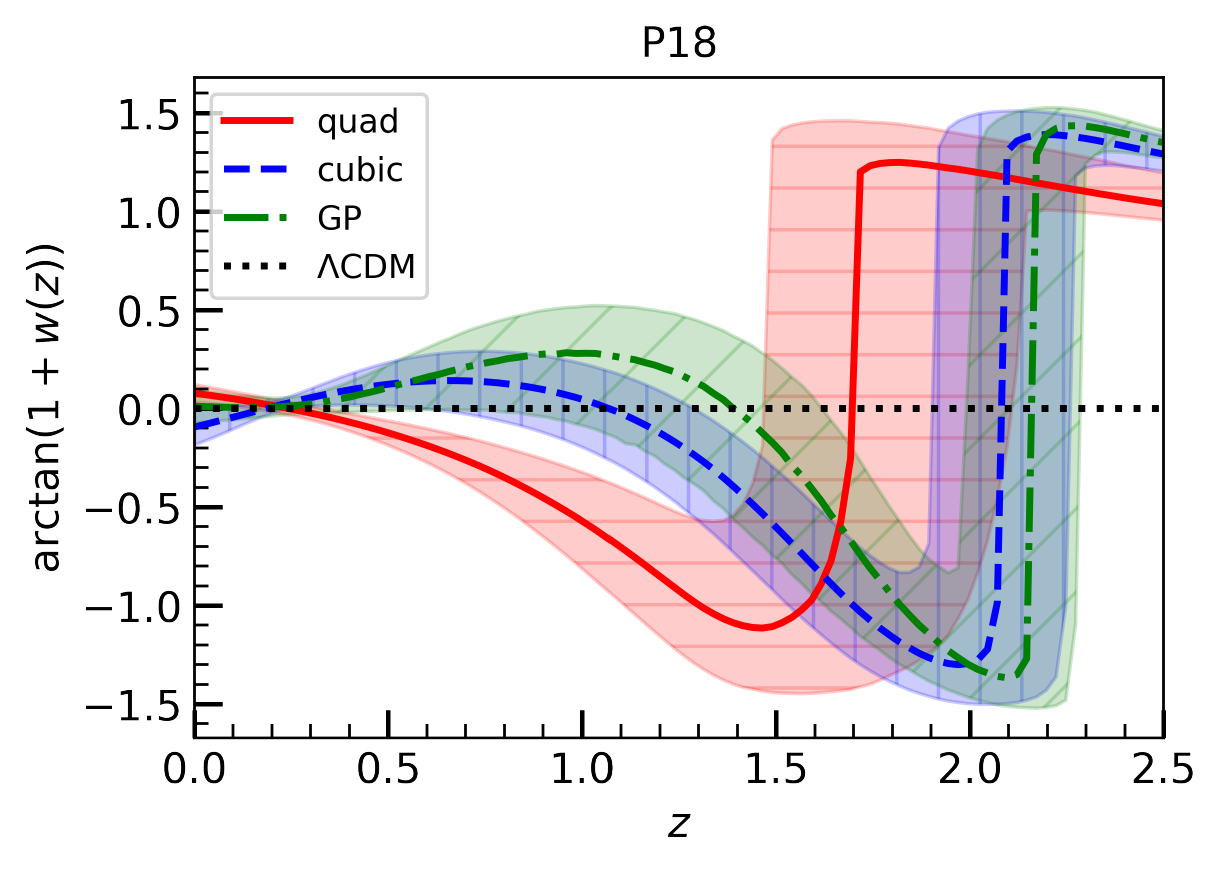}
		}
	\subfigure[ $H_0^{\text{TRGB}} = 69.8 \pm 1.9$ km s$^{-1}$Mpc$^{-1}$ ]{
		\includegraphics[width = 0.475 \textwidth]{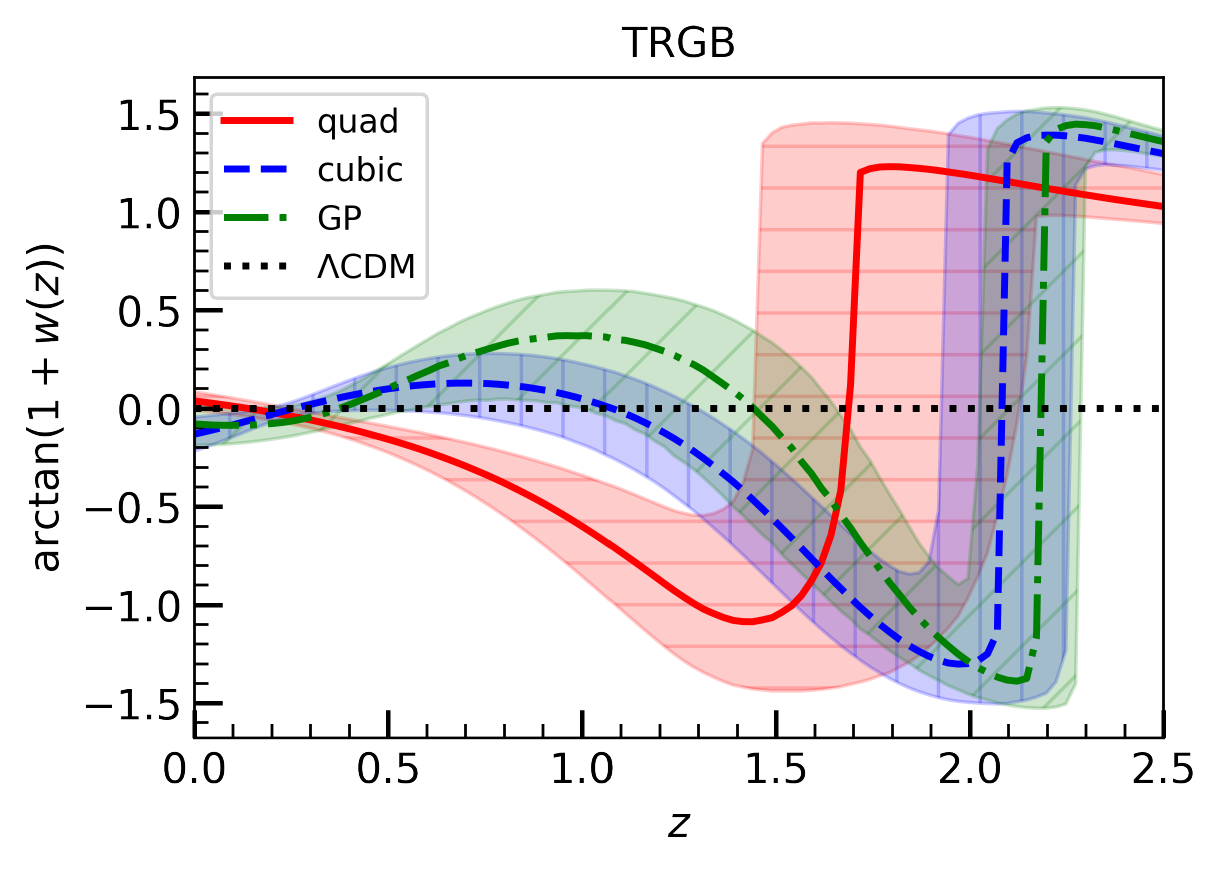}
		}
	\subfigure[ $H_0^{\text{A21}} = 71.5 \pm 1.8$ km s$^{-1}$Mpc$^{-1}$ ]{
		\includegraphics[width = 0.475 \textwidth]{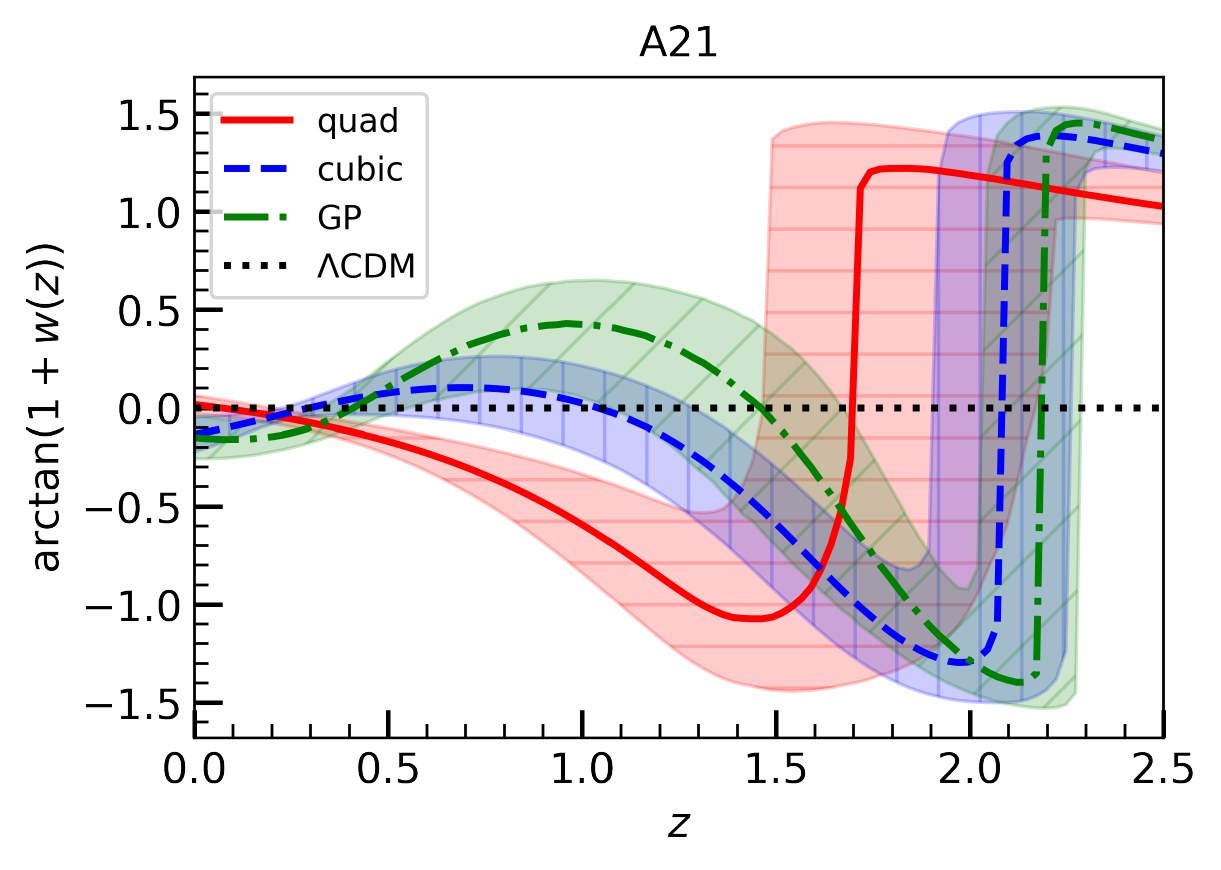}
		}
	\subfigure[ $H_0^{\text{R21}} = 73.04 \pm 1.04$ km s$^{-1}$Mpc$^{-1}$ ]{
		\includegraphics[width = 0.475 \textwidth]{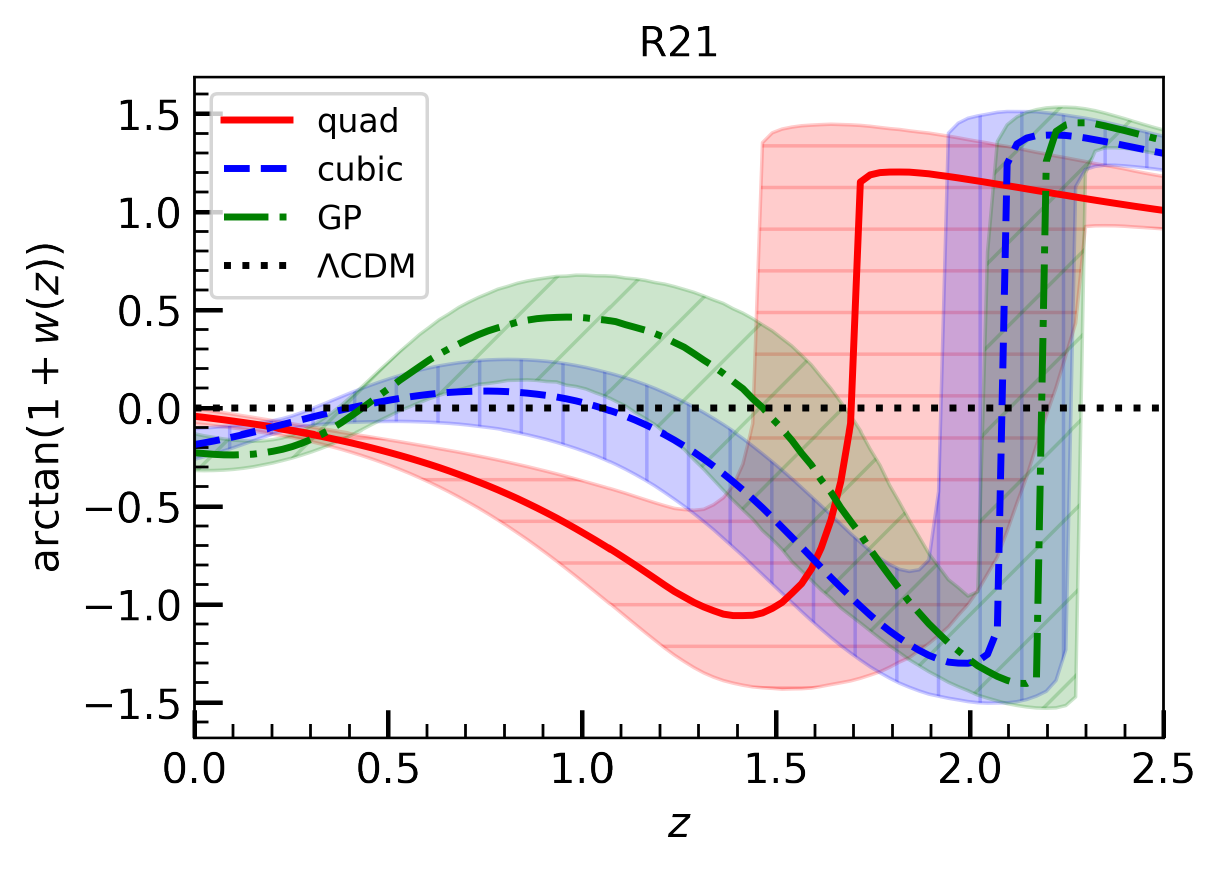}
		}
\caption{The reconstructed compactified DE equation of state per method derived from the base Hubble data (CC + BAO) and supernovae observations (Pantheon/MCT) for each $H_0$ prior: (a) P18, (b) TRGB, (c) A21, and (d) R21. Legends: ``quad'' and ``cubic'' stands for the quadratic and cubic parametrized DE, respectively; ``GP'' for the Gaussian processes. The colored-hatched regions show the median and the surrounding $34.1\%$ probability mass above and below. Hatches: (quad: ``$-$''), (cubic: ``$|$''), (GP: ``$/$'').}
\label{fig:awz_rec_wSNe}
\end{figure}

The reconstructed normalized DE and compactified DE equation of state are shown in Figures \ref{fig:Xz_rec_wSNe} and \ref{fig:awz_rec_wSNe}, respectively, together with the $\Lambda$CDM curves being horizontal dotted lines.

This can be seen to be more or less similar to the reconstructions obtained with only the base Hubble data, especially for the normalized DE posteriors shown in Figure \ref{fig:Xz_rec_wSNe}. A possible explanation to this is that the supernovae apparent brightness are related to the Hubble function (and by extension to the DE density) by means of an integration. The same observations made with the $X(z)$ posteriors with the base Hubble data (CC + BAO) therefore holds in this case including information from supernovae (Pantheon/MCT). That is, there is a striking deviation from the $\Lambda$CDM model of more than $2\sigma$ at high redshifts $z \gtrsim 2.3$ for any method and $H_0$ prior, while when using TRGB, A21, and R21 $H_0$ priors, there are also resolvable deviations from the concordance model of about $2\sigma$ at low redshifts $z \sim 0.3$. On the other hand, the compactified DE equation of state (Figure \ref{fig:awz_rec_wSNe}) turns out to be more sensitive to the addition of supernovae presumably because it is a functional of both $X(z)$ and $X'(z)$ (or alternatively, $H(z)$ and $H'(z)$). The main difference can be seen with respect to the cubic method and the GP posteriors. Whereas both methods generally agreed in the shape, this time with the inclusion of the SNe data set, part of the cubic method posteriors can be seen fall within the corresponding GP posteriors. This again holds independent of the choice of the $H_0$ prior and also preserves the order $z_\text{tr}^\text{quad} < z_\text{tr}^\text{cubic} < z_\text{tr}^\text{GP}$ of the transition redshift.

The overall picture from this is that the deductions that were made in the previous sections about the dynamical nature of DE still hold, or even strengthened, with the inclusion of supernovae observations. Corresponding plots of a diagnostic function $X'(z)$ supporting the same results are shown in \ref{sec:diagnostic}.

Now, we consider for further insight the inclusion of a Horndeski model that is complemented by the GP and by design constructed to match the late-time data \cite{Bernardo:2021qhu}. The gravitational action of theory is given by \cite{Kobayashi:2019hrl}
\begin{equation}
\label{eq:horndeski_action}
	\mathcal{S}_{\rm H} = \int {\rm d}^4 x \sqrt{-g}  \left( K(\phi, X) - G(\phi, X) \Box \phi + \dfrac{M_\text{Pl}^2}{2} R + \cdots \right)\,,
\end{equation}
where $g_{ab}$ is the metric, $g$ is its determinant, $R$ is the Ricci scalar, $M_\text{Pl}^2 = c^4/\left( 8 \pi G \right)$, and $\phi$ is the scalar field. We refer to the functions $K$ and $G$ as the $k$-essence and braiding potentials, respectively. The terms in the ellipses corresponds to conformal coupling terms and those that change the speed of gravitational waves. This action (\ref{eq:horndeski_action}) describes the most-general curvature-based, scalar-tensor theory with only second-order field equations \cite{Horndeski:1974wa,Kobayashi:2019hrl}. Its teleparallel generalization is also worth noting due to a richer phenomenological space allowed by a relaxed Lovelock theorem in torsion-based gravity \cite{Bahamonde:2019shr,Bahamonde:2019ipm,Bahamonde:2020cfv}. The natural way to proceed from Eq. (\ref{eq:horndeski_action}) is to specify the free potentials of the scalar field, leading to various well studied gravity models, derive the field equations by functional differentiation, and then study its phenomenological implications. Many progress have been made in this direction which is making the reasonable trade of functions for a finite number of parameters to be constrained. Alternatively, in Ref. \cite{Bernardo:2021qhu}, a different route was proposed in which the potentials are themselves reconstructed by inverting the Friedmann equations and utilizing the GP with late-time data sets. We shall see this in action for the ``designer Horndeski'' (HDES) model \cite{Arjona:2019rfn}.

We briefly describe the details of the HDES model. The reader uninterested in the derivation may skip ahead. We go about by starting with the modified Friedmann equations of kinetic gravity braiding\footnote{KGB is the limit of Horndeski theory (Eq. (\ref{eq:horndeski_action})) when all of the terms in the ellipsis are ignored. This is a conservative choice nowadays due to the very tight constraint on the speed of gravitational waves.},
\begin{equation}
\label{eq:Feq_hdes}
    3H^2 = \rho - K(X) + 2 X K_X + 3 H \dot{\phi}^2 G_X\,,
\end{equation}
\begin{equation}
\label{eq:Peq_hdes}
    2 \dot{H} + 3 H^2 = -P - K(X) + 2 X \ddot{\phi} G_X\,,
\end{equation}
and the scalar field equation,
\begin{equation}
\label{eq:Seq_hdes}
\ddot{\phi} \left(-\dot{\phi} \left(3 H \left(G_{XX} \dot{\phi}^2+2 G_X\right)+K_{XX} \dot{\phi}\right)-K_X\right)-3 \dot{\phi} \left(G_X \dot{H} \dot{\phi}+3 G_X H^2 \dot{\phi}+H K_X\right) = 0\,,
\end{equation}
where the subscripts in the potentials $K(X)$ and $G(X)$ denote differentiation with respect to $X$. It is most noteworthy for our purposes that the scalar field equation (Eq. (\ref{eq:Seq_hdes})) can be written in terms of a conserved shift current $J$ such that
\begin{equation}
\label{eq:Seq_ss}
    \dot{J} + 3 H J = 0\,,
\end{equation}
where $J$ is explicitly
\begin{equation}
\label{eq:Seq_J}
    J = \dot{\phi} K_X + 3 H \dot{\phi}^2 G_X\,.
\end{equation}
In this way, its solution can be written as
\begin{equation}
\label{eq:J_hdes}
\dot{\phi} K_X + 3 H \dot{\phi}^2 G_X = \dfrac{\mathcal{J}}{a^3} \,,
\end{equation}
where $\mathcal{J}$ is an integration constant which we refer to as the \textit{shift charge}. When the potentials $K$ and $G$ are provided \textit{a priori}, as it usually is in the canon approach, the dynamics of the scalar field can therefore be fully identified with the intersection of this hypersurface (Eq. (\ref{eq:J_hdes})) with the Hamiltonian constraint (Eq. (\ref{eq:Feq_hdes})). However, when the potentials are unknown, the system can instead be closed by providing the explicit functional dependence between the Hubble function and the scalar field though a choice of $H(X)$. This leads to HDES in which the $k$-essence and braiding potentials can be shown to be
\begin{equation}
\label{eq:K_hdes}
K(X) = -3 H_0^2 \Omega_\Lambda + \dfrac{\mathcal{J} \sqrt{2X} H(X)^2}{ H_0^2 \Omega_{m0} } - \dfrac{ \mathcal{J} \sqrt{2X} \Omega_\Lambda }{ \Omega_{m0} }\,,
\end{equation}
and
\begin{equation}
\label{eq:GX_hdes}
G_X(X) = - \dfrac{ 2 \mathcal{J} H'(X) }{ 3 H_0^2 \Omega_{m0} }\,,
\end{equation}
respectively. This is where an improvement can be made using the GP as the right hand sides of Eqs. (\ref{eq:K_hdes}) and (\ref{eq:GX_hdes}) contain the Hubble function and its derivatives. By complementing Eqs. (\ref{eq:K_hdes}) and (\ref{eq:GX_hdes}) with the GP reconstructed Hubble function, we can then predict the data-driven shapes of the $k$-essence and braiding potentials rather than putting them in beforehand. We refer the reader to Refs. \cite{Arjona:2019rfn, Bernardo:2021qhu} for further details.

\begin{figure}[h!]
	\subfigure[ $k$-essence potential ]{
		\includegraphics[width = 0.475 \textwidth]{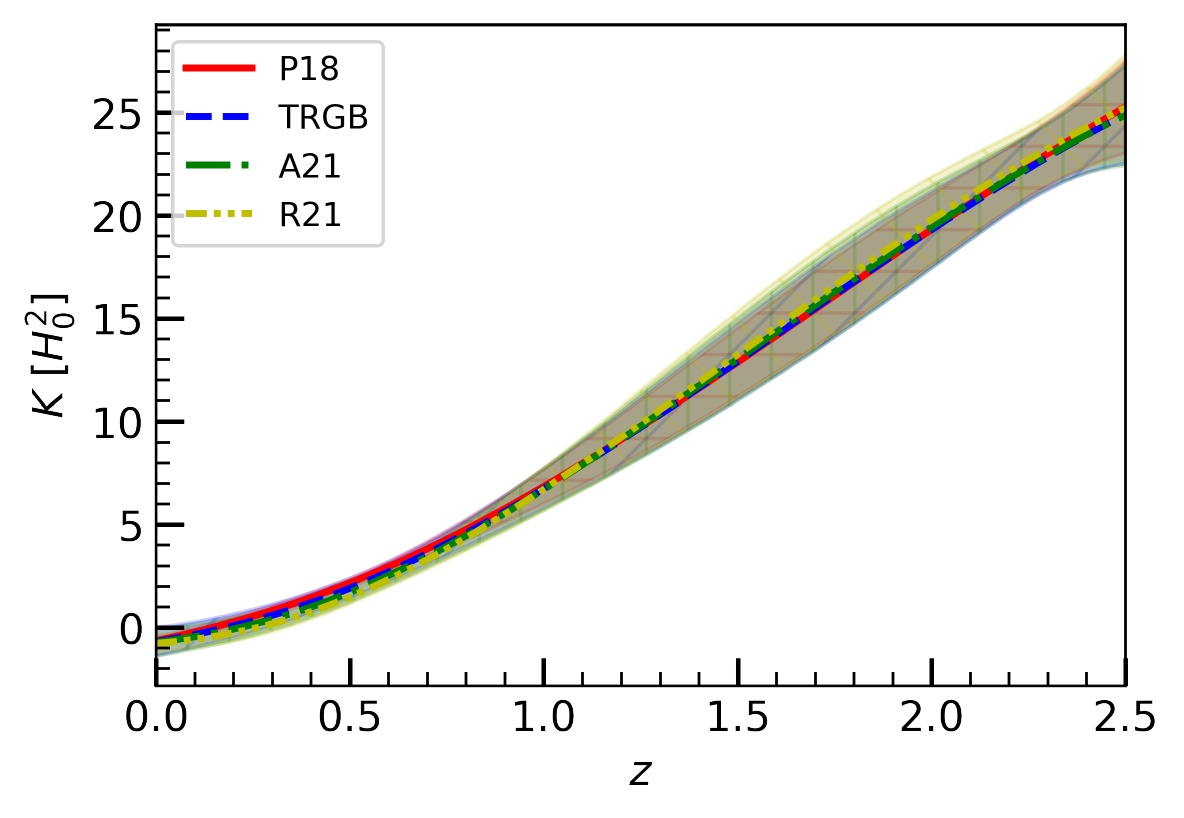}
		}
	\subfigure[ braiding potential ]{
		\includegraphics[width = 0.475 \textwidth]{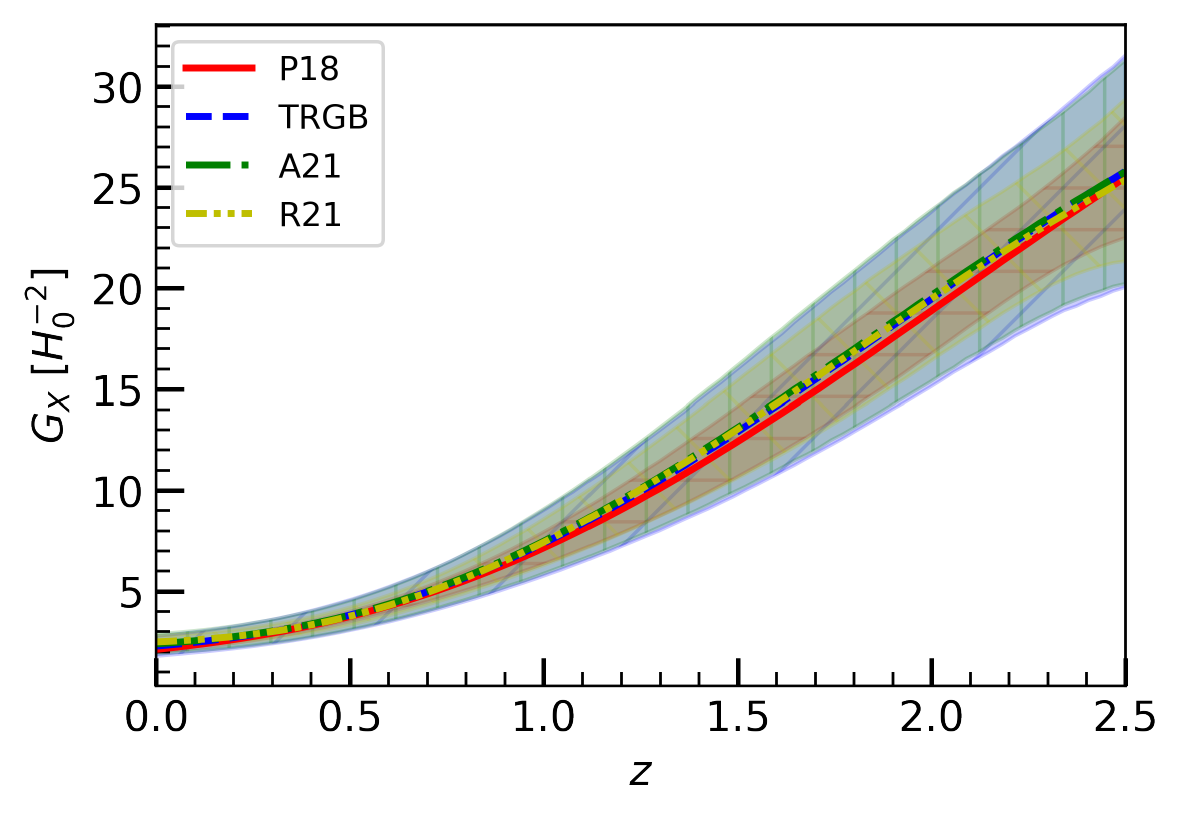}
		}
\caption{The reconstructed $k$-essence and braiding potentials of the designer Horndeski model $\left( c_0 = H_0^{n + 2}, n = 1, \mathcal{J} = H_0\right)$ derived from the base Hubble data (CC + BAO) and SNe (Pantheon/MCT) for each $H_0$ prior. The colored and hatched regions show the $2\sigma$ confidence interval of the reconstructions. Hatches: (P18: ``$-$''), (TRGB: ``$/$''), (A21: ``|''), (R21: ``\textbackslash'').}
\label{fig:hdes_potentials}
\end{figure}

We consider as a prior the tracker ansatz $X = c_0/H(X)^n$ which is motivated in Ref. \cite{Arjona:2019rfn}. In this functional relation of the scalar field and Hubble function, $c_0$ and $n$ are constants, with $c_0$ being in units of $H_0^{n + 2}$. We proceed in this work with the choices $c_0 = H_0^{n + 2}$, $n = 1$, and $\mathcal{J}  = H_0$ of the theory constants scaling with local Universe values. Figure \ref{fig:hdes_potentials} then shows the shape of the scalar field potentials reconstructed with GP and late-time data.

This shows that the resulting Horndeski model is describing modified gravity since neither of the potentials are flat (the $\Lambda$CDM limit of Horndeski gravity). The potentials instead monotonically evolve in redshift to a point when both are significantly larger than their low $z$ values. We should also point out that the reconstructions presented in Figure \ref{fig:hdes_potentials} reflect only a mild influence of the Hubble constant prior, i.e., the mean value of one reconstruction is within reasonable confidence intervals of the others. But, this is due to the potentials being expressed in units of $H_0$ (for numerical convenience). With this said, obviously, when the units are resumed, the influence of the $H_0$ priors will become transparent, either pushing the contours up or down depending on the size of the tension between two $H_0$ values. 

\begin{figure}[h!]
\center
	\subfigure[ $H_0^{\text{P18}} = 67.4 \pm 0.5$ km s$^{-1}$Mpc$^{-1}$ ]{
		\includegraphics[width = 0.475 \textwidth]{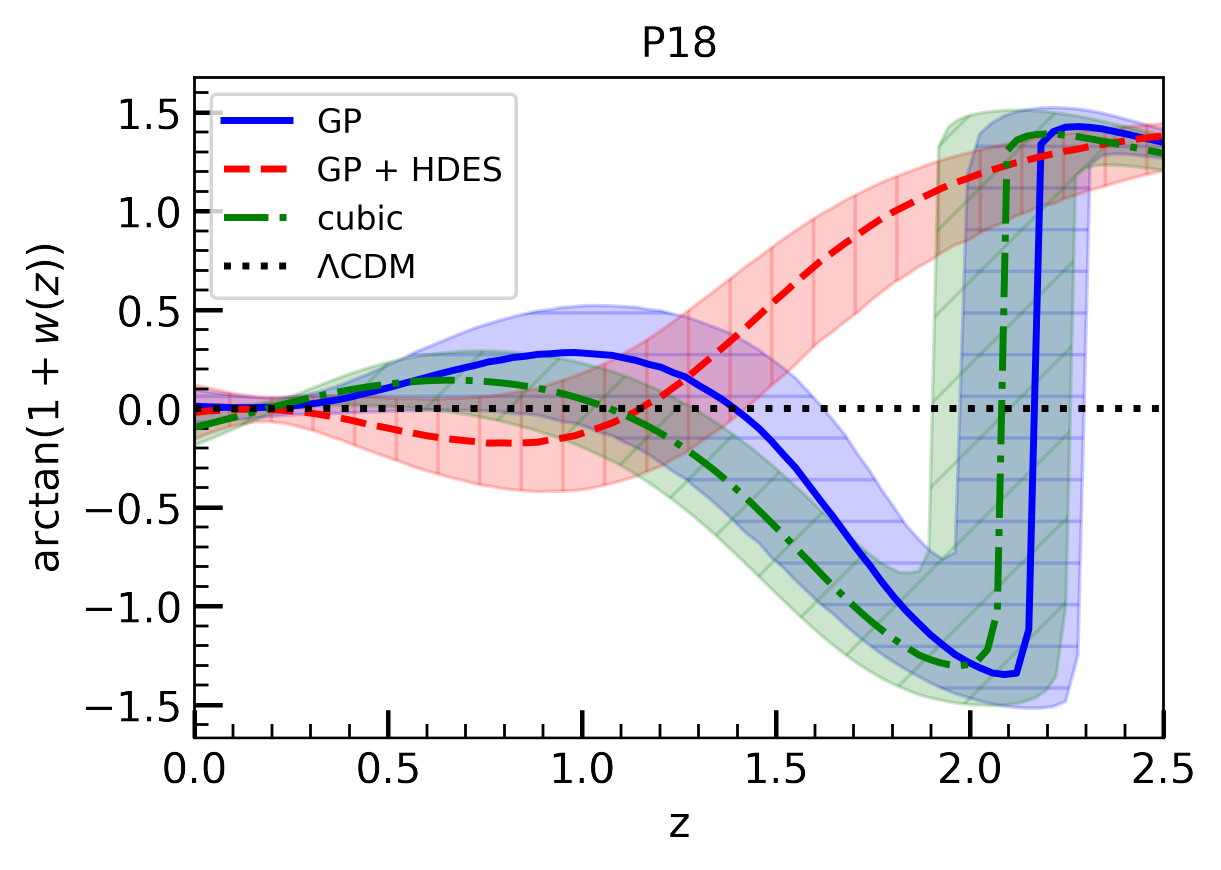}
		}
	\subfigure[ $H_0^{\text{TRGB}} = 69.8 \pm 1.9$ km s$^{-1}$Mpc$^{-1}$ ]{
		\includegraphics[width = 0.475 \textwidth]{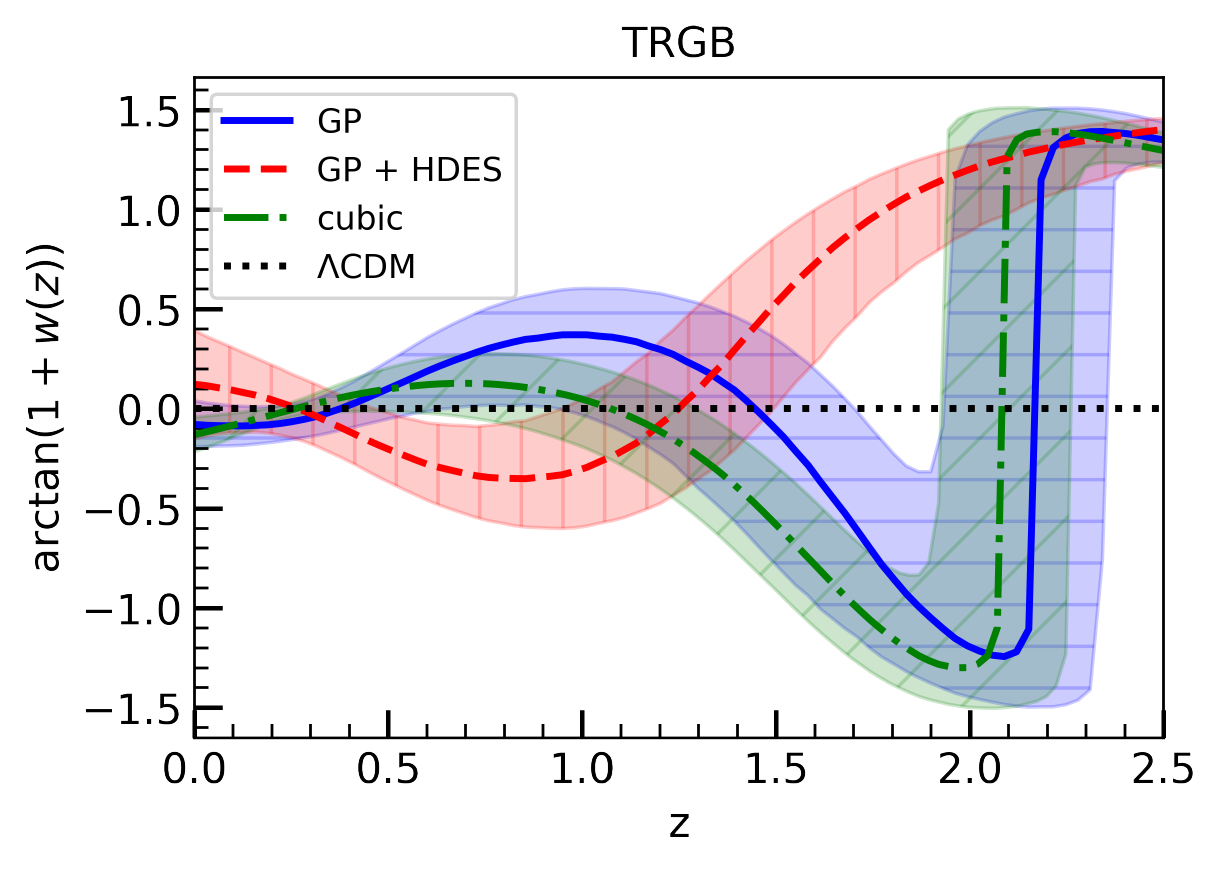}
		}
	\subfigure[ $H_0^{\text{A21}} = 71.5 \pm 1.8$ km s$^{-1}$Mpc$^{-1}$ ]{
		\includegraphics[width = 0.475 \textwidth]{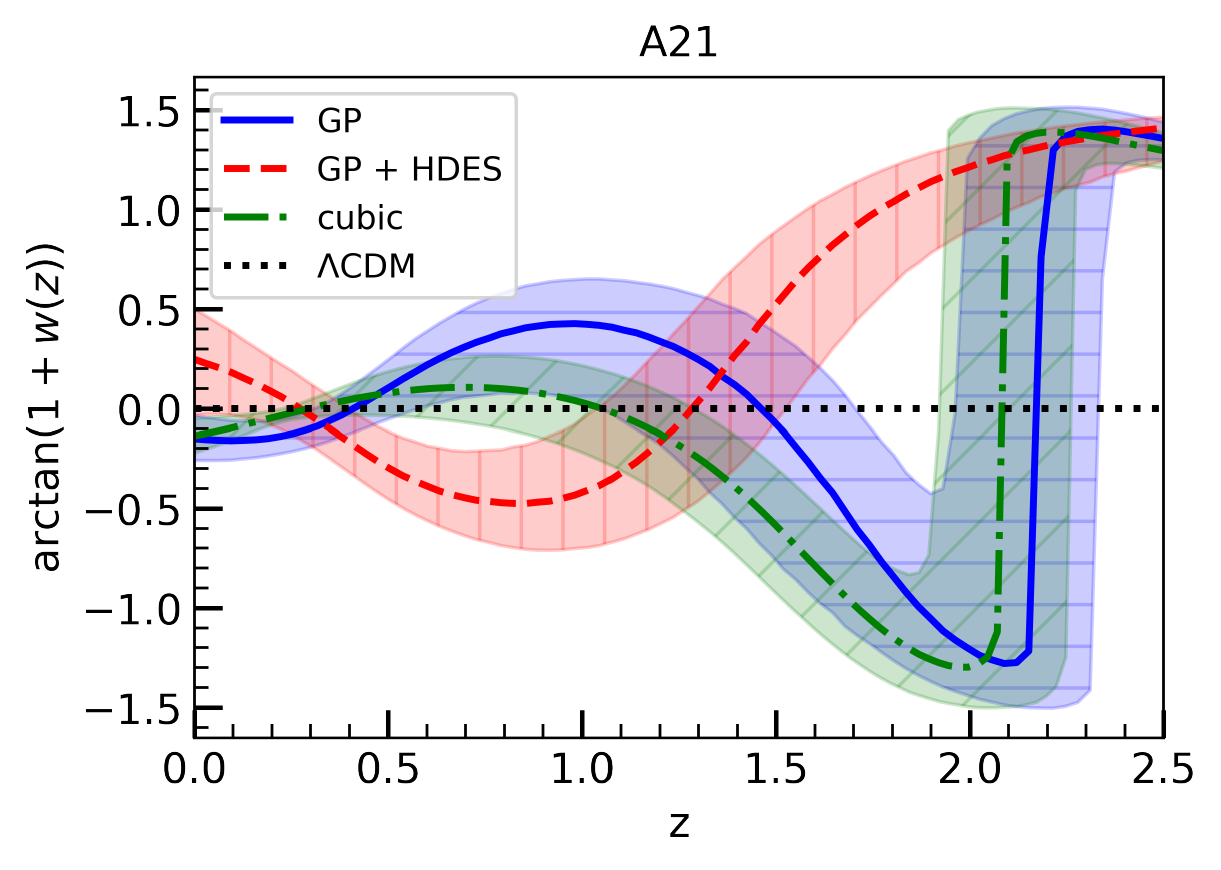}
		}
	\subfigure[ $H_0^{\text{R21}} = 73.04 \pm 1.04$ km s$^{-1}$Mpc$^{-1}$ ]{
		\includegraphics[width = 0.475 \textwidth]{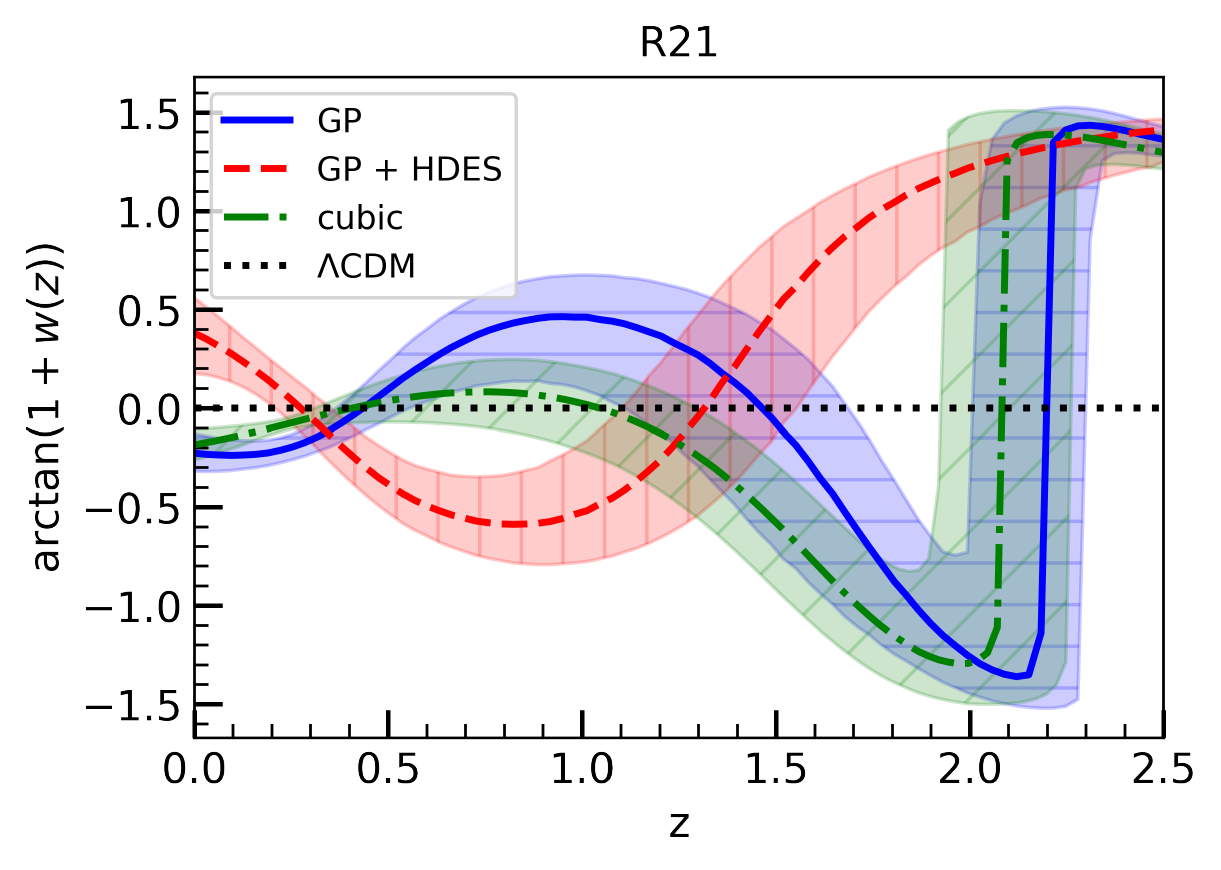}
		}
\caption{The reconstructed compactified DE equation of state for the GP, GP with designer Horndeski, and cubic parametric method using the Hubble data (CC + BAO) for each $H_0$ prior: (a) P18, (b) TRGB, (c) A21, and (d) R21. The colored-hatched regions show the median and the surrounding $34.1\%$ probability mass above and below. Hatches: (GP: ``$-$''), (GP + HDES: ``$|$''), (cubic: ``$/$'').}
\label{fig:awz_rec_wSNE_HDES}
\end{figure}

To get to the point of this theoretical excursion, the DE equation of state in HDES can also be calculated in the following way. In kinetic gravity braiding, this is given by
\begin{equation}
\label{eq:w_de_kgb}
    w_\phi = \dfrac{ - K + \sqrt{2X} \dot{X} G_X }{ K - 2 X \left( K_X + 3 \sqrt{2X} H(X) G_X \right)  }\,.
\end{equation}
Substituting the HDES solution to Eq. (\ref{eq:w_de_kgb}) leads to
\begin{equation}
\label{eq:w_de_hdes}
w_\phi = -1 + \dfrac{ \mathcal{J} \sqrt{2X} \left( H(z)^2 - H_0^2 \Omega_\Lambda \right) }{ 3 H_0^4 \Omega_{m0} \Omega_\Lambda } - \dfrac{ 2 \mathcal{J} \sqrt{2X} (1 + z) H(z) H'(z) }{ 9 H_0^4 \Omega_{m0} \Omega_\Lambda }\,.
\end{equation}
Using Eq. (\ref{eq:w_de_hdes}), we can then see how our reconstructed scalar field potentials can be used to predict the shape of the DE equation of state within this data-driven Horndeski model. We compare the result with our previous reconstructions using GP and the cubic method in Figure \ref{fig:awz_rec_wSNE_HDES}.

There are several interesting results that can be drawn from these plots. First is that within HDES, the DE equation of state turns out to be approximately Gaussian distributed for most redshifts, unlike its theory-agnostic counterparts. Second, the DE equation of state with HDES does not contain a singularity which threatens its convergence for any redshift. Third, and most importantly, the $w(z)$ for HDES can be seen to be almost entirely different in shape from the ones constructed without a Horndeski model. This sounds counter-intuitive taking into consideration that the DE equation of state should represent the macrophysical state of DE, and not whichever microphysical paradigm underlies the observation. This is instead unique in HDES due to how its equations were formulated beginning with a prior functional relation $H(X)$. The same cannot be said for other Horndeski inversion methods such as in quintessence and the tailoring Horndeski model where the exact DE equation of state emerges as in the lone GP reconstruction \cite{Bernardo:2021qhu}. This is the also reason why HDES was singled out in this work. We must also remind that the reconstruction of the HDES potentials and DE equation of state depends on prior values of its internal theory parameters $c_0$, $n$, and $\mathcal{J}$. While there is reason to take these constants to scale with the late Universe, as we did above, different values of these constants are expected to lead to different results.

We emphasize that the GP-assisted HDES results were obtained with the same expansion data considered as the other approaches. The difference is that now this has input from field theory. This naturally leads to incompatibility with the $\Lambda$CDM model (as underscored by the nontrivial scalar potentials in Figure \ref{fig:hdes_potentials}), but it is anchored on background cosmological observations. However, its disagreement with the reconstructed DE EoS from the purely data-driven approaches (as well as $\Lambda$CDM) which is shown in Figure \ref{fig:awz_rec_wSNE_HDES} could potentially be a reason to frown upon it. Of course, what this \textit{does not} mean is that Horndeski gravity (which contains a vast phenomenology) is ruled out since HDES is only one of possible scalar-tensor reconstruction methods \cite{Bernardo:2021qhu}.

To sum up this section, we have strengthened the results of the previous ones by including supernovae observations. We have also constructed the DE equation of state in a GP-assisted Horndeski model. In this way, we were able to see the results of an inherently dynamical DE model from the compromised theory-agnostic ones.

\section{Constraints on the dark energy equation of state}
\label{sec:constraints_de_eos}

We summarize our constraints on the DE equation of state $w_0$ at $z = 0$ in Table \ref{tab:w_de_constraints}. This parameter is useful in constraining DE as $w(z)$ being anything other than $w = -1$ at any time or redshift guarantees DE evolution.

\begin{table}[h!]
\center
\caption{Constraints on the DE equation of state at $z = 0$. $H(Z)$ comprises of the Hubble function measurements from cosmic chronometers and baryon acoustic oscillations. The columns $H_0^\text{P18}$, $H_0^\text{TRGB}$, $H_0^{\text{A21}}$, and $H_0^\text{R21}$ stand for the analysis using the corresponding $H_0$ priors P18, TRGB, A21, and R21. The Planck prior $\Omega_{m0} h^2 = 0.1430 \pm 0.0011$ was considered throughout \cite{Aghanim:2018eyx}. For the designer Horndeski model, $c_0 = H_0^{n + 2}$, $n = 1$, and $\mathcal{J} = H_0$ were considered.}
\resizebox{\textwidth}{!}{
\begin{tabular}{| c | c | c | c | c |}
\hline
\phantom{ $\dfrac{1}{1}$ } \phantom{ $\dfrac{1}{1}$ } & \multicolumn{4}{c}{ $w_0$ } \vline \\
\hline

\phantom{ $\dfrac{1}{1}$ } \textit{Method} + data set \phantom{ $\dfrac{1}{1}$ } & P18 & TRGB & A21 & R21 \\ \hline \hline


\phantom{$\dfrac{1}{1}$} quad + $H(Z)$ \phantom{$\dfrac{1}{1}$} & $-0.91 \pm 0.04$  & $-0.97 \pm 0.06$ & $-1.03 \pm 0.05$ & $-1.09 \pm 0.04$ \\ \hline

\phantom{$\dfrac{1}{1}$} cubic + $H(Z)$ \phantom{$\dfrac{1}{1}$} & $-1.1 \pm 0.1$ & $-1.3 \pm 0.1$ & $-1.4 \pm 0.1$ & $-1.4 \pm 0.1$ \\ \hline

\phantom{$\dfrac{1}{1}$} GP + $H(Z)$ \phantom{$\dfrac{1}{1}$} & $-0.99 \pm 0.09$ & $-1.1 \pm 0.1$ & $-1.2 \pm 0.1$ & $-1.2 \pm 0.1$ \\ \hline \hline


\phantom{$\dfrac{1}{1}$} quad + $H(Z)$ + SNe \phantom{$\dfrac{1}{1}$} & $-0.92 \pm 0.04$  & $-0.96 \pm 0.05$ & $-0.98 \pm 0.05$ & $-1.04 \pm 0.04$ \\ \hline

\phantom{$\dfrac{1}{1}$} cubic + $H(Z)$ + SNe \phantom{$\dfrac{1}{1}$} & $-1.10 \pm 0.09$ & $-1.13 \pm 0.09$ & $-1.14 \pm 0.09$ & $-1.19 \pm 0.08$ \\ \hline

\phantom{$\dfrac{1}{1}$} GP + $H(Z)$ + SNe \phantom{$\dfrac{1}{1}$} & $-0.99 \pm 0.08$ & $-1.1 \pm 0.1$ & $-1.1 \pm 0.1$ & $-1.2 \pm 0.1$ \\ \hline \hline


GP/HDES + $H(Z)$ + SNe & $-1.0 \pm 0.1$ & $-0.9 \pm 0.3$ & $-0.7 \pm 0.3$ & $-0.6 \pm 0.2$ \\ \hline \hline

\phantom{$\dfrac{1}{1}$} $\Lambda$CDM \phantom{$\dfrac{1}{1}$} & \multicolumn{4}{c}{$-1$} \vline \\ \hline

\phantom{$\dfrac{1}{1}$} $w_0$CDM \phantom{$\dfrac{1}{1}$} & \multicolumn{4}{c}{ $-1.03 \pm 0.03$ \cite{Aghanim:2018eyx} } \vline \\ \hline

\phantom{$\dfrac{1}{1}$} $w_0 w_a$CDM \phantom{$\dfrac{1}{1}$} & \multicolumn{4}{c}{ $-0.96 \pm 0.08$ \cite{Aghanim:2018eyx} } \vline \\ \hline
\end{tabular}
}
\label{tab:w_de_constraints}
\end{table}

Several observations can be made. The first one related to the influence of an $H_0$ prior is that $w_0$ tends to an increasingly negative direction for increasing $H_0$ regardless of the method, i.e., $w_0^{\text{R21}} < w_0^{\text{A21}} < w_0^{\text{TRGB}} < w_0^{\text{P18}}$ which is reverse to the average order of the Hubble constant, $H_0^{\text{P18}} < H_0^{\text{TRGB}} < H_0^{\text{A21}} < H_0^{\text{R21}}$. This is with the exception of GP-designer Horndeski (HDES) model which we have also included in Table \ref{tab:w_de_constraints} for completeness. Another observation, this time related to the method, is that the supernovae data has the most effect on the cubic method together with the TRGB, A21, and R21 $H_0$ priors. This enhanced sensitivity to the addition of data may be due the cubic method being arguably the most flexible among the three, having a total of five parameters (including the $H_0$ and $\Omega_{m0}$ priors) to be statistically determined. Obviously, the cubic method has more parameters than the quadratic method, while the GP has only two hyperparameters to be optimized. Lastly, we note that the GP and HDES results are the only ones among all the rows in Table \ref{tab:w_de_constraints} which includes $w_0 = - 1$ within $1\sigma$ in three of four of the $H_0$ priors. In the case of HDES, it can be seen that the trend toward positive $w_0$ for increasing $H_0$ is compensated by larger uncertainties when it comes to $H_0^{\text{TRGB}}$ and $H_0^{\text{A21}}$. The closest to this is the GP in which $w_0 = -1$ is included within $1\sigma$ for P18, TRGB, and A21 $H_0$ priors. In the quadratic method, $w_0 = -1$ is within $1\sigma$ only for the TRGB and A21 $H_0$ priors, while for the cubic method $w_0 = -1$ is within $1\sigma$ only for the P18 $H_0$ prior without supernovae data. The cubic method disfavors a constant $\Lambda$ to support the late-time cosmic acceleration.

The trend of decreasing $w_0$ for increasing $H_0$ was also shown to be the case in $w$CDM and $w_0 w_a$CDM (CPL) models as well as quintessence \cite{Banerjee:2020xcn}. This points to a potential problematic regime for quintessence models as this implies that it always make the Hubble tension worse. Nonetheless, this analysis remains to be extended to broader sectors of scalar-tensor gravity.

\section{Conclusions}
\label{sec:conclusions}

DE remains elusive of a theoretical understanding despite decades since its discovery. Nonetheless, there is optimism that with the abundance of available, and forthcoming, data at various distances \cite{Bargiacchi:2021hdp, DeSimone:2021rus, Fanizza:2021tiv, Fanizza:2021tuh, Virgo:2021bbr} together with progress in fundamental physics, then it should be possible to illuminate this longstanding puzzle. Meanwhile, in this work, we explored the interplay between fundamental physics in cosmology and the late Universe in a theory agnostic approach. In this way, by taking in the features that survive regardless of the choice of the reconstruction method and of the $H_0$ priors (reflective of the current Hubble tension), we were lead to the conclusion that DE evolves, or rather is dynamical.

The use of priors admittedly compromise the notion of model-independence, as we have reiterated a few times in this paper. Model-independence is a tricky concept in the first place and this work does not claim to be faithful to this theme. Priors depend on how they were obtained and what assumptions entered their measurements. Our analysis involving various $H_0$ priors reflecting the Hubble tension and the Planck prior on $\Omega_{m0} h^2$ instead lead to a compromised model-independent approach. We found that even with the Planck priors on the Hubble constant and the matter density, depending on the $\Lambda$CDM model, the parametric and nonparametric approaches employed in this work supported a deviation from standard cosmology through an evolving DE component. We have shown that other $H_0$ priors obtained via the distance-ladder, regardless of their principled differences, also supported the existence of an evolving dark energy component (recall Figures \ref{fig:quad_bestfit_zm2} and \ref{fig:cubic_bestfit_zm2} where the dark energy parameters $x_i$ can be seen to be consistent regardless of $H_0$ choice).

An important observation is that the strongest hints of dark energy evolution come at the high redshifts ($z \sim 2.3$) near the Lyman alpha BAO measurements. These could be more influenced by the systematics and were reportedly in tension with $\Lambda$CDM. So it is reasonable to ask whether the conclusion of this paper holds when these high redshift data points are dropped. Our investigation answers affirmative, but with a weaker evidence for dark energy evolution, or rather a stronger support for the standard cosmological model. But there are other caveats, the first being that the shapes of the reconstructions with the parametric methods and Gaussian processes do not agree. Another is that the reconstructions have much larger uncertainties as expected since the BAO were the most precise among the Hubble function observations. We take these as motivation to include the BAO in the analysis. The stringent data set considered in this work lead to the best agreement between the different reconstruction methods and priors. The departures from a constant dark energy that were obtained in this work reinforces previous reports of the BAO measurements being in tension with the $\Lambda$CDM model \cite{BOSS:2013igd, BOSS:2014hwf, Bautista:2017zgn, Busca36, Bengaly:2021wgc}.

The evolving dark energy feature which is revealed for all the methods and $H_0$ priors considered is the highlight of this paper. If this feature appeared with only either parametric or nonparametric approaches, it may easily questioned as an artefact of the methodology. This time, however, we have shown that despite the stark contrast between how parametric and nonparametric methods treat the data, dark energy evolution continues to be generally supported.

Granted, the methods employed in our paper have their issues, e.g., the parametric methods are often criticized for the unavoidable arbitrariness by which the phenomenological functions can be parametrized \cite{Colgain:2021pmf} while the GP, among other machine learning methods, is known to underestimate uncertainties, among other quirks \cite{Escamilla-Rivera:2021rbe, Colgain:2021ngq, Bernardo:2021mfs}. A natural extension of this work which overcomes this is to incorporate various more approaches \cite{AlbertoVazquez:2012ofj, Zhao:2017cud, Wang:2018fng, Escamilla:2021uoj} and see whether the same features of DE can still be observed.

It should be emphasized as well that dynamical DE does not necessarily imply modified gravity as alluded in Ref. \cite{Wen:2021bsc}. Particularly, it may just so happen that a better fit can be obtained with data even if the underlying physical model is standard cosmology, or unmodified gravity. To make robust conclusions about modified gravity, for example, one can instead extend this work to accommodate observations of redshift space distorsions. Relevant progress in this direction are Refs. \cite{Pogosian:2021mcs, Raveri:2021dbu}.

On the data sets themselves, we should mention that there potentially may be added complications due to the impact of potential systematics in the data, which may have a greater impact for higher redshift data points \cite{BOSS:2013igd, BOSS:2014hwf, Bautista:2017zgn, Busca36}. While the specific amount of tension changes, the main conclusions of our work do not alter drastically when these high redshift points are removed.

We also wanted to draw more attention to the compactified DE equation of state which was introduced in Ref. \cite{Bernardo:2021qhu}. This simple phenomenological tool overcomes challenges in reconstructing the canon DE equation state and allows us to look more deeply into DE evolution, particularly near and beyond the redshift at which DE becomes singular. This can be realized to become even more useful with future distance indicators and cosmological data sets such as gamma ray bursts and standard sirens which probe the Universe at much higher redshifts. Improvements to this are welcome future work.

As a concluding note, we mention that DE not only affects the expansion but also the perturbations that grow on top of this background. It would be interesting to see how the methods employed in this paper perform when dealing with linear observables such as the growth data, among others.

\section*{Acknowledgments}
The authors thank Eoin Colg\'ain for helpful comments on an earlier draft. DG acknowledges support by project ANIDPFCHA/Doctorado Nacional/2019-21191886. JLS would like to acknowledge networking support by the COST Action CA18108 and funding support from Cosmology@MALTA which is supported by the University of Malta.

\appendix

\section{Hubble expansion data}
\label{sec:hubble_data}

We present the Hubble expansion data used in this paper in Table \ref{tab:expansion_data}.

\begin{table}[h!]
    \centering
    \caption{Expansion data from cosmic chronometers (31 points, left column) and baryon acoustic oscillations (26 points, right column).} \resizebox*{!}{\dimexpr\textheight-3\baselineskip\relax}{%
    \begin{tabular}{| c | c | c | c | c | c |} \hline
         \phantom{$\dfrac{1}{1}$} $z$ \phantom{$\dfrac{1}{1}$} & \phantom{$\dfrac{1}{1}$} $H$ [km s$^{-1}$ Mpc$^{-1}$] \phantom{$\dfrac{1}{1}$} & \phantom{$\dfrac{1}{1}$} Ref. \phantom{$\dfrac{1}{1}$} & \phantom{$\dfrac{1}{1}$} $z$ \phantom{$\dfrac{1}{1}$} & \phantom{$\dfrac{1}{1}$} $H$ [km s$^{-1}$ Mpc$^{-1}$] \phantom{$\dfrac{1}{1}$} & \phantom{$\dfrac{1}{1}$} Ref. \phantom{$\dfrac{1}{1}$} \\ \hline \hline
         $0.07$ & $69 \pm 19.6$ & \cite{2014RAA....14.1221Z} & $0.24$ & $79.69 \pm 2.99$ & \cite{Gaz34}\\ \hline
         $0.09$ & $69 \pm 12$ & \cite{2010JCAP...02..008S} & $0.3$ & $81.7 \pm 6.22$ & \cite{Oka37}\\ \hline
         $0.12$ & $68.6 \pm 26.2$ & \cite{2014RAA....14.1221Z} & $0.31$ & $78.18 \pm 4.74$ & \cite{Wang33}\\ \hline
         $0.17$ & $83 \pm 8$ & \cite{2010JCAP...02..008S} & $0.34$ & $83.8 \pm 3.66$ & \cite{Gaz34}\\ \hline
         $0.1791$ & $75 \pm 4$ & \cite{2012JCAP...08..006M} & $0.35$ & $82.7 \pm 9.1$ & \cite{Chuang28}\\ \hline
         $0.1993$ & $75 \pm 5$ & \cite{2012JCAP...08..006M} & $0.36$ & $79.94 \pm 3.38$ & \cite{Wang33}\\ \hline
         $0.2$ & $72.9 \pm 29.6$ & \cite{2014RAA....14.1221Z} & $0.38$ & $81.5 \pm 1.9$ & \cite{Alam38}\\ \hline
         $0.27$ & $77 \pm 14$ & \cite{2010JCAP...02..008S} & $0.4$ & $82.04 \pm 2.03$ & \cite{Wang33}\\ \hline
         $0.28$ & $88.8 \pm 36.6$ & \cite{2014RAA....14.1221Z} & $0.43$ & $86.45 \pm 3.97$ & \cite{Gaz34}\\ \hline
         $0.3519$ & $83 \pm 14$ & \cite{2012JCAP...08..006M} & $0.44$ & $84.81 \pm 1.83$ &\cite{Wang33}\\ \hline
         $0.3802$ & $83\pm 13.5$ & \cite{Moresco:2016mzx} & $0.44$ & $82.6 \pm 7.8$ &\cite{2012MNRAS.425..405B}\\ \hline
         $0.4$ & $95 \pm 17$ & \cite{2010JCAP...02..008S}  & $0.48$ & $87.79 \pm 2.03$ & \cite{Wang33}\\ \hline
         $0.4004$ & $77 \pm 10.2$ & \cite{Moresco:2016mzx} & $0.51$ & $90.4 \pm 1.9$ & \cite{Alam38}\\ \hline
         $0.4247$ & $87.1 \pm 11.2$ & \cite{Moresco:2016mzx} & $0.52$ & $94.35 \pm 2.64$ & \cite{Wang33}\\ \hline
         $0.4497$ & $92.8 \pm 12.9$ & \cite{Moresco:2016mzx}  & $0.56$ & $93.34 \pm 2.3$ & \cite{Wang33}\\ \hline
         $0.47$ & $89 \pm 34$ & \cite{Ratsimbazafy:2017vga} & $0.57$ & $87.6 \pm 7.8$ & \cite{Chuang:2013hya}\\ \hline
         $0.4783$ & $80.9 \pm 9$ & \cite{Moresco:2016mzx}& $0.57$ & $96.8 \pm 3.4$ & \cite{Anderson32}\\ \hline
         $0.48$ & $97 \pm 62$ & \cite{2010JCAP...02..008S} & $0.59$ & $98.48 \pm 3.18$ & \cite{Wang33}\\ \hline
         $0.5929$ & $104 \pm 13$ & \cite{2012JCAP...08..006M}& $0.6$ & $87.9 \pm 6.1$ & \cite{2012MNRAS.425..405B}\\ \hline
         $0.6797$ & $92 \pm 8$ & \cite{2012JCAP...08..006M} & $0.61$ & $97.3 \pm 2.1$ & \cite{Alam38}\\ \hline
         $0.7812$ & $105 \pm 12$ & \cite{2012JCAP...08..006M} & $0.64$ & $98.82 \pm 2.98$ & \cite{Wang33}\\ \hline
         $0.8754$ & $125 \pm 17$ & \cite{2012JCAP...08..006M} & $0.73$ & $97.3 \pm 7.0$ & \cite{2012MNRAS.425..405B}\\ \hline
         $0.88$ & $90 \pm 40$ & \cite{2010JCAP...02..008S} & $2.3$ & $224 \pm 8.6$ & \cite{Busca36} \\ \hline
         $0.9$ & $117 \pm 23$ & \cite{2010JCAP...02..008S} & $2.33$ & $224 \pm 8$ & \cite{Bautista:2017zgn}\\ \hline
         $1.037$ & $154 \pm 20$ & \cite{2012JCAP...08..006M}& $2.34$ & $222 \pm 8.5$ & \cite{BOSS:2014hwf}\\ \hline
         $1.3$ & $168 \pm 17$ & \cite{2010JCAP...02..008S} & $2.36$ & $226 \pm 9.3$ & \cite{BOSS:2013igd}\\ \hline
         $1.363$ & $160 \pm 33.6$ & \cite{Moresco:2015cya} \\ \cline{1-3}
         $1.43$ & $177 \pm 18$ & \cite{2010JCAP...02..008S} \\ \cline{1-3}
         $1.53$ & $140 \pm 14$ & \cite{2010JCAP...02..008S} \\ \cline{1-3}
         $1.75$ & $202 \pm 40$ & \cite{2010JCAP...02..008S} \\ \cline{1-3}
         $1.965$ & $186.5 \pm 50.4$ & \cite{Moresco:2015cya} \\ \cline{1-3}
    \end{tabular} }
    \label{tab:expansion_data}
\end{table}

\section{Marginalized posteriors for the matter density and Hubble constant}
\label{sec:Om0h2}

\begin{figure}[h!]
\center
	\subfigure[ quad : best fit $\Omega_{m0}h^2 = 0.143 \pm 0.001$ ]{
		\includegraphics[width = 0.475 \textwidth]{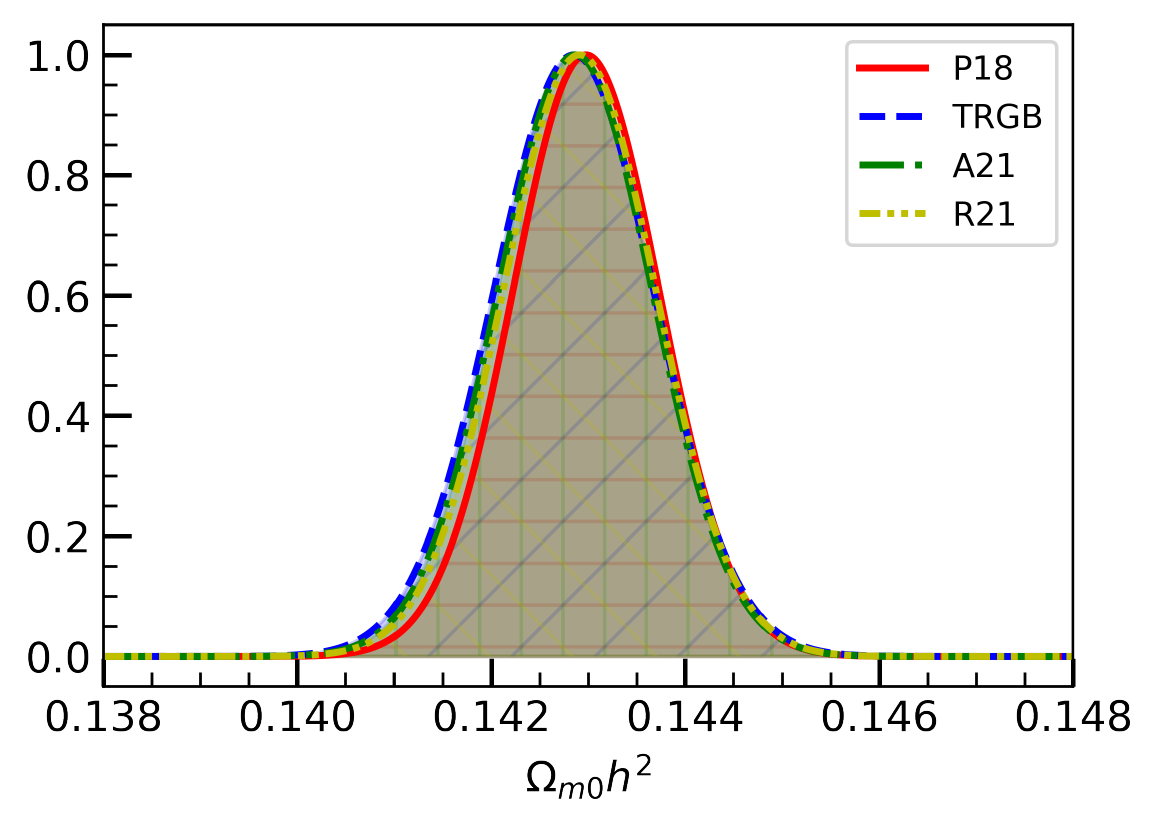}
		}
	\subfigure[ cubic : best fit $\Omega_{m0}h^2 = 0.143 \pm 0.001$ ]{
		\includegraphics[width = 0.475 \textwidth]{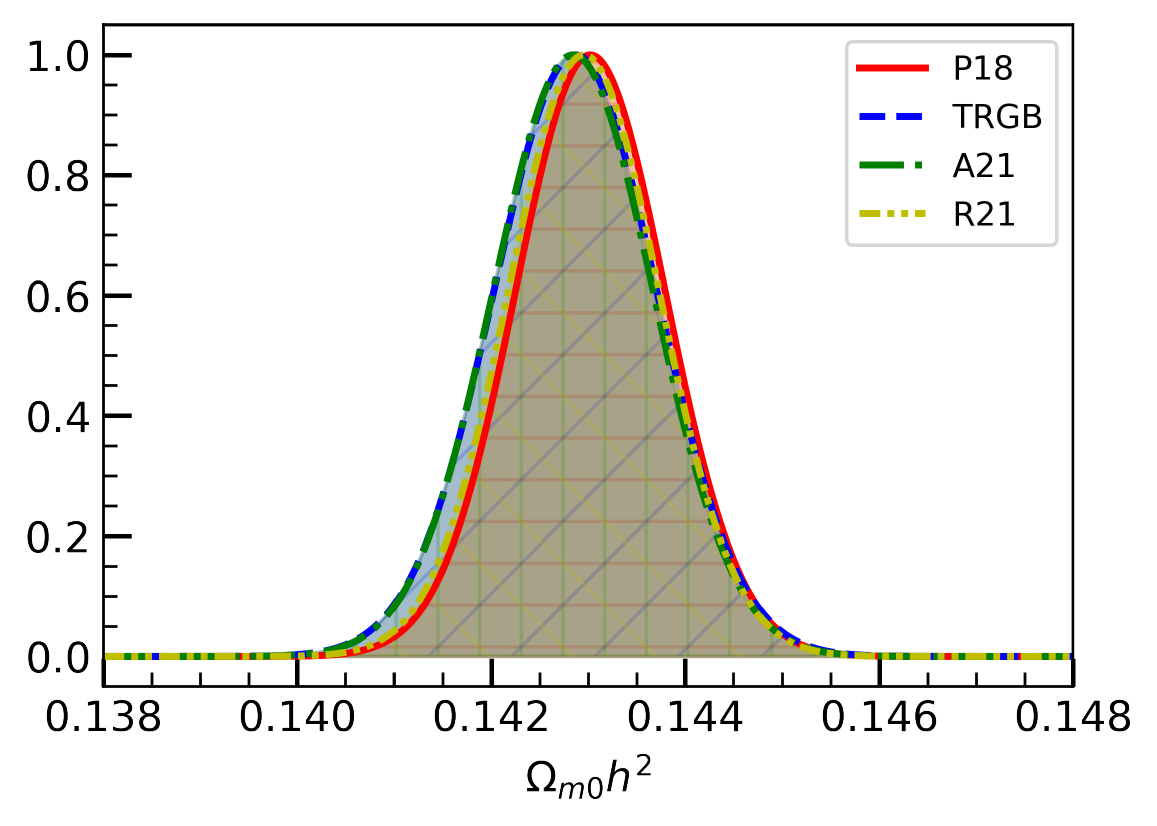}
		}
\caption{Marginalized, one-dimensional posteriors of $\Omega_{m0} h^2$ obtained using the quadratic (Figure \ref{fig:quad_bestfit_zm2}) and cubic (Figure \ref{fig:cubic_bestfit_zm2}) methods.}
\label{fig:Om0h2}
\end{figure}

We present the marginalized, one-dimensional posteriors for the combination $\Omega_{m0} h^2$ of the matter density $\Omega_{m0}$ and the Hubble constant $H_0$. This is obtained by the full covariance matrix coming from the Bayesian analysis in the parametric methods. The results are presented in Figure \ref{fig:Om0h2}.

In all cases, the best fit value $\Omega_{m0}h^2 = 0.143 \pm 0.001$ was obtained. This is of course confirming that the Planck prior on $\Omega_{m0} h^2$ was completely respected in the analysis, and supporting the behavior of the Hubble constant and the matter density presented in Figures \ref{fig:quad_bestfit_zm2} and \ref{fig:cubic_bestfit_zm2}.

\section{A dark energy diagnostic function}
\label{sec:diagnostic}

We present reconstructions of a diagnostic function $X'(z)$ which probes further the redshift dependence of $\Lambda$, i.e., $X'(z) = 0$ for $\Lambda$CDM. This can be obtained within the parametric analysis by differentiating Eqs. (\ref{inter1}) and (\ref{eq:cubic}), and then using the resulting expression together with the best fit values and associated covariances. On the other hand, by means of the GP, one can reconstruct $X'(z)$ by differentiating the Friedmann constraint to obtain a functional $X'\left[ H(z), H'(z) \right]$ where $H(z)$, the first derivative $H'(z)$, and their covariance $C\left( H(z), H'(z) \right)$ are produced by the usual optimization of the GP marginal likelihood.

\begin{figure}[h!]
\center
	\subfigure[ $H_0^{\text{P18}} = 67.4 \pm 0.5$ km s$^{-1}$Mpc$^{-1}$ ]{
		\includegraphics[width = 0.475 \textwidth]{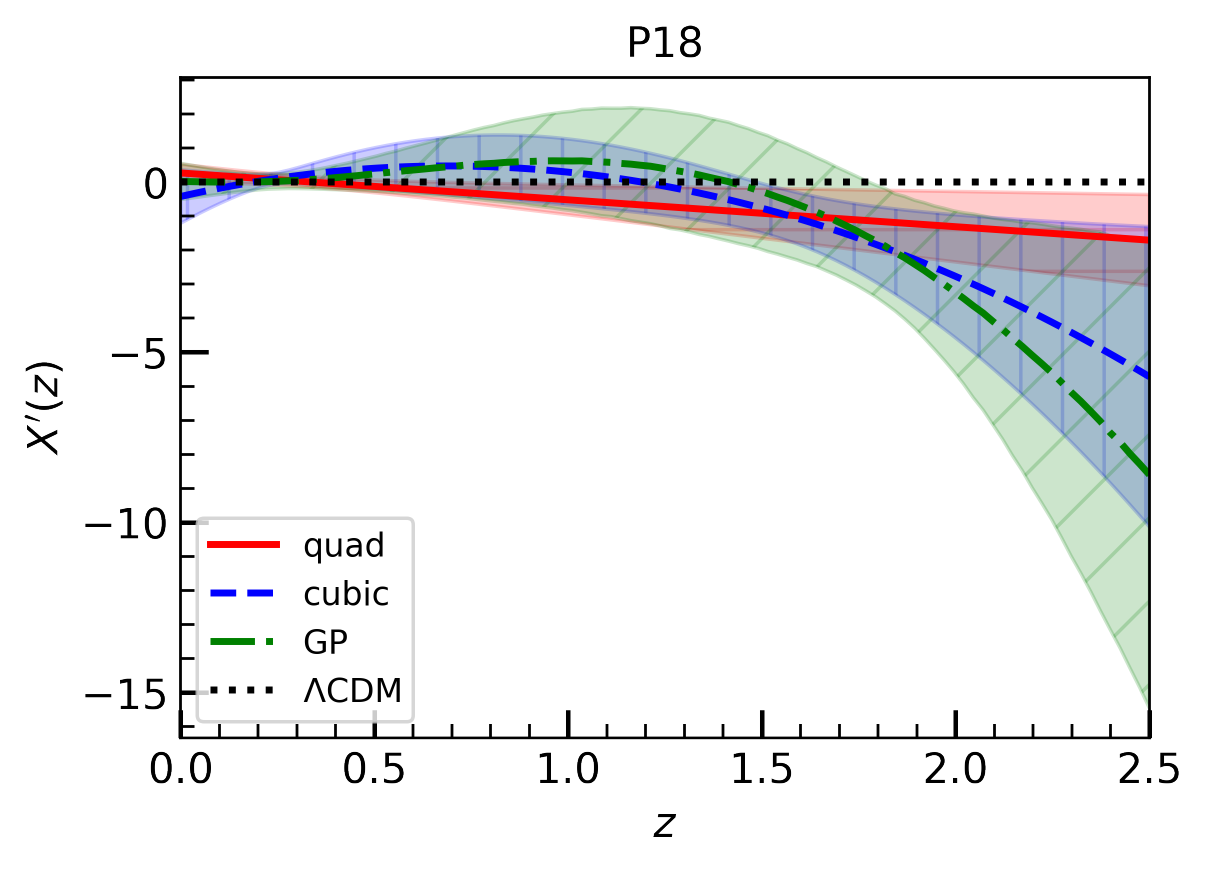}
		}
	\subfigure[ $H_0^{\text{TRGB}} = 69.8 \pm 1.9$ km s$^{-1}$Mpc$^{-1}$ ]{
		\includegraphics[width = 0.475 \textwidth]{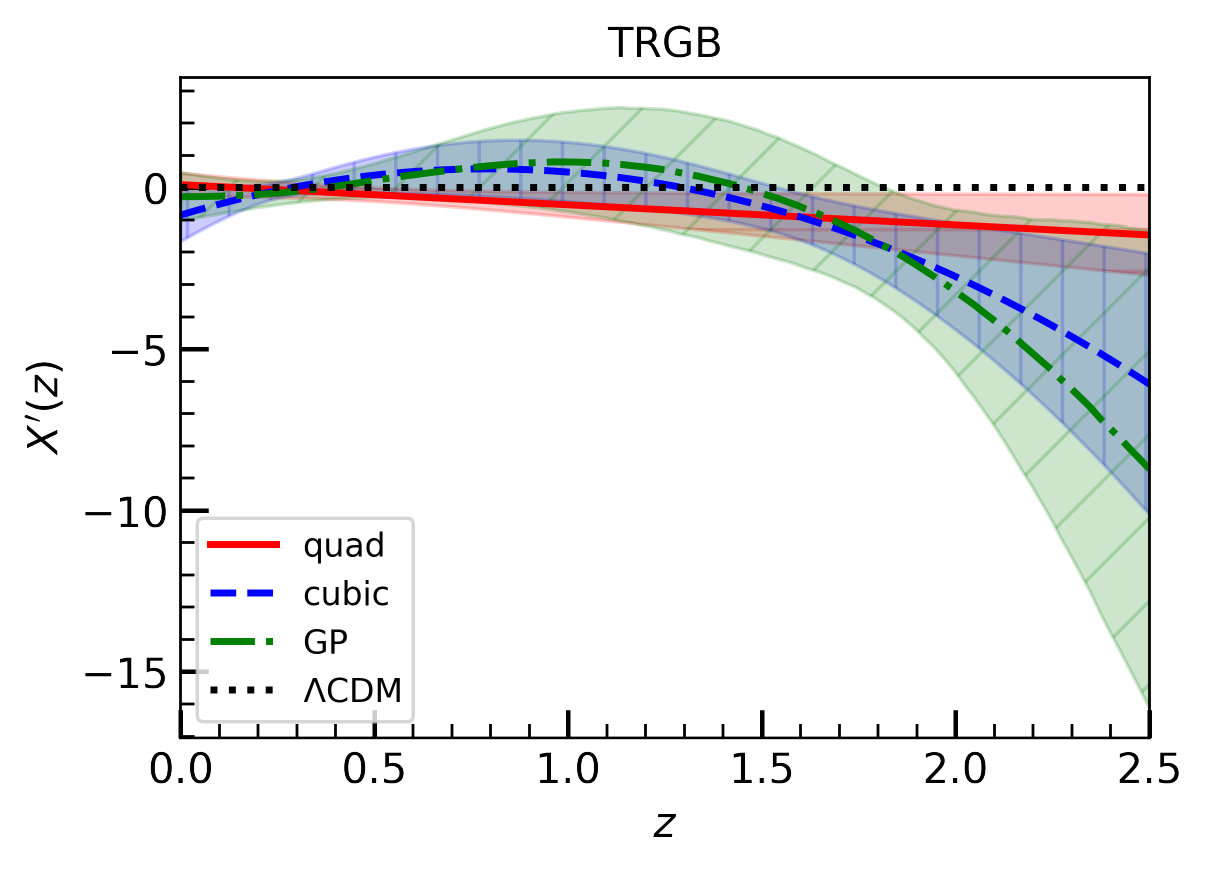}
		}
	\subfigure[ $H_0^{\text{R21}} = 71.5 \pm 1.8$ km s$^{-1}$Mpc$^{-1}$ ]{
		\includegraphics[width = 0.475 \textwidth]{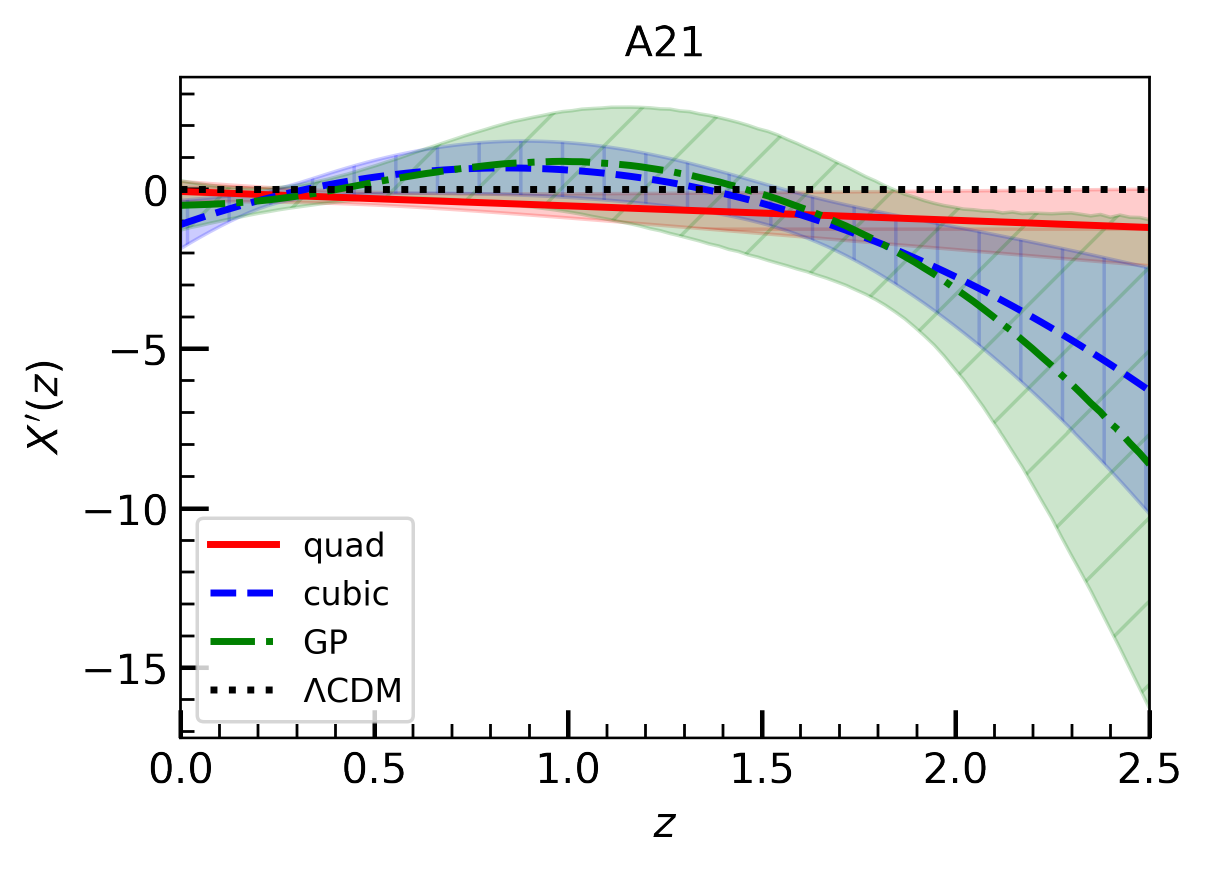}
		}
	\subfigure[ $H_0^{\text{R21}} = 73.04 \pm 1.04$ km s$^{-1}$Mpc$^{-1}$ ]{
		\includegraphics[width = 0.475 \textwidth]{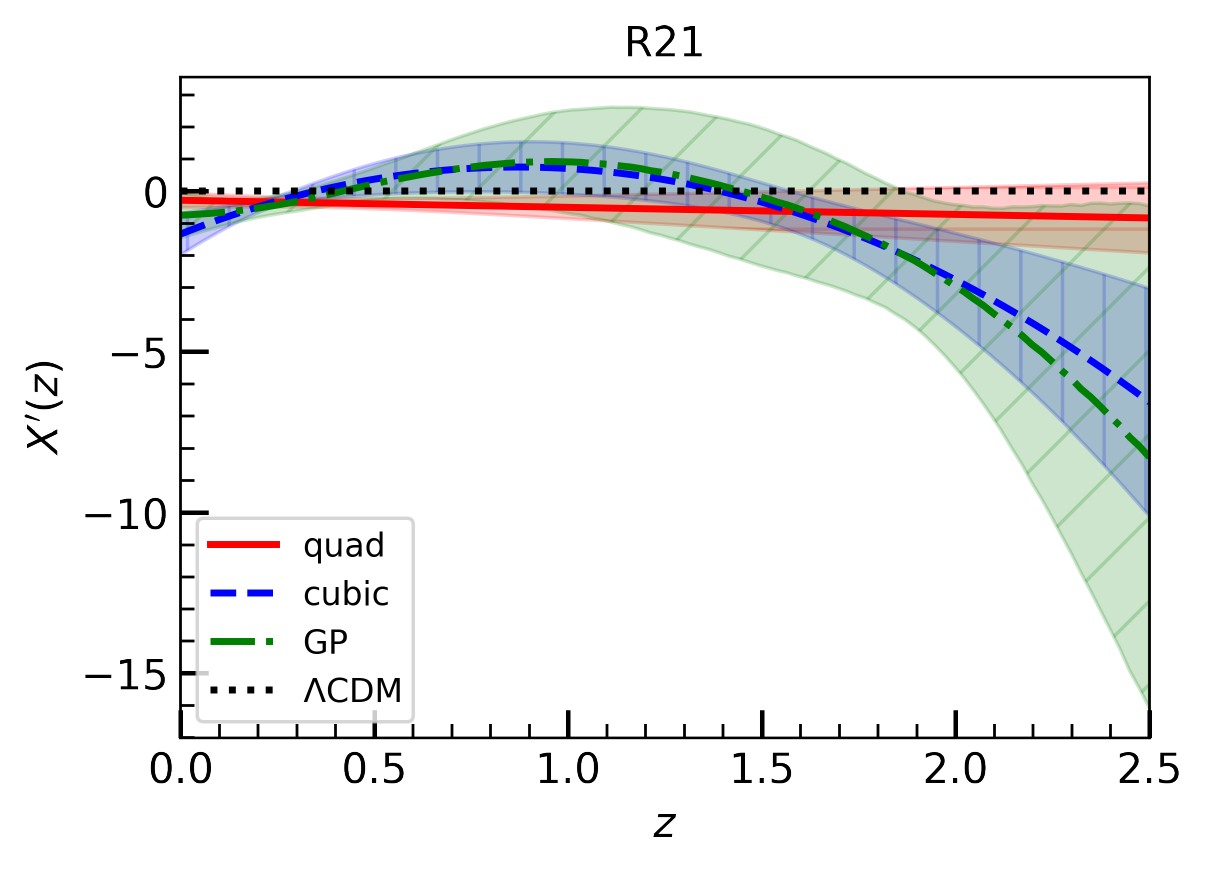}
		}
\caption{The $X'(z)$ diagnostic function per method derived from the base Hubble data (CC + BAO) for each $H_0$ prior: (a) P18, (b) TRGB, (c) A21, and (d) R21. Legends: ``quad'' and ``cubic'' stands for the quadratic and cubic parametrized DE, respectively; ``GP'' for the Gaussian processes. The colored and hatched regions show the $2\sigma$ confidence interval of the reconstructions. Hatches: (quad: ``$-$''), (cubic: ``$|$''), (GP: ``$/$'').}
\label{fig:Xpz_rec_per_method}
\end{figure}

The $X'(z)$ diagnostic obtained with the base Hubble data (CC + BAO) is shown in Figure \ref{fig:Xpz_rec_per_method}. The $\Lambda$CDM curves are shown as the horizontal dotted line at $X'= 0$.

A careful inspection shows that the quadratic approach excludes the $\Lambda$CDM model (at $2\sigma$) for the majority of the redshifts. The deviation also remains consistent throughout regardless of the $H_0$ priors, but subtle changes could be attributed to the use of the Planck matter fraction prior. The GP and cubic approach posteriors generally are in agreement in the overall shape throughout as can be seen. A notable difference is that at very low redshifts ($z \ll 1$), the cubic method nearly excludes $\Lambda$CDM at $2\sigma$, and in the case of the higher ones ($z = O(1)$), the $\Lambda$CDM limit starts to be excluded from within the $2\sigma$ posterior at a lower redshift in the cubic method than in the GP. The trend of increasing deviation from $\Lambda$CDM with increasing $H_0$ priors can also be numerically inspected. Echoing the theme of the present paper, the observation to takeaway from this is that regardless of the methodology and the $H_0$ prior, there is a clear deviation from the standard model that is supported by the present Hubble data sets. As will be shown, this conclusion by using the $X'(z)$ diagnostic holds even with the inclusion of the supernovae in the analysis.

\begin{figure}[h!]
\center
	\subfigure[ $H_0^{\text{P18}} = 67.4 \pm 0.5$ km s$^{-1}$Mpc$^{-1}$ ]{
		\includegraphics[width = 0.475 \textwidth]{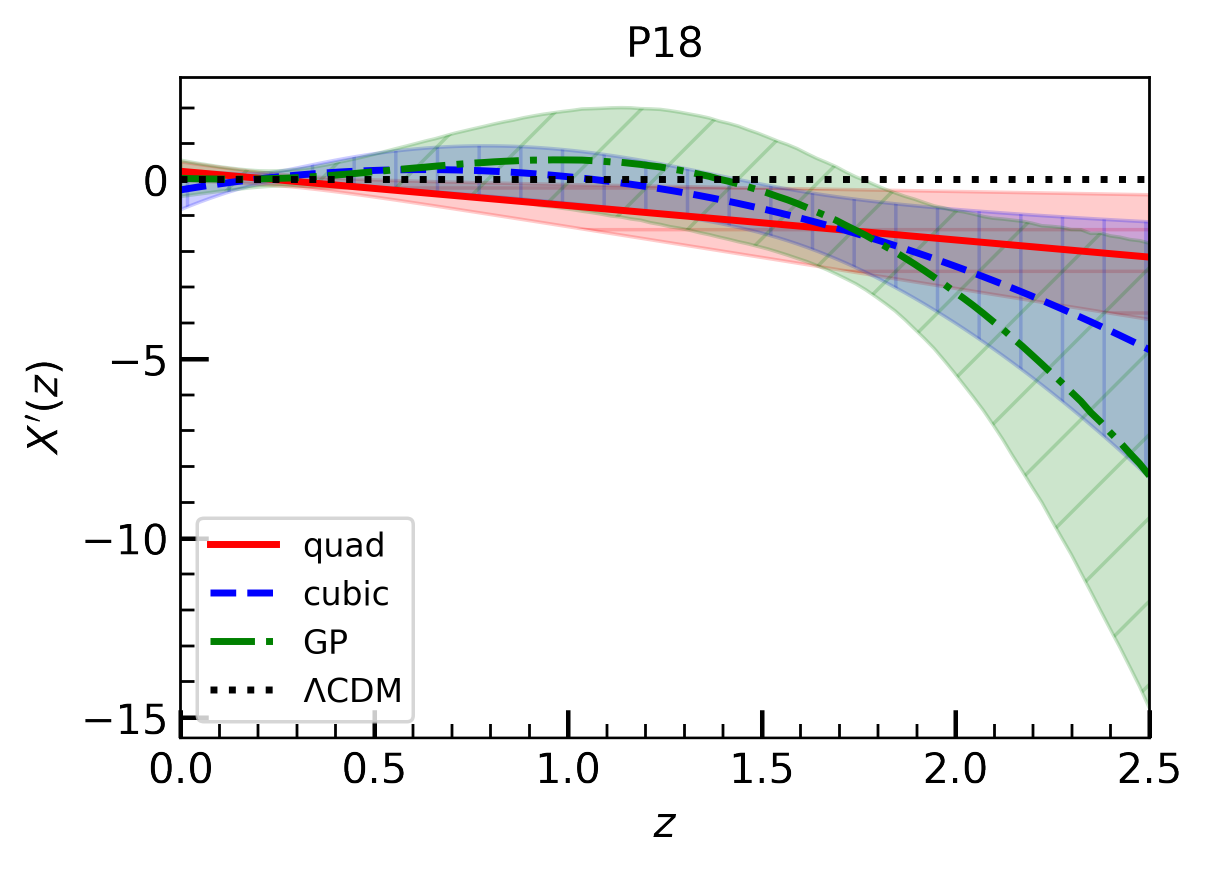}
		}
	\subfigure[ $H_0^{\text{TRGB}} = 69.8 \pm 1.9$ km s$^{-1}$Mpc$^{-1}$ ]{
		\includegraphics[width = 0.475 \textwidth]{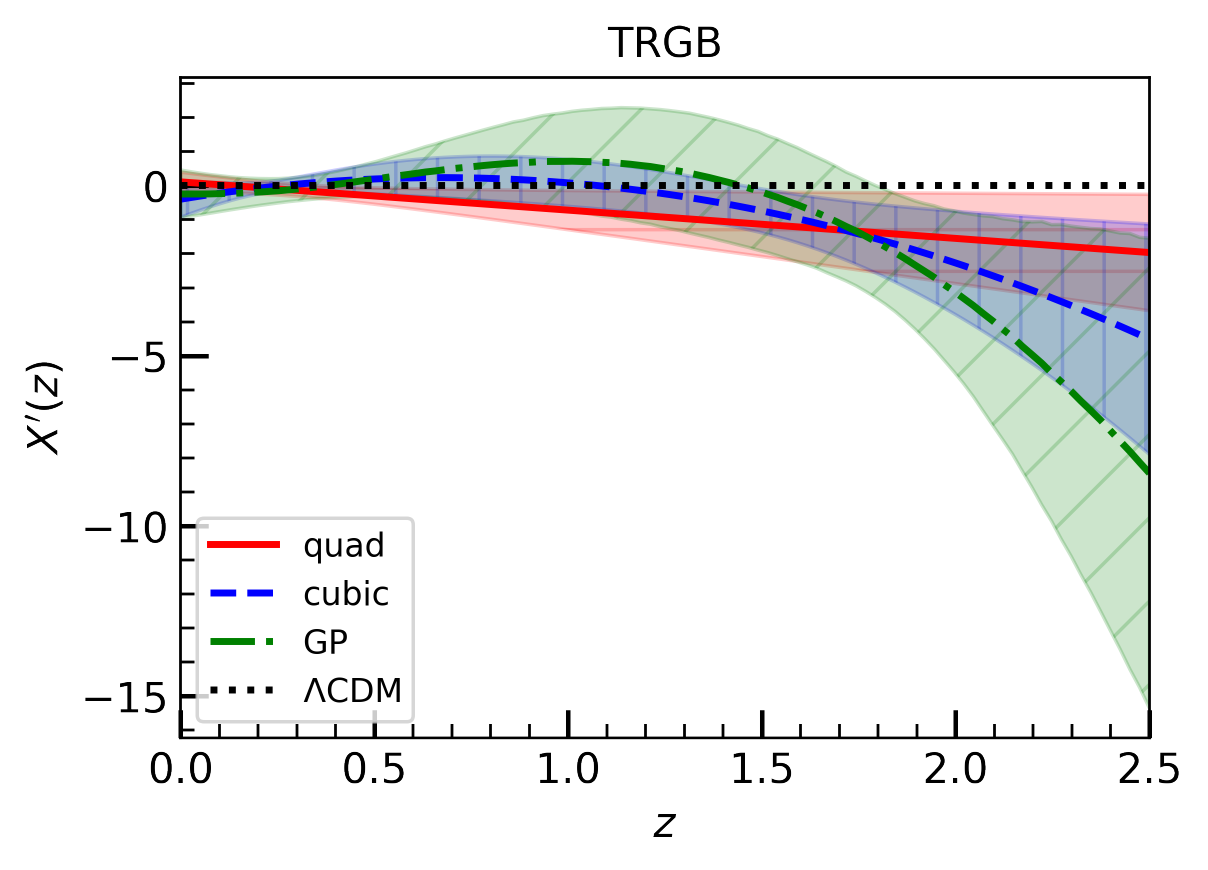}
		}
	\subfigure[ $H_0^{\text{A21}} = 71.5 \pm 1.8$ km s$^{-1}$Mpc$^{-1}$ ]{
		\includegraphics[width = 0.475 \textwidth]{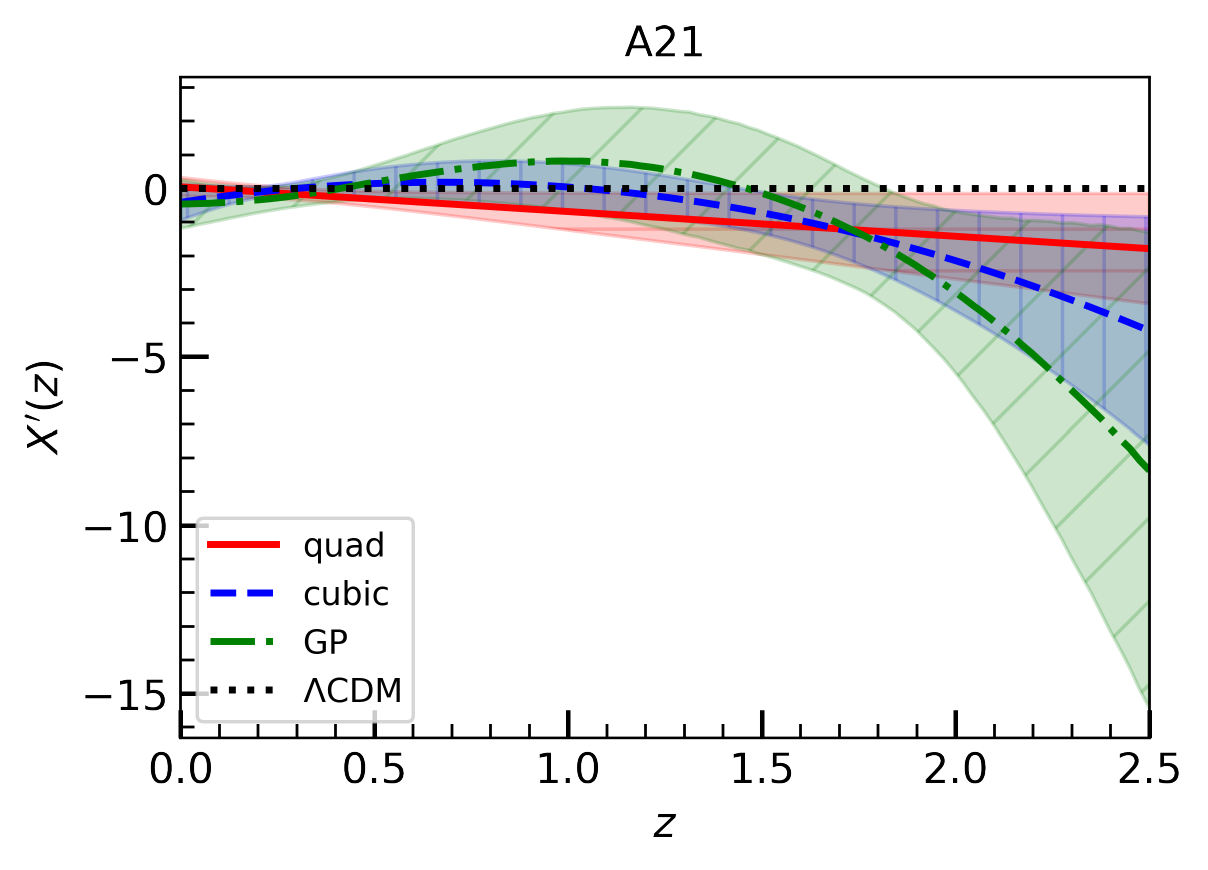}
		}
	\subfigure[ $H_0^{\text{R21}} = 73.04 \pm 1.04$ km s$^{-1}$Mpc$^{-1}$ ]{
		\includegraphics[width = 0.475 \textwidth]{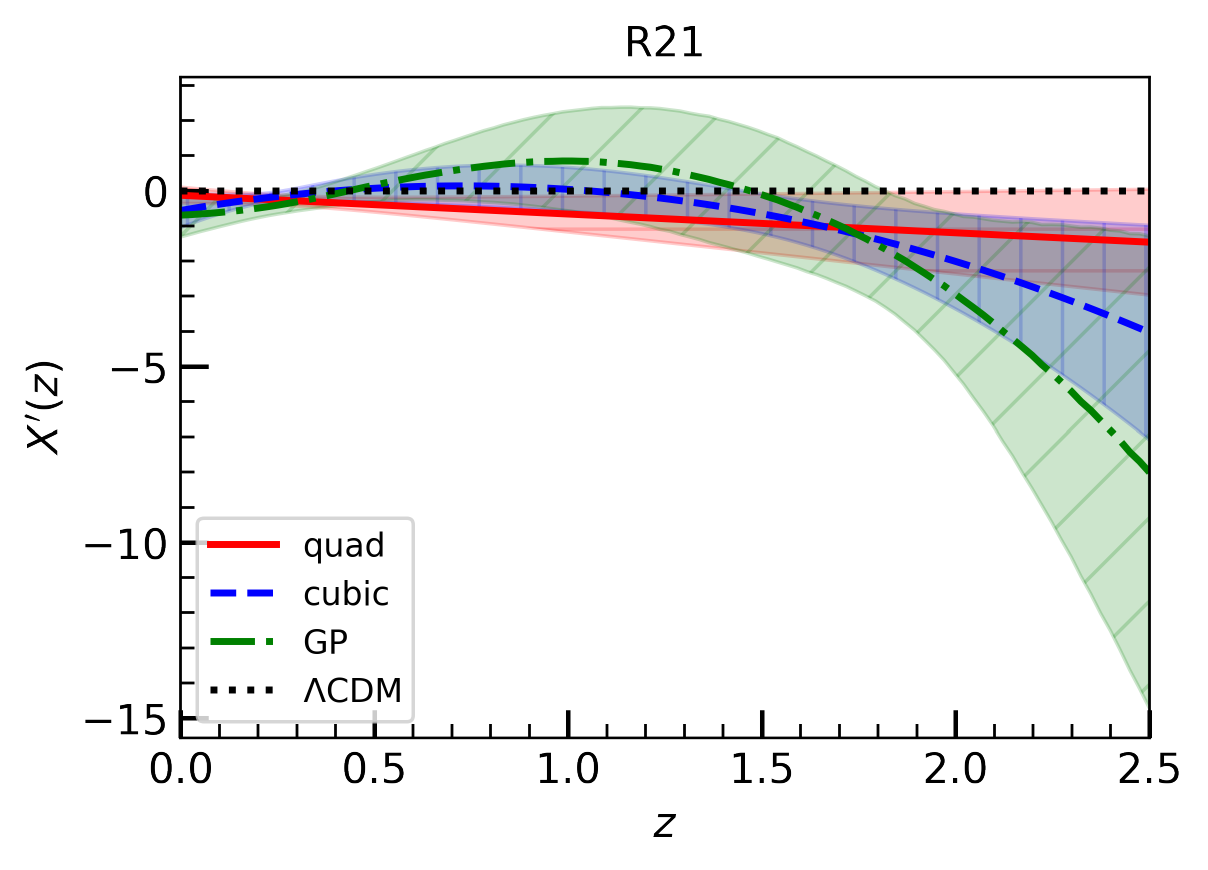}
		}
\caption{The $X'(z)$ diagnostic function per method derived from the base Hubble data (CC + BAO) and SNe for each $H_0$ prior: (a) P18, (b) TRGB, (c) A21, and (d) R21. Legends: ``quad'' and ``cubic'' stands for the quadratic and cubic parametrized DE, respectively; ``GP'' for the Gaussian processes. The colored and hatched regions show the $2\sigma$ confidence interval of the reconstructions. Hatches: (quad: ``$-$''), (cubic: ``$|$''), (GP: ``$/$'').}
\label{fig:Xpz_rec_per_method_wSNe}
\end{figure}

The $X'(z)$ reconstructions with the base Hubble data and supernovae observations from Pantheon are shown in Figure \ref{fig:Xpz_rec_per_method_wSNe} with the $\Lambda$CDM expectation appearing as the horizontal dotted line.

Visually, the results look identical to the ones constructed with only the base Hubble data. The subtle differences can in fact be observed only at the extreme redshifts in the figures and by numerical inspection. For example, with the supernovae, the cubic method can be seen to find itself in better agreement with the $\Lambda$CDM model at $z \ll 1$. Outside these fine details and based on the similarities with previous plot, this supports a departure from the standard model that is anchored on cosmological data. 



\end{document}